\documentclass[%
 reprint,
 superscriptaddress,
nofootinbib,
 amsmath,amssymb,
 aps, 
 prl,
]{revtex4-2}

\usepackage{graphicx}
\usepackage{dcolumn}
\usepackage{bm}
\usepackage{hyperref}
\hypersetup{
    colorlinks=true,
    citecolor=blue,
    linkcolor=blue,
    filecolor=blue,  
    urlcolor=blue,
    pdftitle={Position measurement and the nonlinear regime of cavity quantum optomechanics},
    pdfpagemode=FullScreen,
    }
\usepackage[mathlines]{lineno}
\usepackage{color}
\usepackage{physics}
\usepackage{dsfont}
\usepackage{mathtools}
\usepackage{lipsum}
\usepackage[top=2cm, bottom=2cm, left=1.5cm, right=1.5cm]{geometry}
\usepackage{hyperref}
\usepackage[normalem]{ulem} 
\usepackage[dvipsnames]{xcolor} 

\newcommand{\rme}{{e}}
\newcommand{\rmi}{{i}}
\newcommand{\rmd}{{d}}
\newcommand{\Pl}{P_{{l}}}
\newcommand{\Xl}{X_{{l}}}
\newcommand{\nl}{n_{{l}}}
\newcommand{\cl}{c_{{l}}}
\newcommand{\kb}{k_{{b}}}
\newcommand{\kl}{k_{{l}}}
\newcommand{\Xm}{X_{{m}}}
\newcommand{\Pm}{P_{{m}}}
\newcommand{\Ym}{Y_{{m}}}
\newcommand{\Xmr}{\tilde{X}_{{m}}}
\newcommand{\Pmr}{\tilde{P}_{{m}}}

\newcommand{\wm}{\omega_{{m}}}
\newcommand{\wc}{\omega_{{c}}}
\newcommand{\wl}{\omega_{{l}}}
\newcommand{\ain}{a_{{in}}}
\newcommand{\aout}{a_{{out}}}

\newcommand{\Xa}{X_{\alpha}}
\newcommand{\Np}{N_{{p}}}
\newcommand{\rhor}{\tilde{\rho}}
\newcommand{\rhoin}{\rho_{{i}}}
\newcommand{\rhof}{\rho_{{f}}}
\newcommand{\rhol}{\rho_{{l}}}
\newcommand{\Ph}{\mathcal{P}}
\newcommand{\sigmaf}{\sigma_{f}}
\newcommand{\Pin}{P_{i}}
\newcommand{\Pf}{P_{f}}
\newcommand{\Ubs}{B}
\newcommand{\rv}{\expval{\mathbf{r}}}
\newcommand{\A}{f_{R}}
\newcommand{\B}{f_{I}}

\begin{document}    

\preprint{APS/123-QED}

\title{Cavity quantum optomechanical nonlinearities \\ and position measurement beyond the breakdown of the linearized approximation}

\author{J.~Clarke}
 \email{jack.clarke@imperial.ac.uk}
 \affiliation{QOLS, Blackett Laboratory, Imperial College London, London SW7 2BW, United Kingdom}
\author{P.~Neveu}
\affiliation{Center for Nanophotonics, AMOLF, Science Park 104, 1098 XG Amsterdam, The Netherlands}%
\author{K.~E.~Khosla}
 \affiliation{QOLS, Blackett Laboratory, Imperial College London, London SW7 2BW, United Kingdom}
\author{E.~Verhagen}
\affiliation{Center for Nanophotonics, AMOLF, Science Park 104, 1098 XG Amsterdam, The Netherlands}%
\author{M.~R.~Vanner}
\email{m.vanner@imperial.ac.uk}
\homepage{www.qmeas.net}
\affiliation{QOLS, Blackett Laboratory, Imperial College London, London SW7 2BW, United Kingdom}

\date{\today}

\begin{abstract}
Several optomechanics experiments are now entering the highly sought nonlinear regime where optomechanical interactions are large even for low light levels. Within this regime, new quantum phenomena and improved performance may be achieved, however, a corresponding theoretical formalism of cavity quantum optomechanics that captures the nonlinearities of both the radiation-pressure interaction and the cavity response is needed to unlock these capabilities. Here, we develop such a nonlinear cavity quantum optomechanical framework, which we then utilize to propose how position measurement can be performed beyond the breakdown of the linearized approximation. Our proposal utilizes optical general-dyne detection, ranging from single to dual homodyne, to obtain mechanical position information imprinted onto both the optical amplitude and phase quadratures and enables both pulsed and continuous modes of operation. These cavity optomechanical nonlinearities are now being confronted in a growing number of experiments, and our framework will allow a range of advances to be made in e.g. quantum metrology, explorations of the standard quantum limit, and quantum measurement and control.
\end{abstract}

\keywords{Suggested keywords}
\maketitle

\textit{Introduction.}---Cavity quantum optomechanics utilizes the cavity-enhanced radiation-pressure interaction to enable precision control of mechanical degrees of freedom. Significant progress has been made in the linearized regime where, for large coherent drives, mechanical displacements give rise to small optical phase shifts. However, there is significant current experimental drive to increase single-photon coupling rates, which necessarily gives rise to the nonlinear response of the cavity and the breakdown of the linearized approximation. Importantly, such behaviour occurs well before the strong single-photon-coupling regime where qualitatively different phenomena may be explored~\cite{kronwald2013optomechanically, borkje2013signatures, lemonde2013nonlinear}. Crucially, due to the nonlinearities of radiation pressure and the cavity response, mechanical displacements give rise to significant rotations to the optical field in phase space and thus the mechanical signal is transduced onto both the phase and amplitude quadratures. This radiation-pressure nonlinearity allows for selective position-squared measurements~\cite{vanner2011selective}, the generation of macroscopic superposition states~\cite{bose1999scheme,ringbauer2018generation,clarke2018growing}, non-Gaussian entanglement~\cite{marshall2003towards,vacanti2008optomechanical,akram2013entangled,kanari2021two}, and even optomechanical photon blockade for strong single-photon coupling~\cite{rabl2011photon,nunnenkamp2011single}. Recent experimental progress has seen rapid improvements in coupling rates, with cold-atom implementations reaching the strong single-photon coupling regime~\cite{brennecke2008cavity,purdy2010tunable} and, notably even well within the weak single-photon coupling regime, optical rotations have now been observed, which cannot be described by the linearized approximation~\cite{brawley2016nonlinear,  leijssen2017nonlinear, fedorov2020thermal}.

Closely connected to the radiation-pressure interaction, optomechanical position measurement is a critical component for many research areas ranging from mechanical quantum state engineering to tests of fundamental physics. Prominent example position-measurement applications include ultrasensitive accelerometry~\cite{krause2012high,guzman2014high}, yoctogram-resolution mass sensing~\cite{chaste2012nanomechanical}, zeptonewton weak-force sensing~\cite{ranjit2016zeptonewton}, single-spin detection~\cite{rugar2004single}, and even biomedical sensing~\cite{longo2013rapid}. The precision enabled by cavity optomechanical position measurement also provides a valuable tool for fundamental physics experiments such as gravitational wave detection~\cite{abbott2016observation}, and searches for new physics~\cite{moore2021searching} including dark matter~\cite{carney2021mechanical} and non-Newtonian gravity~\cite{geraci2008improved}.

Owing to the interplay between radiation-pressure back-action and optical quantum noise, optomechanical position measurement is a rich field of study. Prominently, this interplay leads to the standard quantum limit (SQL)~\cite{braginsky1995quantum}, which describes restrictions on position measurement sensitivity and is now well understood experimentally~\cite{murch2008observation, purdy2013observation, cripe2019measurement}. Furthermore, within the SQL, weak continuous position measurement~\cite{doherty1999feedback} combined with feedback~\cite{mancini1998optomechanical,cohadon1999cooling} can be utilized to cool mechanical oscillators towards their ground state~\cite{wilson2015measurement,rossi2018measurement}. To go beyond the SQL, back-action evading (BAE) measurements, such as two-tone drive~\cite{braginsky1980quantum,clerk2008back,suh2014mechanically,shomroni2019optical} or pulses much shorter than the mechanical period~\cite{braginskii1978optimal, vanner2011pulsed, vanner2013cooling, muhonen2019state}, may be employed. Thus far, the aforementioned position measurement techniques operate in the linearized regime but advances in coupling rates are now forcing experiments to confront the cavity optomechanical nonlinearities.


\begin{figure*}
    \centering
    \includegraphics[width=\textwidth]{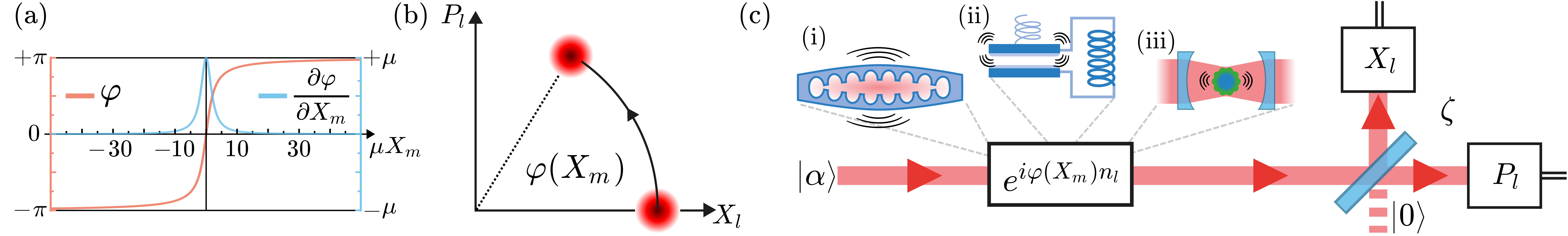}
    \caption{Cavity quantum optomechanical nonlinearities. (a) Plot of the optical phase $\varphi(\Xm)$ and the dimensionless mechanical momentum kick per photon $\partial{\varphi}(\Xm)/\partial\Xm$ arising from the cavity-enhanced nonlinear optomechanical interaction. Here, $\mu\propto g_{0}/\kappa$ is the dimensionless coupling strength, $\Xm$ is the mechanical position quadrature, and the detuning is $\Delta=0$. (b) In optical phase space there is a position-dependent rotation of an input coherent state owing to the nonlinearities of the radiation-pressure interaction and the cavity response. Our framework goes well beyond both the linearized regime, which approximates the rotation as a displacement, and treatments where the interaction is modelled as the unitary $\rme^{\rmi \mu \nl \Xm}$, which neglects the cavity response. (c) Position measurement scheme. An optical coherent state $\ket{\alpha}$ interacts with an optomechanical cavity and is then measured by general-dyne detection, which may be realized with a beamsplitter of variable transmission coefficient $\zeta$, vacuum on one input, and amplitude and phase homodyne measurements at the outputs. Our scheme and theoretical formalism are broadly applicable with example nonlinear optomechanical systems including (i) photonic crystals, (ii) microwave LC resonators, and (iii) levitation-based or cold-atom implementations.} \label{fig_1_main}
\end{figure*}


In this Letter, we develop a framework of optomechanics that accounts for the nonlinearities of both the radiation-pressure interaction and the cavity response. Secondly, we utilize this framework to propose how both pulsed and continuous position measurement can be performed in the nonlinear regime. To measure the mechanical signal transduced onto the optical field  we employ the established technique of general-dyne detection~\cite{walker1986multiport,furusawa1998unconditional,vidrighin2014joint}, which ranges from single to dual homodyne. Our formalism allows the optomechanical interaction to be described throughout the entire cavity phase response and enables new regimes of operation, such as the stochastic Gaussian regime we identify and term here. Our framework will enable a broad range of further studies and applications including the important tools of position measurement, squeezing by measurement, and mechanical state tomography to be performed beyond the linearized approximation.

\textit{Nonlinear position measurement scheme.}---Radiation pressure is described by the cubic Hamiltonian $H/\hbar=-{g}_{0}{a}^{\dagger}a(b+b^{\dagger})$, where $g_{0}$ is the optomechanical coupling rate, and $a$ ($b$) is the annihilation operator of the cavity field (mechanical mode). In addition to this nonlinearity, the full response of the cavity itself is also nonlinear, i.e. the optical phase shift depends nonlinearly on the mechanical position and asymptotes to $\pm \pi$ for large displacements. The combination of these nonlinearities are captured by the Heisenberg-Langevin and input-output equations, which yields the relation $\aout=f(\Xm)\ain$, where $\ain$ and $\aout$ are the optical input and output fields and $f(\Xm)=\left[1+\rmi\left(\frac{\mu}{2}{\Xm}+\frac{\Delta}{\kappa}\right)\right]/\left[1-\rmi\left(\frac{\mu}{2}{\Xm}+\frac{\Delta}{\kappa}\right)\right]$ is the nonlinear response function. Here, $\Xm$ ($\Pm$) is the dimensionless mechanical position (momentum) quadrature, $\mu=\sqrt{8}g_{0}/\kappa$ is the nonlinear coupling strength, $\kappa$ is the cavity amplitude decay rate, which is dominated by external coupling and $\Delta$ is the detuning from the cavity's resonance at zero mechanical displacement. In Fig.~\ref{fig_1_main}(a), we plot the optical phase $\varphi(\Xm)=\arg(f)$ and the dimensionless mechanical momentum kick per photon $\partial{\varphi}(\Xm)/\partial\Xm$ to {{help}} visualize these nonlinearities. {{We then}} obtain the nonlinear optomechanical unitary $U=\rme^{\rmi\varphi(\Xm)\nl}$, where $\nl$ is the photon number operator of the field entering or leaving the cavity. Notably, for this model, the only approximation made here is that the cavity is adiabatic, i.e. $\dot{a} \simeq 0$~\cite{supp}. Thus, our framework is applicable beyond the linearized regime, and furthermore, overcomes the limitations of previous works that model the optomechanical interaction as $\rme^{\rmi\mu\nl\Xm}$, e.g. Refs~\cite{pikovski2012probing,wang2017enhancing}, by accounting for the nonlinear response of the cavity. Fig.~\ref{fig_1_main}(b) shows the rotation of an optical coherent state $\ket{\alpha}$ by angle $\varphi(\Xm)$ following the nonlinear optomechanical interaction, where information about the mechanical position is encoded on both the optical phase $\Pl$ and amplitude $\Xl$ quadratures. The optical field is then measured by general-dyne detection (cf.~Fig.~\ref{fig_1_main}(c)), which uses two homodyne detectors to measure both optical quadratures in a proportion controlled by a variable beamsplitter $\Ubs$ with transmission coefficient $\zeta$.


\begin{figure}
    \centering
    \includegraphics[width=\columnwidth]{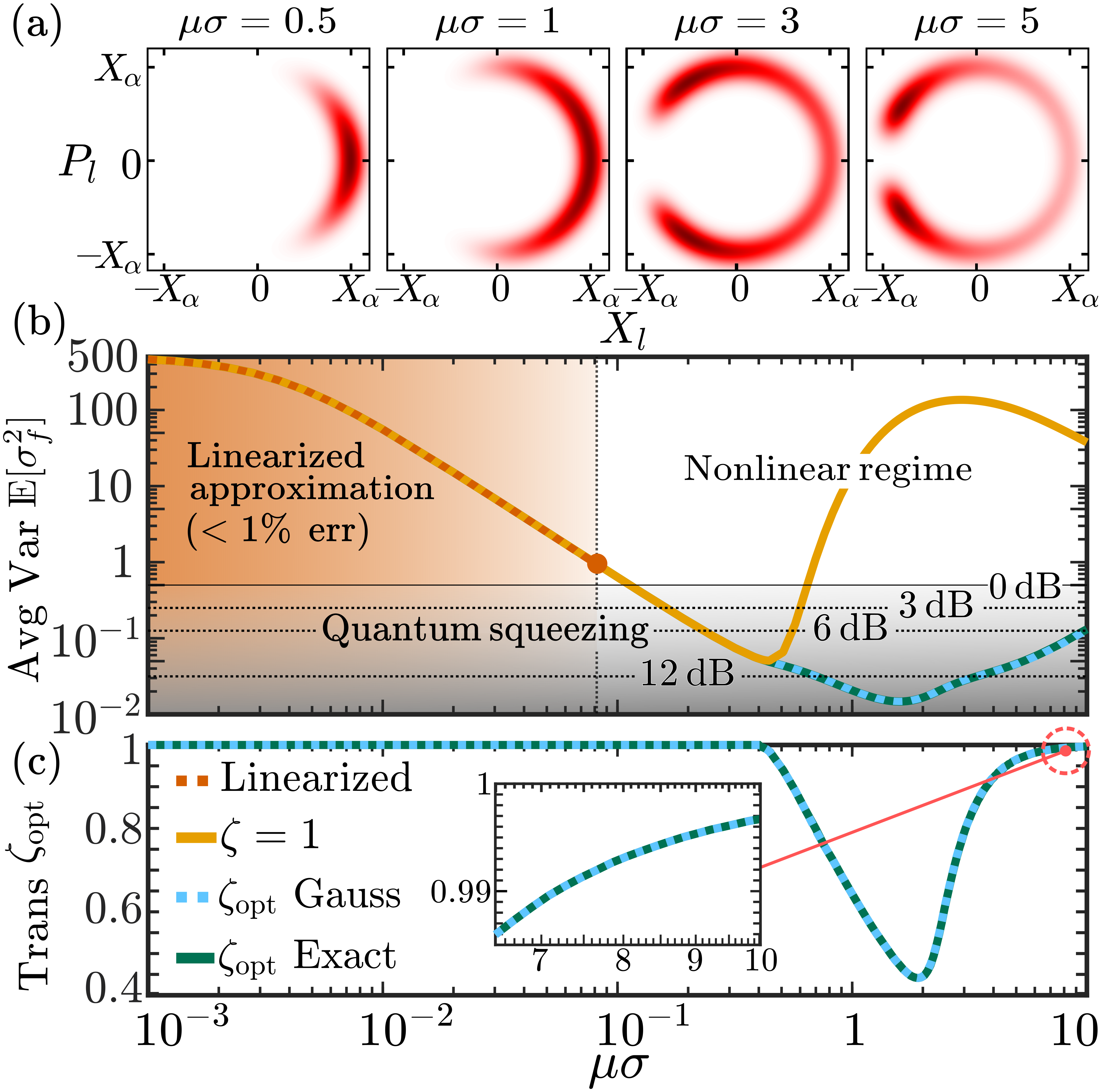}
    \caption{Nonlinear pulsed position measurement. (a) $Q$ functions of the optical pulse after the optomechanical interaction for $\Xa=10$ and increasing $\mu\sigma$. For an animation, see the Supplemental Material~\cite{supp}. (Color scale: white to dark red corresponds to $Q=0$ to $Q=\mathrm{max}(Q)$.) (b) Plot of the mechanical position variance averaged over all measurement outcomes, using either the optimal general-dyne or a phase-homodyne measurement ($\zeta=1$), as a function of $\mu\sigma$ for $X_{\alpha}=200$, and $\sigma^2=500$. The grey shaded area indicates quantum squeezing and for these parameters squeezing below $12\,\mathrm{dB}$ is achievable using the optimal general-dyne. The orange shaded area indicates where the error from the linearized approximation is less than $1\%$, which ends at $\mu\sigma=0.08$. (c) Plot of the optimal beamsplitter transmission coefficient $\zeta_{\mathrm{opt}}$. Agreement is seen between the exact case and the Gaussian approximation.}
    \label{fig_2_main}
\end{figure}


\textit{Nonlinear pulsed position measurement.}---A pulsed interaction much shorter than a mechanical period enables a BAE position measurement~\cite{braginskii1978optimal, vanner2011pulsed} as the mechanical free evolution (and, indeed, mechanical dissipation) is negligible during this timescale. Such operations require the unresolved sideband regime ($\kappa\gg\wm$) to accommodate the pulse within the cavity bandwidth. Here, we consider such a pulsed interaction taking into account the cavity optomechanical nonlinearities. Following the nonlinear interaction $U$, the general-dyne measurement outcomes are then used to estimate the mechanical position. The interaction and measurement are described by the measurement operator $\Upsilon=\bra{\Xl}\bra{\Pl}{\Ubs}U\ket{\alpha}\ket{0}$, which localizes the mechanical state according to $\rho\rightarrow\Upsilon\rho\Upsilon^{\dagger}/{\Ph}$, where $\Ph=\tr\left(\Upsilon^{\dagger}\Upsilon\rho\right)$ is the probability for the general-dyne outcome $(\Xl,\Pl)$. More explicitly, $\Upsilon$ is 
\begin{eqnarray}
\Upsilon&&=\frac{1}{\sqrt{\pi}}\exp\Big[-\frac{1}{2}\left(\Xl-\sqrt{1-\zeta^2}\Xa{\A}\right)^2-\frac{1}{2}\left(\Pl-\zeta\Xa{\B}\right)^2
\nonumber\\
&&-\rmi\zeta\Xa\Pl{\A}+\rmi\sqrt{1-\zeta^2}\Xa\Xl{\B}-\frac{\rmi}{2}(1-2\zeta^2)\Xa^2\A\B\Big]\label{krauss},
\end{eqnarray} 
where $\alpha=\Xa/\sqrt{2}$ is the coherent optical amplitude, assumed to be real without loss of generality, and {{$\A=\mathrm{Re}\left(f(\Xm)\right)$ and $\B=\mathrm{Im}\left(f(\Xm)\right)$}}~\cite{supp}. 

We consider a resonant pulsed drive ($\Delta=0$) and an initial thermal mechanical state with position variance $\sigma^2=\expval{\Xm^2}=\bar{n}+1/2$ for mean occupation $\bar{n}$. The pulsed interaction gives rise to optical phase-shifts depending on the mechanical position distribution and the nonlinearity of $\varphi(\Xm)$. Such optical states, with spread determined by $\mu\sigma$, are plotted in Fig.~\ref{fig_2_main}(a) using the Husimi-$Q$ function. Note that, with increasing $\mu\sigma$, the maxima of $Q$ move away from $\Xl=\Xa$ towards $\Xl=-\Xa$ and the probability of landing within the linear response quickly vanishes. In the limit $\mu\sigma\rightarrow\infty$, the input coherent state rotates through an angle $\pm\pi$ in phase space and the mechanical state is unchanged.

Optical loss, including detection inefficiencies, are modelled by a beamsplitter of transmission $\eta$ placed before the general-dyne detector~\cite{leonhardt1997measuring}. As the pulsed unitary $U$ is a function of mechanical position $\Xm$ alone, optical loss will induce changes in the mechanical momentum~\cite{supp}, while the only effect to the final position marginal is to reduce the strength of the measurement via $\Xa\rightarrow\eta\Xa$ in Eq.~\eqref{krauss}. 

The Bayesian inference associated with the general-dyne measurement reduces the mechanical variance from $\sigma^2$ to $\sigmaf^2$, which, in contrast to the linearized regime, depends on the measurement outcome $(\Xl,\Pl)$. As $\Xl$ and $\Pl$ are continuous variables, we average over all measurement outcomes to quantify the performance of reducing the position variance $\mathds{E}[{\sigmaf^2}]=\int\rmd{\Xl}\int\rmd{\Pl}\Ph(\Xl,\Pl)\sigmaf^2$~\cite{footnoteWindow}. Fig.~\ref{fig_2_main}(b) shows the averaged variance $\mathds{E}[{\sigmaf^2}]$ as a function of $\mu\sigma$ for the optimal beamsplitter transmission coefficient $\zeta_{\mathrm{opt}}$ demonstrating squeezing well below the ground state width. Here, $\zeta_{\mathrm{opt}}$ is computed numerically and is plotted in Fig.~\ref{fig_2_main}(c) up to a maximum value of $\mu\sigma$ corresponding to $g_{0}/\kappa\approx0.16$ for the fixed $\sigma^2=500$. At small and large $\mu\sigma$, the $Q$ function is aligned mostly along the $\Pl$ axis, and hence, the optimal beamsplitter transmission coefficient is close to $\zeta=1$, which agrees with the predictions of the linearized approximation for small $\mu\sigma$. While for intermediate $\mu\sigma$, $Q$ is aligned mostly parallel to the $\Xl$ axis, so the optimal setting is closer to $\zeta=0$. For comparison, Fig.~\ref{fig_2_main}(b) also plots $\mathds{E}[{\sigmaf^2}]$ for a phase-homodyne measurement $\zeta=1$, which misses the mechanical information encoded on the optical amplitude quadrature causing the averaged variance to grow after $\mu\sigma\simeq0.5$. Thus, despite $\zeta_{\mathrm{opt}}$ being close to 1 at high $\mu\sigma$, the $\mathds{E}[{\sigmaf^2}]$ curves for $\zeta_{\mathrm{opt}}$ and $\zeta=1$ still differ and we explore their convergence for even higher values of $\mu\sigma$ in the Supplemental Material~\cite{supp}. The nonlinear pulsed model we introduce here enables BAE position measurement well beyond the validity of the linearized regime, even when only the phase quadrature is measured. Moreover, Fig.~\ref{fig_2_main}(b) shows that the amount of quantum squeezing continues to increase as one exits the linearized regime. A Gaussian approximation in $\Xm$ may also be made to the position measurement~\cite{supp}, which agrees with the exact theory for all $\mu\sigma$, and with the linearized approximation within its range of applicability, as shown in Fig.~\ref{fig_2_main}. 

\textit{Nonlinear continuous position measurement.}---In addition to pulsed measurements, our scheme in Fig.~\ref{fig_1_main}(c) can be used for continuous position measurement, where the record of general-dyne outcomes can be used to best estimate the quantum trajectory of the mechanical motion in the presence of the nonlinearities via a stochastic master equation (SME). The SME is derived by writing the optical drive as $|\alpha|^2=2{k}\rmd{t}$, where ${k}$ describes the photon flux, and expanding Eq.~\eqref{krauss} to first order in $\rmd{t}$~\cite{supp}. The full SME, including mechanical open-system dynamics, is
\begin{eqnarray}
&&\rmd\rho=-\frac{\rmi}{\hbar}[H_{0},\rho]\rmd{t}+\mathcal{D}[c]\rho\rmd{t}+\mathcal{D}[L]\rho\rmd{t}\nonumber\\&+&\sqrt{(1-\zeta^2)\eta}\mathcal{H}[c]\rho\rmd{W_{\Xl}}+\zeta\sqrt{\eta}\mathcal{H}[-\rmi c]\rho\rmd{W_{\Pl}},\label{SME}
\end{eqnarray}
where $H_{0}/\hbar=\wm b^{\dagger}b$,  $\wm$ is the mechanical angular frequency, the superoperators are given by  $\mathcal{D}[O]\rho=O\rho{O^{\dag}}-\frac{1}{2}\{O^{\dag}O,\rho\}$ and $\mathcal{H}[O]\rho=O\rho+\rho{O^{\dag}}-\expval{O+O^{\dagger}}\rho$, the measurement output operator is $c=\sqrt{2{k}}f(\Xm)$, and $L=\sqrt{\left({4\gamma \kb T}\right)/\left({\hbar\wm}\right)}\Xm+\rmi\sqrt{\left({\hbar\wm \gamma}\right)/\left({4 \kb T}\right)}\Pm$, which models quantum Brownian motion~\cite{schlosshauer2007decoherence}. Here, $T$ is the environmental temperature, $\gamma$ is the mechanical decay rate, and $\eta$ is the measurement efficiency. Further, the Wiener increments for the $\Xl$ and $\Pl$ measurements obey $\rmd{W}_{i}\rmd{W}_{j}=\delta_{i,j}\rmd{t}$ and $\mathds{E}[\rmd{W}_{i}]=0$, for $i,j=\Xl,\Pl$, where $\mathds{E}$ represents the stochastic average. The SME of Eq.~\eqref{SME} is valid provided the cavity can be adiabatically eliminated, which is readily achieved in the unresolved-sideband regime, and for resolved-sideband systems, $\kappa$ must be larger than the interaction rate~\cite{supp}.
 
For a continuous-input drive, the measurement records of the $\Xl$ and $\Pl$ homodyne detectors are $\rmd{y}_{\Xl}=\expval{\A}\rmd{t}+{\rmd{W}_{\Xl}}/{\sqrt{8\eta(1-\zeta^2){k}}}$ and $\rmd{y}_{\Pl}=\expval{\B}\rmd{t}+{\rmd{W}_{\Pl}}/{\sqrt{8\eta\zeta{k}}}$, respectively~\cite{jacobs2006straightforward,*jacobs2014quantum,wiseman2009quantum}. And, the general-dyne measurement currents are defined by $\Xl(t)=\sqrt{{4{k}\eta(1-\zeta^2)}/{\tau}}\int_{t-\tau}^{t}\,\rmd{y}_{\Xl}$ and $\Pl(t)=\sqrt{{4{k}\eta\zeta^2}/{\tau}}\int_{t-\tau}^{t}\,\rmd{y}_{\Pl}$, where $\tau$ is the integration time of the homodyne detectors. Here, $1/\tau$ must be much faster than all other relevant rates, such that the SME in Eq.~\eqref{SME} accurately models the continuous measurement~\cite{caves1987quantum}. {{Also}}, at $\mu=0$ the measurement currents are normalized to give $\mathbb{E}(\Xl^2)-\mathbb{E}({\Xl})^2=\mathbb{E}(\Pl^2)-\mathbb{E}({\Pl})^2=1/2$, $\mathbb{E}(\Pl)=0$, and $\mathbb{E}(\Xl)=\Xa\sqrt{\eta(1-\zeta^2)}$, with $\Xa^2=4{k}\tau$.

For small optical rotations ($\mu\sigma\ll1$) Eq.~\eqref{SME} reduces to the standard SME of linearized optomechanics as $c=\sqrt{2{k}}f(\Xm)\approx\sqrt{2{k}}+\rmi\sqrt{2\mu^2 k}\Xm$ and $\mu^2 k$ recovers the linearized measurement rate $2g^2/\kappa$, where $g$ is the linearized coupling rate{{~\cite{rossi2019observing,meng2020mechanical}}}. However, we find that a distinct Gaussian limit of Eq.~\eqref{SME} exists beyond linearized optomechanics for arbitrarily large optical rotations. Curiously, in this regime, the evolution of the mechanical covariance matrix is \emph{stochastic} as the variances depend on the measurement outcomes. We thus term this new Gaussian regime of operation the \emph{stochastic Gaussian regime}. To derive this regime, at every time step $t\rightarrow t+\rmd{t}$ we write $\mu\Xm=\mu\expval{\Xm}+\mu \Ym$ in Eq.~\eqref{SME}. Provided $\mathrm{Var}(\mu \Ym)\ll1$, which implies weak single-photon coupling, we then expand the SME to first order in $\mu \Ym$ and the small optical phase shifts $\mu \Ym$ may be integrated to obtain arbitrarily large optical rotations over a finite duration. Introducing $\mathbf{r}=(\Ym,\Pm)^{\mathrm{T}}$, the dynamics of the first moments ${\rv}=\tr\left(\rho \mathbf{r}\right)$ and the covariance matrix elements $V_{ij}=\expval{\left\{r_{i},r_{j}\right\}}/2-\expval{r_{i}}\expval{r_{j}}$ can be computed from {{the Gaussian approximation to}} Eq.~\eqref{SME}. In the stochastic Gaussian regime, the dynamics of an initial Gaussian mechanical state are completely described by a stochastic differential equation for $\rv$ and a stochastic Riccati equation for $V$
\begin{eqnarray}
\rmd\rv&=&\left(M\rv+\mathbf{d}\right)\rmd{t}+N{\sqrt{\bm{\eta}}}\rmd{\mathbf{W}},\label{SDEmean}\\
\dot{V}&=&MV+V{M}^{\mathrm{T}}+D-N\bm{\eta}N^{\mathrm{T}}\label{Riccati}.
\end{eqnarray}
Here, $\bm{\eta}=\mathrm{diag}\left(\eta,\eta,0\right)$ and $\rmd\mathbf{W}=\left(\rmd{W}_{\Xl},\rmd{W}_{\Pl},0\right)^{\mathrm{T}}$. We have also introduced the displacement vector $\mathbf{d}$ and the drift $M$, diffusion $D$, and noise $N$ matrices, given in the Supplemental Material~\cite{supp}. See e.g. Ref.~\cite{zhang2017prediction} for other usage of the Riccati equation. Note, Eq.~\eqref{Riccati} is stochastic as $M$, $D$, and $N$ depend on the mean position $\expval{\Xm}$.


\begin{figure}
    \raggedright{
    \includegraphics[width=\columnwidth]{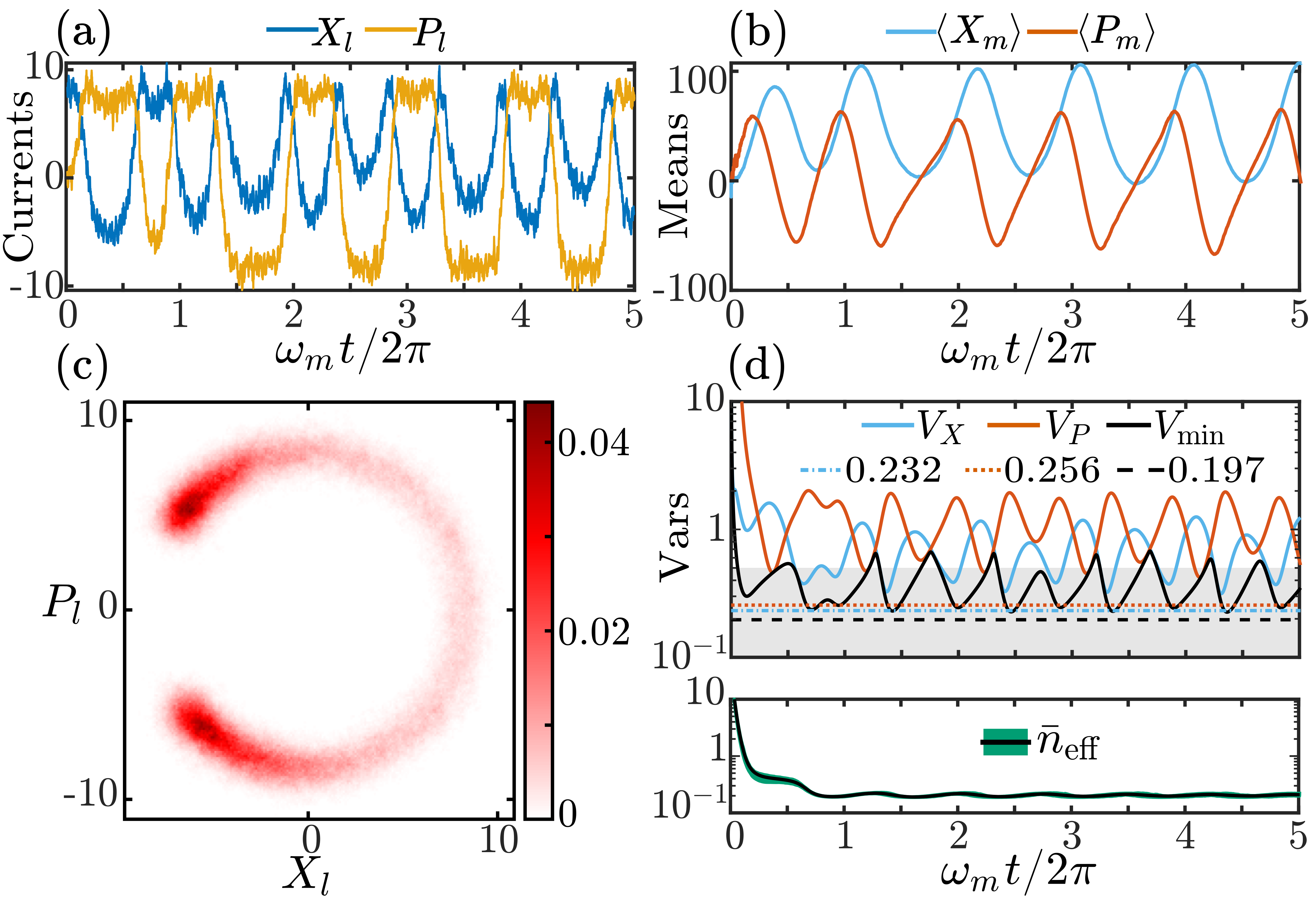}}
    \caption{Nonlinear continuous position measurement in the stochastic Gaussian regime. Here, $\Xa/\sqrt{2}=10$, $\wm/2\pi=10\,\mathrm{MHz}$, $\mu=0.05$, $\mu^2k/\wm=2$, $\eta=0.7$, $\gamma/2\pi=10\,\mathrm{Hz}$, $\zeta=1/\sqrt{2}$, $T=4\,\mathrm{K}$ ($\bar{n}=8.4\times10^3$), and the initial mechanical state is precooled to $100\,\mathrm{mK}$ ($\bar{n}=2.1\times10^2$). (a) The general-dyne measurement currents $\Xl$ and $\Pl$ plotted as a function of time for a given trajectory. (b) The corresponding trajectories of the mechanical means $\expval{\Xm}$ and $\expval{\Pm}$. (c) The same general-dyne currents plotted as a normalized $2$D histogram. (d) Top: The position variance $V_{X}$, the momentum variance $V_{P}$, and the minimum eigenvalue $V_{\mathrm{min}}$ for the same trajectory. Here, the dashed lines correspond to the mean of the minimum of each variance quantity over 100 trajectories and the grey shaded area indicates quantum squeezing. Bottom: The effective thermal occupation $\bar{n}_{\mathrm{eff}}$ averaged over 100 trajectories, where the shaded green area indicates the upper and lower quartile range. 
    }
    \label{fig_3_main}
\end{figure}


For a fixed drive frequency at $\Delta=0$, the optical phase shift averaged over a given stochastic trajectory is positive because the radiation-pressure force causes {{a non-zero}} time-averaged mean mechanical position. Hence, to maximize the amount of light that enters the cavity, and ensure the time-averaged optical phase shift is zero, we model a lock of the pump field to the mean cavity resonance frequency using a third-order Butterworth filter with a cutoff frequency at $0.5\wm$. This locking effectively cancels the slowly varying component of $\expval{\Xm}$, while the dynamics and fluctuations of the mechanical position are still measured.

The dynamics of the mechanical Gaussian state governed by Eqs~\eqref{SDEmean} and \eqref{Riccati} are solved numerically using an Euler-Maruyama method~\cite{jacobs2010stochastic}. Here, we use a parameter set (see Fig.~\ref{fig_3_main} caption) based on sliced-photonic crystal structures~\cite{leijssen2017nonlinear, muhonen2019state} and more parameter sets are explored in the Supplemental Material~\cite{supp}. {{For this continuous measurement, we choose a beamsplitter coefficient of $\zeta=1/\sqrt{2}$ corresponding to heterodyne detection, which is readily implemented using a single balanced detector and a frequency-offset local oscillator.}} Fig.~\ref{fig_3_main}(a) shows the general-dyne measurement currents with time for a given stochastic trajectory, {{and the corresponding}} means $\expval{\Xm}$ and $\expval{\Pm}$ are plotted in Fig.~\ref{fig_3_main}(b). We see the evolution of the means is anharmonic resulting from the optomechanical nonlinearities. Meanwhile, Fig.~\ref{fig_3_main}(c) plots the general-dyne currents as a $2$D histogram in optical phase space, which is symmetric due to the drive-locking. Fig.~\ref{fig_3_main}(d) shows the position variance $V_{X}$, the momentum variance $V_{P}$, and the minimum eigenvalue of the covariance matrix $V_{\mathrm{min}}$ for the same trajectory and the dashed lines indicate the mean of the minimum value of these quantities over $100$ runs, each lasting $100$ mechanical periods. We also plot the effective thermal occupation $\bar{n}_{\mathrm{eff}}=\sqrt{\det\left(V\right)}-1/2=\left(1/\tr(\rho^2)-1\right)/2$ averaged over the $100$ runs. These plots show that for the {{modest}} parameters used {{here}}, quantum squeezing below $3\,\mathrm{dB}$ ($V_{\mathrm{min}},V_{X}<0.25$) and effective ground-state cooling are achievable in the stochastic Gaussian regime.  

\textit{Conclusions and Outlook.}---We introduce a framework for cavity quantum optomechanics that captures the nonlinearities of both radiation-pressure and the cavity response. As a natural first application of this framework, we study mechanical position measurement in the presence of these nonlinearities, which enables mechanical information encoded on both optical quadratures throughout the full cavity phase response of $\pm\pi$ to be measured via optical general-dyne detection. Our scheme can be employed with back-action-evading pulsed measurements or with continuous measurements and we derive a measurement operator and a stochastic master equation for each mode of operation, respectively. As well as recovering the linearized theory, our framework is capable of describing the three regimes where the linearized approximation fails: (i) the weak single-photon coupling regime but with significant optical rotations, (ii) the strong single-photon coupling regime, and (iii) systems operating at low light levels. Thus, this treatment vastly extends the range that cavity optomechanical position measurement can be performed and enables the identification of new regimes of operation, such as the stochastic Gaussian regime we identify and term here. In addition to describing how to perform mechanical quantum squeezing by measurement in nonlinear systems, our formalism will also lead to a broad range of experimental and theoretical advances in sensing, mechanical quantum state engineering, and quantum measurement and control.

\begin{acknowledgments}
\textit{Acknowledgements.}---We acknowledge useful discussions with G.~A.~Brawley, R.~Clarke, G.~Enzian, L.~A.~Kanari-Naish, G.~J.~Milburn, B.~A.~Stickler, and S.~Qvarfort. This project was supported by the Engineering and Physical Sciences Research Council (EP/T031271/1), UK Research and Innovation (MR/S032924/1), the Royal Society, and the European Research Council (ERC Starting Grant No. 759644-TOPP). This work is part of the research programme of the Netherlands Organisation for Scientific Research (NWO) and supported by an NWO Vidi grant.
\end{acknowledgments}


\providecommand{\noopsort}[1]{}\providecommand{\singleletter}[1]{#1}%
%




\onecolumngrid
\clearpage
\newgeometry{left=1.5cm,right=1.5cm,top=1.5cm,bottom=1.5cm}
\setcounter{equation}{0}
\def\theequation{S\arabic{equation}}
\setcounter{figure}{0}
\renewcommand{\thefigure}{S\arabic{figure}}
\pagenumbering{roman}

\begin{center}
\textbf{\large{Cavity quantum optomechanical nonlinearities \\ and position measurement beyond the breakdown of the linearized approximation:\\ Supplemental Material}}
\end{center}
\vspace{10pt}

\begin{centering}
J.~Clarke$^{1}$,~  
P.~Neveu$^{2}$,~
K.~E.~Khosla$^{1}$,~
E.~Verhagen$^{2}$,
and M.~R.~Vanner$^{1}$\\
\end{centering}

\vspace{16pt}

\begin{centering}
\textit{\small
 $^1$QOLS, Blackett Laboratory, Imperial College London, London SW7 2BW, United Kingdom\\
 $^2$Center for Nanophotonics, AMOLF, Science Park 104, 1098 XG Amsterdam, The Netherlands\\
}
\end{centering}

\begin{quote}
{\small{Here, we provide further details on our nonlinear theoretical framework of cavity quantum optomechanics and approach for position measurement in the nonlinear regime. Firstly, we give a step-by-step derivation of the nonlinear optomechanical unitary, the measurement operator, and the SME. Following this, we detail the calculations that show how these tools may be used to achieve pulsed and continuous measurements of mechanical position in the nonlinear regime via optical general-dyne detection. We describe the Gaussian approximation for pulsed operation and show how the equations for the stochastic Gaussian regime may be derived from the SME. Finally, we study more parameter sets based on current state-of-the-art sliced-photonic crystal devices.}}
\end{quote}

\vspace{16pt}
\twocolumngrid
\noindent\textbf{Contents}\\
{\small{
\noindent I. \hyperlink{S1}{Nonlinear cavity-enhanced unitary}\dotfill{\pageref{sec:S1}} \\
\hspace*{10mm} A. \hyperlink{S1A}{Pulsed interaction} \dotfill{\pageref{sec:S1A}} \\
\hspace*{10mm} B. \hyperlink{S1B}{Continuous infinitesimal unitary} \dotfill{\pageref{sec:S1B}} \\
II. \hyperlink{S2}{Measurement operator} \dotfill{\pageref{sec:S2}} \\
III. \hyperlink{S3}{Stochastic master equation} \dotfill{\pageref{sec:S3}} \\
\hspace*{10mm} A. \hyperlink{S3A}{Infinitesimal measurement operator \\
\hspace*{10mm} in a rotating frame} \dotfill{\pageref{sec:S3A}} \\
\hspace*{10mm} B. \hyperlink{S3B}{Stochastic master equation in a rotating frame} \dotfill{\pageref{sec:S3B}} \\
\hspace*{10mm} C. \hyperlink{S3C}{Stochastic master equation in the lab frame} \dotfill{\pageref{sec:S3C}} \\
IV. \hyperlink{S4}{Pulsed measurement} \dotfill{\pageref{sec:S4}} \\
\hspace*{10mm} A. \hyperlink{S4A}{Krauss map and heralding probability} \dotfill{\pageref{sec:S4A}} \\
\hspace*{10mm} B. \hyperlink{S4B}{The Husimi-$Q$ function and $\pi$-phase limit} \dotfill{\pageref{sec:S4B}} \\
\hspace*{10mm} C. \hyperlink{S4C}{Gaussian approximation to pulsed measurement} \dotfill{\pageref{sec:S4C}}  \\
\hspace*{10mm} D. \hyperlink{S4D}{Optical loss} \dotfill{\pageref{sec:S4D}} \\
\hspace*{10mm} E. \hyperlink{S4E}{Phase quadrature measurements} \dotfill{\pageref{sec:S4E}} \\
\hspace*{10mm} F. \hyperlink{S4F}{Dependence on the detuning} \dotfill{\pageref{sec:S4F}} \\
V. \hyperlink{S5}{Continuous measurement} \dotfill{\pageref{sec:S5}}\\
\hspace*{10mm} A. \hyperlink{S5A}{Agreement with the linearized regime of \\
\hspace*{10mm} optomechanics and the linearized measurement rate} \dotfill{\pageref{sec:S5A}}  \\
\hspace*{10mm} B. \hyperlink{S5B}{Stochastic Gaussian measurement regime} \dotfill{\pageref{sec:S5B}} \\
\hspace*{10mm} C. \hyperlink{S5C}{Drive-locking} \dotfill{\pageref{sec:S5C}} \\
\hspace*{10mm} D. \hyperlink{S5D}{Experimental parameters for sliced-photonic \\
\hspace*{10mm} crystal devices} \dotfill{\pageref{sec:S5D}} \\
}}

\onecolumngrid

\hypertarget{S1}{}\section{I.  Nonlinear cavity-enhanced unitary}\label{sec:S1}
\hypertarget{S1A}{}\subsection{A. Pulsed interaction}\label{sec:S1A}
In the unresolved sideband regime ($\kappa\gg\wm$), which is used for pulsed optomechanics, the Heisenberg-Langevin equations describing the dynamics generated by $H/\hbar=-\Delta a^{\dagger}a-{g}_{0}{a}^{\dagger}a(b+b^{\dagger})$ are 
\begin{eqnarray}
\dot{a}&=&\rmi(\sqrt{2}g_{0}\Xm+\Delta)a-\kappa{a}+\sqrt{2\kappa}\ain\label{HLcav}\\
\dot{X}_{m}&=&0\\
\dot{P}_{m}&=&\sqrt{2}g_{0}a^{\dagger}a\label{HLmomentum},
\end{eqnarray}
for an input pulse at $\wl$ in a frame rotating at the cavity frequency $\wc$. Here, the detuning is $\Delta=\wl-\wc$ and we neglect free mechanical evolution and dissipation as the interaction time is much less than the mechanical period. The operator $\ain(t)$ describes the time-dependent input to the cavity mode and $\Xm=(b+b^{\dag})/\sqrt{2}$. Furthermore, in pulsed optomechanics the cavity field can be adiabatically eliminated $\dot{a}=0$ if $\kappa\gg\tau_{{in}}^{-1}$, where $\tau_{{in}}$ is the characteristic timescale over which the input $\ain$ changes. In these limits, the equations of motions may be solved to obtain 
\begin{eqnarray}
a(t)&=&\sqrt{2\kappa}\int_{-\infty}^{t}\rmd{t'}~\ain(t')\rme^{-[\kappa-\rmi(\sqrt{2}g_{0}{\Xm}+\Delta)]|t-t'|}.\label{aoft}
\end{eqnarray}
Generally, if $h(t')$ is a slowly varying function, such that $\kappa\gg\tau_{h}^{-1}$, where $\tau_{h}$ is the characteristic timescale over which the function $h(t')$ changes,
\begin{eqnarray}
h(t')\rme^{-\kappa|t-t'|}&\simeq&h(t')\frac{2}{\kappa}\delta(t-t')\nonumber\\
&=&h(t)\frac{2}{\kappa}\delta(t-t')\nonumber\\
&\simeq&h(t)\rme^{-\kappa|t-t'|}\label{slowdelta}.
\end{eqnarray}
Here, we used that $\lim_{\kappa\rightarrow\infty}\frac{\kappa}{2}\rme^{-\kappa|t-t'|}=\delta(t-t')$. This property \eqref{slowdelta}  implies that
$\ain(t')\rme^{-[\kappa-\rmi(\sqrt{2}g_{0}{\Xm}+\Delta)]|t-t'|}=\ain(t)\rme^{-[\kappa-\rmi(\sqrt{2}g_{0}{\Xm}+\Delta)]|t-t'|}$
if $\kappa\gg\tau_{{in}}^{-1}$, regardless of $g_0$, $\Xm$, and $\Delta$, which contribute only to the phase term. The expression for $a(t)$ in Eq.~\eqref{aoft} then becomes
\begin{eqnarray}
a(t)&=&\sqrt{2\kappa}\ain(t)\int_{-\infty}^{t}\rmd{t'}~\rme^{-[\kappa-\rmi(\sqrt{2}g_{0}{\Xm}+\Delta)](t-t')}\nonumber\\
&=&\dfrac{\sqrt{2\kappa}}{\kappa-\rmi(\sqrt{2}g_{0}{\Xm}+\Delta)}\ain(t),\label{acav}
\end{eqnarray}
which amounts to adiabatic elimination of the cavity field $\dot{a}=0$ in Eq. \eqref{HLcav}.

Using the input-output relations for the cavity $\aout(t)=\sqrt{2\kappa}a(t)-\ain(t)$, and defining the nonlinear optomechanical coupling strength $\mu=\sqrt{8}g_{0}/\kappa$, then gives
\begin{eqnarray}
\aout(t)=\dfrac{1+\rmi\left(\frac{\mu}{2}{\Xm}+\frac{\Delta}{\kappa}\right)}{1-\rmi\left(\frac{\mu}{2}{\Xm}+\frac{\Delta}{\kappa}\right)}\ain(t)\label{aouteqn}.
\end{eqnarray}
While defining the photon number operator $\nl$ to be
\begin{eqnarray}
\nl&=&\int_{0}^{\tau_{p}} \ain^{\dagger}\ain(t)\rmd{t}=\int_{0}^{\tau_{p}} \aout^{\dagger}\aout(t)\rmd{t}\label{numop},
\end{eqnarray}
allows us the integrate Eq.~\eqref{HLmomentum}, and using \eqref{acav}, we arrive at
\begin{eqnarray}
\Pm(\tau_{p})-\Pm(0)&=&\dfrac{\mu}{1+\left(\frac{\mu}{2}\Xm+\frac{\Delta}{\kappa}\right)^2}\nl\label{Ptau},
\end{eqnarray}
where $\tau_{p}$ is the duration of the pulsed optomechanical interaction.

Introducing the nonlinear response function $f(\Xm)=\left[1+\rmi\left(\frac{\mu}{2}{\Xm}+\frac{\Delta}{\kappa}\right)\right]/\left[1-\rmi\left(\frac{\mu}{2}{\Xm}+\frac{\Delta}{\kappa}\right)\right]$ and the nonlinear phase $\varphi(\Xm)=\arg(f)$, allows us to write Eqs~\eqref{aouteqn} and \eqref{Ptau} as
\begin{eqnarray}
\aout&=&f(\Xm)\ain, \label{aout1}\\
\Pm(\tau_{p})&=&\Pm(0)+\dfrac{\partial\varphi(\Xm)}{\partial \Xm}\nl\label{Pout1},
\end{eqnarray}
which are equivalent to Heisenberg transformations
\begin{eqnarray}
\aout&=&U^{\dagger}\ain U, \label{aout2}\\
\Pm(\tau_{p})&=&U^{\dagger}\Pm(0)U \label{Pout2},
\end{eqnarray}
with the nonlinear pulsed optomechanical unitary given by $U=\rme^{\rmi\varphi(\Xm)\nl}$. The equivalence between Eqs~\eqref{aout1} and \eqref{aout2}, may be understood via the equation for phase-space rotations $\aout=\rme^{\rmi\phi\nl}\ain \rme^{-\rmi\phi\nl}=\left(\cos\phi-\rmi\sin\phi\right)\ain$. Furthermore, the Hadamard lemma 
\begin{eqnarray*}
\rme^{X}Y\rme^{-X}=Y+[X,Y]+\frac{1}{2!}[X,[X,Y]]+\ldots+\frac{1}{n!}[X,[X,\ldots,[X,Y]\ldots]]+\ldots,
\end{eqnarray*}
along with $[\varphi(\Xm),\Pm]=\rmi\dfrac{\partial \varphi(\Xm)}{\partial \Xm}$, can be used to derive the equivalence between Eqs~\eqref{Pout1} and \eqref{Pout2}.

When $\Delta=0$, expanding $U=\rme^{\rmi\varphi(\Xm)\nl}$ to second order in $\Xm$ agrees with the nonlinear unitary $\rme^{\rmi\mu\Xm\nl}$ of~[\href{https://doi.org/10.1038/nphys2262}{I. Pikovski \emph{et al.}, Nat. Phys. \textbf{8}, 393 (2012)}; \href{https://doi.org/10.1038/ncomms15886}{Z. Wang and A. H. Safavi-Naeini, Nat. Commun. \textbf{8}, 1
(2017)}]. The higher orders in $\Xm$ account for the nonlinearity of the cavity response. Furthermore, the linearized unitary is recovered in the limit of small optical rotations and large intracavity amplitude: $a\rightarrow\alpha_{c}+\delta{a}$ and $|\alpha_{{c}}|\gg\expval{|\delta{a}|}$~[\href{https://doi.org/10.1073/pnas.1105098108}{M. R. Vanner \emph{et al.}, Proc. Natl. Acad. Sci. \textbf{108}, 16182 (2011)}]. Finally, we note that for our single cavity mode treatment to be valid the change in the cavity frequency due to a displacement $\Xm$, which for $\Delta=0$ is $|\wc(\Xm)-\wc|=\sqrt{2}g_{0}\Xm\simeq\sqrt{2}g_{0}\sigma$, must be much smaller than the distance in frequency space to the next optical mode $\omega_{fsr}$. This condition ensures neighbouring cavity modes can be ignored and division by $\kappa$ allows one to write this condition as $\mu\sigma\ll2\mathcal{F}_{c}$, where $\mathcal{F}_{c}=\omega_{fsr}/\kappa$ is the cavity finesse.

\hypertarget{S1B}{}\subsection{B. Continuous infinitesimal unitary}\label{sec:S1B}
The optomechanical Hamiltonian is $H_{{om}}=\hbar\wc{a}^{\dagger}a+\hbar\wm{b}^{\dagger}b-\hbar{g}_{0}{a}^{\dagger}a(b+b^{\dagger})=H_{0}+H$, with $H_{0}=\hbar\wc{a}^{\dagger}a+\hbar\wm{b}^{\dagger}b$ and $H=-\hbar{g}_{0}{a}^{\dagger}a(b+b^{\dagger})$. Transforming from the lab frame to a rotating frame using $\tilde{H}=U_{{F}}^{\dagger}H_{{om}}U_{{F}}+\rmi\hbar\dot{U}_{{F}}^{\dagger}U_{{F}}$, with $U_{{F}}=\rme^{-\rmi\wl{a}^{\dagger}at-\rmi\wm{b}^{\dagger}bt}$ gives
\begin{eqnarray}
\tilde{H}/\hbar&=& -\Delta a^{\dagger}a -g_{0}a^{\dagger}a(b\rme^{-\rmi\wm{t}}+b^{\dagger}\rme^{+\rmi\wm{t}}),\nonumber\\
&=&-\Delta a^{\dagger}a-\sqrt{2}g_{0}{a}^{\dagger}a{\Xmr},\\
\end{eqnarray}
where $\Xmr=\Xm\cos\wm{t}+\Pm\sin\wm{t}$ and $\Pmr=\Pm\cos\wm{t}-\Xm\sin\wm{t}$. In this rotating frame, the Heisenberg-Langevin equations with optical driving of the cavity are given by
\begin{eqnarray}
\dot{X}_{m}&=&-\sqrt{2}g_{0}a^{\dagger}a\sin\wm{t},\\
\dot{P}_{m}&=&\sqrt{2}g_{0}a^{\dagger}a\cos\wm{t},\\
\dot{a}&=&\rmi \left[\sqrt{2}g_{0}(\Xm\cos\wm{t}+\Pm\sin\wm{t})+\Delta\right]a-\kappa{a}+\sqrt{2\kappa}\ain
\end{eqnarray}
We have not included mechanical interactions with the bath as they aren't a part of the construction of the infinitesimal unitary. These mechanics-bath interactions will be included in the SME via a Lindblad dissipator. Solving these equations over an arbitrary infinitesimal time window $t\rightarrow t+\rmd{t}$ gives
\begin{eqnarray}
\Xm(t+\rmd{t})&=&\Xm(t)-\sqrt{2}g_{0}a^{\dagger}a\sin(\wm{t})\rmd{t},\\
\Pm(t+\rmd{t})&=&\Pm(t)+\sqrt{2}g_{0}a^{\dagger}a\cos(\wm{t})\rmd{t},\\
a(t+\rmd{t})&=&a(t)+\rmi \left(\sqrt{2}g_{0}{\Xmr}+\Delta\right)a\rmd{t}-\kappa{a}\rmd{t}+\sqrt{2\kappa}\ain\rmd{t}\label{adt}.
\end{eqnarray}

We now consider a `good position measurement' where the cavity field adiabatically follows the mechanical oscillator~[\href{https://doi.org/10.1103/PhysRevA.60.2700}{A. C. Doherty and K. Jacobs, Phys. Rev. A \textbf{60}, 2700 (1999)}]. Such a measurement is readily achieved in the unresolved sideband regime $\kappa\gg\wm$, as the cavity field responds quickly to the mechanical motion and adiabatically follows the mechanical oscillator~[\href{https://doi.org/10.1038/ncomms10988}{G. Brawley, \emph{et al.}, Nat. Commun. \textbf{7}, 10988 (2016)}]. Outside the unresolved sideband regime, adiabatic elimination of the cavity mode is also valid, provided the cavity decay rate exceeds other rates in the Hamiltonian~[\href{https://doi.org/10.1103/PhysRevA.47.642}{H. M. Wiseman and G. J. Milburn, Phys. Rev. A \textbf{47}, 642 (1993)}].

In the adiabatic regime, the cavity field rapidly reaches the steady state $\dot{a}=0$ on the order of $1/\kappa$ and so together with Eq.~\eqref{adt}, the cavity input-output relation gives
\begin{eqnarray}
\aout(t)&=&\dfrac{1+\rmi\left(\frac{\mu}{2}{\Xmr}(t)+\frac{\Delta}{\kappa}\right)}{1-\rmi\left(\frac{\mu}{2}{\Xmr}(t)+\frac{\Delta}{\kappa}\right)}\ain(t)\label{aoutdt}.
\end{eqnarray}
The definition of the photon number operator in Eq.~\eqref{numop} implies that over a time increment $\rmd{t}$, $\nl(\rmd{t})=\ain^{\dagger}\ain\rmd{t}$, and therefore 
\begin{eqnarray}
\Xm(t+\rmd{t})&=&\Xm(t)-\dfrac{\mu}{1+\left(\frac{\mu}{2}{\Xmr}+\frac{\Delta}{\kappa}\right)^2}\sin(\wm{t})\nl(\rmd{t})\label{Xmrdt},\\
\Pm(t+\rmd{t})&=&\Pm(t)+\dfrac{\mu}{1+\left(\frac{\mu}{2}{\Xmr}+\frac{\Delta}{\kappa}\right)^2}\cos(\wm{t})\nl(\rmd{t})\label{Pmrdt}.
\end{eqnarray}
We make an ansatz for the infinitesimal unitary in this interaction picture to be $U_{{I}}(\rmd{t})=\rme^{\rmi\varphi({\Xmr})\nl(\rmd{t})}$. Here, we explicitly labelled the dependence on $\rmd{t}$. Similarly to the derivation of the pulsed nonlinear unitary, this ansatz can be shown to give the same expression as Eqs~\eqref{aoutdt}, \eqref{Xmrdt} and \eqref{Pmrdt} through
\begin{eqnarray}
\aout(t)&=&U_{{I}}^{\dagger}(\rmd{t})\ain(t) U_{{I}}(\rmd{t}),\\
\Xm(t+\rmd{t})&=&U_{{I}}^{\dagger}(\rmd{t})\Xm(t) U_{{I}}(\rmd{t}),\\
\Pm(t+\rmd{t})&=&U_{{I}}^{\dagger}(\rmd{t})\Pm(t) U_{{I}}(\rmd{t}).
\end{eqnarray}
To derive the last two equations the Hadamard lemma was again employed and the following property was used: for a function of two the conjugate variables $f(\Xm,\Pm)$, we have $[f(\Xm,\Pm),\Xm]=-\rmi\dfrac{\partial f}{\partial \Pm}$ and $[f(\Xm,\Pm),\Pm]=+\rmi\dfrac{\partial f}{\partial \Xm}$.

\hypertarget{S2}{}\section{II. Measurement operator}\label{sec:S2}
The nonlinear pulsed unitary $U=\rme^{\rmi\varphi(\Xm)\nl}$ acts on the input coherent state $\ket{\alpha}$ via $U\ket{\alpha}=\ket{\alpha\left(\A(\Xm)+\rmi \B(\Xm)\right)}$, where the nonlinear response function has been decomposed into its real and imaginary parts $f(\Xm)=\A(\Xm)+\rmi \B(\Xm)$,
\begin{eqnarray}
\A(\Xm)&=&\dfrac{1-\left(\frac{\mu}{2}{\Xm}+\frac{\Delta}{\kappa}\right)^2}{1+\left(\frac{\mu}{2}{\Xm}+\frac{\Delta}{\kappa}\right)^2},\\
\B(\Xm)&=&\dfrac{2\left(\frac{\mu}{2} \Xm+\frac{\Delta}{\kappa}\right)}{1+\left(\frac{\mu}{2}{\Xm}+\frac{\Delta}{\kappa}\right)^2}.\\
\end{eqnarray}
Furthermore, note that $\A^2+\B^2=1$ and $\varphi(\Xm)=\arg(f)=\arg(\A+\rmi \B)$, so $\A(\Xm)=\cos\left[\varphi(\Xm)\right]$ and $\B(\Xm)=\sin\left[\varphi(\Xm)\right]$,
which clarifies that $U$ rotates the initial coherent state through an angle $\varphi(\Xm)$ for a given value of $\Xm$. 

The action of the nonlinear pulsed optomechanical interaction, followed by optical general-dyne measurement is described by the measurement operator
\begin{eqnarray}
\Upsilon&=&\bra{\Xl}\bra{\Pl}{\Ubs}U\ket{\alpha}\ket{0}\nonumber\\
&=&\bra{\Xl}\ket{\sqrt{1-\zeta^2}\alpha(\A(\Xm)+\rmi \B(\Xm))}\bra{\Pl}\ket{\zeta\alpha(\A(\Xm)+\rmi \B(\Xm))}\nonumber\\
&=&\frac{1}{\sqrt{\pi}}\exp\Big[-\frac{1}{2}\left(\Xl-\sqrt{1-\zeta^2}\Xa{\A}(\Xm)\right)^2-\frac{1}{2}\left(\Pl-\zeta\Xa{\B}(\Xm)\right)^2-\rmi\zeta\Xa\Pl{\A}(\Xm)\nonumber\\
&&+\rmi\sqrt{1-\zeta^2}\Xa\Xl{\B}(\Xm)-\frac{\rmi}{2}(1-2\zeta^2)\Xa^2\A(\Xm)\B(\Xm)\Big]\label{pulsedupsilon},
\end{eqnarray}
where $\alpha=\Xa/\sqrt{2}$ and $\Xa$ is assumed real without loss of generality. At the output of the optomechanical cavity, the optical mode mixes with vacuum $\ket{0}$ at a beamsplitter, described by ${\Ubs}$. While at the beamsplitter outputs, light is sent towards the phase $\Pl$ or amplitude $\Xl$ homodyne, in a proportion determined by the beamsplitter parameter $\zeta$. If the annihilation operator of the optical mode in the vacuum state is given by $a_{v}$, the Heisenberg transformations at the beamsplitter are given by $\Ubs^{\dagger}a \Ubs=\zeta a+\sqrt{1-\zeta^2}a_{v}$ and $\Ubs^{\dagger}a_{v} \Ubs=\zeta a_{v}-\sqrt{1-\zeta^2}a$.

\hypertarget{S3}{}\section{III. Stochastic master equation}\label{sec:S3}
\hypertarget{S3A}{}\subsection{A. Infinitesimal measurement operator in a rotating frame}\label{sec:S3A}
The infinitesimal measurement operator $\Upsilon_{I}(\Delta{t})$ describes the action of the continuous general-dyne measurement over a small time increment $\Delta{t}$. In the rotating frame, it is given by
\begin{eqnarray}
\Upsilon_{I}(\Delta{t})&=&\bra{\Xl}\bra{\Pl}{\Ubs}U_{I}(\Delta{t})\ket{\alpha}\ket{0}\nonumber\\
&=&\bra{\Xl}\ket{\sqrt{1-\zeta^2}\alpha\left(\A({\Xmr})+\rmi \B({\Xmr})\right)}\bra{\Pl}\ket{\zeta\alpha\left(\A({\Xmr})+\rmi \B({\Xmr})\right)}\nonumber\\
&=&\Upsilon_{I}(\Xl)\Upsilon_{I}(\Pl),\\
\Upsilon_{I}(\Xl)&=&\dfrac{1}{{\pi}^{\frac{1}{4}}}\exp\Big[-\frac{a_{\alpha}^2}{2}(\A({\Xmr})-\frac{\Xl}{a_{\alpha}})^2+\rmi{a_{\alpha}^2}\frac{\Xl}{a_{\alpha}}\B({\Xmr})-\frac{\rmi}{2}a_{\alpha}^2\A({\Xmr})\B({\Xmr})\Big],\\
\Upsilon_{I}({\Pl})&=&\dfrac{1}{{\pi}^{\frac{1}{4}}}\exp\Big[-\frac{b_{\alpha}^2}{2}(\B({\Xmr})-\frac{\Pl}{b_{\alpha}})^2-\rmi{b_{\alpha}^2}\frac{\Pl}{b_{\alpha}}\A(\tilde{\Xm})+\frac{\rmi}{2}b_{\alpha}^2\A({\Xmr})\B({\Xmr})\Big],
\end{eqnarray}
where we have split up the measurement operator into a part corresponding to the $\Xl$ measurement and the $\Pl$ measurement. We have also introduced $a_{\alpha}=\sqrt{1-\zeta^2}\Xa$ and $b_{\alpha}=\zeta\Xa$. 

The $\Delta{t}$ dependence in $\Upsilon_{I}(\Delta{t})$ comes in as the mean number of input photons in a $\Delta{t}$ is $\Np=|\alpha|^2=\frac{1}{2}\Xa^2=|\alpha_{{in}}|^2\Delta{t}$. Here, $\alpha$ is the dimensionless input amplitude, which depends on $\Delta{t}$, whereas $\alpha_{{in}}$ is the input field amplitude in units of $\mathrm{s}^{-1/2}$ and does not depend on $\Delta{t}$. For later convenience we define $k$ to be half of the input photon flux ${k}=|\alpha_{{in}}|^2/2$ so $\Xa^2=4{k}\Delta{t}$.


\hypertarget{S3B}{}\subsection{B. Stochastic master equation in a rotating frame}\label{sec:S3B}

\subsubsection{A general continuous homodyne measurement}\label{SMEbaby}
To derive the SME for our continuous nonlinear cavity optomechanical general-dyne-based position measurement, we first consider the SME that results from a continuous homodyne measurement of the Hermitian observable $X=X^{\dag}$. Over a time increment $\Delta{t}$, the measurement operator for such a measurement is given by the following
\begin{eqnarray}
\upsilon(\Delta{t})&=&\dfrac{1}{\pi^{\frac{1}{4}}}\exp\left[-2k\Delta{t}(X-\bar{X})^2+4\rmi{k}\Delta{t}\bar{X}Y-2\rmi{k}\Delta{t}XY\right]\label{upsilonbaby1},\\
\bar{X}&=&\expval{X}+\dfrac{1}{\sqrt{8k}}\dfrac{\Delta W}{\Delta t}\label{upsilonbaby2},\\
c_{\upsilon}&=&\sqrt{2k}(X+\rmi Y)\label{upsilonbaby3}.
\end{eqnarray}
This measurement operator is similar to the operator used in~[\href{https://doi.org/10.1080/00107510601101934}{K. Jacobs and D. A. Steck, Contemp. Phys. \textbf{47}, 279 (2006)}] to derive continuous position measurement but with additional unitary terms that describe homodyne-measurement back action. Here, $k$ is the measurement rate, $\bar{X}$ is the measurement record, $Y$ is another Hermitian operator, satisfying $[X,Y]=0$, that leads to back action, and $c_{\upsilon}$ will turn out to be the measurement output operator. The Wiener increment $\Delta{W}$ is a Gaussian random variable with zero mean and variance $\Delta{t}$, which obeys the It\^{o} rule $(\Delta{W})^2=\Delta{t}$. The It\^{o} rule is strictly only true in the limit $\Delta{t}\rightarrow\rmd{t}$, $\Delta{W}\rightarrow\rmd{W}$, but we will take this limit shortly in order to derive the SME.

We may write the homodyne measurement operator as $\upsilon(\Delta{t})\propto\rme^{-2k\Delta t[X^2-2\expval{X}(X+\rmi Y)+\rmi XY]+\sqrt{2k}(X+\rmi Y)\Delta W}$, ignoring terms in the exponential which are not operators and so will only effect the state normalization. Then considering an infinitesimal change in the state $\ket{\psi(t+\rmd t)}=\ket{\psi(t)}+\rmd\ket{\psi}=\mathcal{N}\lim_{\Delta{t}\rightarrow\rmd{t}}\upsilon(\Delta t)\ket{\psi(t)}$, and expanding to first order in $\Delta t$ with the It\^{o} rule, gives
\begin{eqnarray}
&&\ket{\psi(t+\rmd t)}=\ket{\psi(t)}+\rmd\ket{\psi}\nonumber\\
&=&\lim_{\Delta{t}\rightarrow\rmd{t}}\mathcal{N}\left\{1-k\Delta t[X^2+Y^2-4\expval{X}(X+\rmi Y)]+\sqrt{2k}(X+\rmi Y)\Delta W\right\}\ket{\psi(t)}.
\end{eqnarray}
Here, the normalization of the state is given by $\mathcal{N}=1-k\Delta t\expval{X}^2-\sqrt{2k}\expval{X}\Delta W$. This leads to the stochastic Schr\"{o}dinger equation for a general continuous homodyne measurement
\begin{eqnarray}
\rmd\ket{\psi}&=&\left\{-k[(X-\expval{X})^2+Y^2-2\rmi\expval{X}Y]\rmd t+\sqrt{2k}[X-\expval{X}+\rmi Y]\rmd W\right\}\ket{\psi}.
\end{eqnarray}
Following the derivation in~[\href{https://doi.org/10.1080/00107510601101934}{K. Jacobs and D. A. Steck, Contemp. Phys. \textbf{47}, 279 (2006)}], we may use  $\rmd\rho=\rmd\ket{\psi}\bra{\psi}+\ket{\psi}\rmd\bra{\psi}+\rmd\ket{\psi}\rmd\bra{\psi}$ to derive the SME for general homodyne measurement
\begin{eqnarray}
\rmd\rho&=&-k[X^2\rho+\rho X^2 -2X\rho X+Y^2\rho+\rho Y^2 -2Y\rho Y-2\rmi Y\rho X+2\rmi X\rho Y]\rmd{t}+\sqrt{2k}[(X+\rmi Y)\rho+\rho(X-\rmi Y)-2\expval{X}\rho]\rmd{W},\nonumber\\
&=&\mathcal{D}[c_{\upsilon}]\rho\rmd{t}+\mathcal{H}[c_{\upsilon}]\rho\rmd{W}\label{SMEbabyeqn}.
\end{eqnarray}
Here, we see that $c_{\upsilon}$ is the measurement output operator. The first term in Eq.~\eqref{SMEbabyeqn} represents measurement back action and the Lindblad superoperator is given by $\mathcal{D}[O]\rho=O\rho{O^{\dag}}-\frac{1}{2}\{O^{\dag}O,\rho\}$. Furthermore, the second term in Eq.~\eqref{SMEbabyeqn} describes localization of $X$, with the measurement superoperator being $\mathcal{H}[O]\rho=O\rho+\rho{O^{\dag}}-\expval{O+O^{\dagger}}\rho$.

\subsubsection{Continuous general-dyne measurement}
Writing out the individual homodyne measurement operators $\Upsilon_{I}(\Xl)$ and $\Upsilon_{I}(\Pl)$ in the form of Eqs~\eqref{upsilonbaby1}, \eqref{upsilonbaby2}, and \eqref{upsilonbaby3} gives
\begin{eqnarray}
\Upsilon_{I}({\Xl})&=&\dfrac{1}{{\pi}^{\frac{1}{4}}}\exp\Big[-\frac{a_{\alpha}^2}{2}(\A(\Xmr)-\frac{\Xl}{a_{\alpha}})^2+\rmi{a_{\alpha}^2}\frac{\Xl}{a_{\alpha}}\B(\Xmr)-\frac{\rmi}{2}a_{\alpha}^2\A(\Xmr)\B(\Xmr)\Big],\\
\frac{\Xl}{a_{\alpha}}&=&\expval{\A(\Xmr)}+\dfrac{1}{\sqrt{8{k}(1-\zeta^2)}}\dfrac{\Delta W_{\Xl}}{\Delta t},\\
\tilde{c}_{\Xl}&=&\sqrt{2{k}(1-\zeta^2)}f(\Xmr),
\end{eqnarray}
and 
\begin{eqnarray}
\Upsilon_{I}({\Pl})&=&\dfrac{1}{{\pi}^{\frac{1}{4}}}\exp\Big[-\frac{b_{\alpha}^2}{2}(\B(\Xmr)-\frac{\Pl}{b_{\alpha}})^2-\rmi{b_{\alpha}^2}\frac{\Pl}{b_{\alpha}}\A(\Xmr)+\frac{\rmi}{2}b_{\alpha}^2\A(\Xmr)\B(\Xmr)\Big],\\
\frac{\Pl}{b_{\alpha}}&=&\expval{\B(\Xmr)}+\dfrac{1}{\sqrt{8{k}\zeta^2}}\dfrac{\Delta W_{\Pl}}{\Delta t},\\
\tilde{c}_{\Pl}&=&-\rmi\sqrt{2{k}\zeta^2}f(\Xmr).
\end{eqnarray}
Here, the $\Delta{t}$ dependence comes in through $\Xa^2=4{k}\Delta{t}$ to give $a_{\alpha}^2=4{k}(1-\zeta^2)\Delta{t}$ and $b_{\alpha}^2=4{k}\zeta^2\Delta{t}$. The Wiener increments from the two homodyne measurements are independent $\Delta{W}_{i}\Delta{W}_{j}=\delta_{i,j}\Delta{t}$.

Then using the result of Eq.~\eqref{SMEbabyeqn}, we have that in the rotating frame, the continuous general-dyne measurement produces the nonlinear SME
\begin{eqnarray}
\rmd\rhor&=&\mathcal{D}[\tilde{c}_{\Xl}]\rhor\rmd{t}+\mathcal{D}[\tilde{c}_{\Pl}]\rhor\rmd{t}+\mathcal{H}[\tilde{c}_{\Xl}]\rhor\rmd{W_{\Xl}}+\mathcal{H}[\tilde{c}_{\Pl}]\rhor\rmd{W_{\Pl}}\\
&=&2{k}\left(f{(\Xmr)}\rhor f{(-\Xmr)}-\rhor\right)\rmd t \nonumber\\
&+&\sqrt{2{k}(1-\zeta^2)}\left(f{(\Xmr)}\rhor+\rhor f{(-\Xmr)}-2\expval{\A(\Xmr)}\rhor\right)\rmd{W_{\Xl}}\nonumber\\
&+&\sqrt{2{k}\zeta^2}\left(-\rmi f(\Xmr)\rhor+\rmi\rhor f{(-\Xmr)}-2\expval{\B(\Xmr)}\rhor\right)\rmd{W_{\Pl}},
\end{eqnarray}
where $\rhor$ is the mechanical state in the rotating frame. Then introducing $\tilde{c}=\sqrt{2{k}}f{(\Xmr)}$, $\tilde{c}_{\Xl}=\sqrt{1-\zeta^2}~\tilde{c}$, and $\tilde{c}_{\Pl}=-\rmi\zeta \tilde{c}$, allows us to write the nonlinear SME as
\begin{eqnarray}
\rmd\rhor&=&\mathcal{D}[\tilde{c}]\rhor\rmd{t}+\sqrt{1-\zeta^2}\mathcal{H}[\tilde{c}]\rhor\rmd{W_{\Xl}}+\zeta\mathcal{H}[-\rmi \tilde{c}]\rhor\rmd{W_{\Pl}},
\end{eqnarray}
with $\rmd{W}_{i}\rmd{W}_{j}=\delta_{i,j}\rmd{t}$, $\mathds{E}[\rmd{W}_{i}]=0$, for $i,j=\Xl,\Pl$. Here, the stochastic average is represented by $\mathds{E}$ to differentiate from the quantum expectation value.

\hypertarget{S3C}{}\subsection{C. Stochastic master equation in the lab frame}\label{sec:S3C}
The state in the lab frame is $\rho(t)=U_{{F}}(t)\tilde\rho(t) U_{{F}}^{\dagger}(t)$, so then
\begin{eqnarray}
\rho(t+\rmd{t})&=&U_{{F}}(t+\rmd{t})\tilde\rho(t+\rmd{t}) U_{{F}}^{\dagger}(t+\rmd{t})\nonumber\\
&=&U_{{F}}(t)U_{{F}}(\rmd{t})\tilde\rho(t)U_{{F}}^{\dagger}(t)U_{{F}}^{\dagger}(\rmd{t})+U_{{F}}(t)\rmd\tilde\rho(t)U_{{F}}^{\dagger}(t)\nonumber\\
&=&\left(1-\rmi{H_0}/\hbar\rmd{t}\right)\rho(t)\left(1+\rmi{H_0}/\hbar\rmd{t}\right)+U_{{F}}(t)\rmd\tilde\rho(t)U_{}^{\dagger}(t),
\end{eqnarray}
where $H_0=\hbar\wm b^{\dagger}b$ is the free mechanical Hamiltonian and terms higher order than $\rmd{t}$ have been dropped. Note, the cavity part of $H_{0}$ does not act on the mechanical subspace so is ignored here. Therefore, as $\rho(t+\rmd{t})=\rho(t)+\rmd{\rho}$, we can identify
\begin{eqnarray}
\rmd{\rho}=-\frac{\rmi}{\hbar}[H_0,\rho]\rmd{t}+U_{{F}}(t)\rmd\tilde\rho(t)U_{{F}}^{\dagger}(t).
\end{eqnarray}
by using 
$U_{{F}}^{\dagger}U_{{F}}=\mathds{1}$, $\Xm^{n}=U_{{F}}{\Xmr}^{n}U_{{F}}^{\dagger}$, and $c=U_{{F}}\tilde{c}U_{{F}}^{\dagger}=\sqrt{2{k}}f{(\Xm)}$, we have that $
U_{{F}}\mathcal{D}[\tilde{c}]\tilde\rho U_{{F}}^{\dagger}=\mathcal{D}[c]\rho$. Then using that expectation values are the same in the rotating and the lab frame $\expval{\tilde{c}}=\tr\left(\tilde{c}\tilde\rho\right)=\tr\left(U_{{F}}^{\dagger}{c}U_{{F}}U_{{F}}^{\dagger}\rho U_{{F}}\right)
=\tr\left({c}\rho\right)=\expval{c}$,
we have that $U_{{F}}\mathcal{H}[\tilde{c}]\tilde\rho U_{{F}}^{\dagger}=\mathcal{H}[c]\rho$. Hence, we find that 
\begin{eqnarray}
U_{{F}}(t)\rmd\tilde\rho(t)U_{{F}}^{\dagger}(t)=\mathcal{D}[c]\rho\rmd{t}+\sqrt{1-\zeta^2}\mathcal{H}[c]\rho\rmd{W_{\Xl}}+\zeta\mathcal{H}[-\rmi c]\rho\rmd{W_{\Pl}},
\end{eqnarray}
which shows that the nonlinear SME for an ideal continuous optomechanical general-dyne position measurement is 
\begin{eqnarray}
\rmd\rho&=&-\frac{\rmi}{\hbar}[H_{0},\rho]\rmd{t}+\mathcal{D}[c]\rho\rmd{t}+\sqrt{1-\zeta^2}\mathcal{H}[c]\rho\rmd{W_{\Xl}}+\zeta\mathcal{H}[-\rmi c]\rho\rmd{W_{\Pl}}\label{SMEnoloss}.
\end{eqnarray}

\subsubsection{Mechanical dissipation, inefficient detection and the total stochastic master equation}
Inefficient detection is accounted for by the transformation $\mathcal{H}[O]\rightarrow\sqrt{\eta}\mathcal{H}[O]$, where $\eta$ is the detection efficiency [{\href{https://www.google.co.uk/books/edition/Quantum_Measurement_and_Control/8H8hAwAAQBAJ?hl=en&gbpv=0}{H. M. Wiseman and G. J. Milburn, \emph{Quantum measurement and control} (Cambridge university press, 2009)}}]. In general, the efficiencies of the $\Pl$ and $\Xl$ homodyne measurement may be different, so we label these efficiencies $\eta_{\Xl}$ and $\eta_{\Pl}$, respectively. While mechanical interactions with the environment may be accounted for with an additional Lindblad dissipator $L=\sqrt{\dfrac{4\gamma \kb T}{\hbar\wm}}\Xm+\rmi\sqrt{\dfrac{\hbar\wm \gamma}{4 \kb T}}\Pm$, which models quantum Brownian motion~[{\href{https://www.google.co.uk/books/edition/Decoherence/URAchzQIsTgC?hl=en&gbpv=0}{M.~A.  Schlosshauer, \emph{Decoherence: and  the quantum-to-classical transition} (Springer  Science  \&  Business  Media, 2007)}}]. Here, $T$ is the effective temperature of the environment and $\gamma$ is the mechanical decay rate. To summarize, the total SME for our general-dyne measurement is
\begin{eqnarray}
&&\rmd\rho=-\frac{\rmi}{\hbar}[H_{0},\rho]\rmd{t}+\mathcal{D}[c]\rho\rmd{t}+\mathcal{D}[L]\rho\rmd{t}+\sqrt{(1-\zeta^2)\eta_{\Xl}}\mathcal{H}[c]\rho\rmd{W_{\Xl}}+\zeta\sqrt{\eta_{\Pl}}\mathcal{H}[-\rmi c]\rho\rmd{W_{\Pl}}.\label{SMEloss}
\end{eqnarray}

\hypertarget{S4}{}\section{IV. Pulsed measurement}\label{sec:S4}
For a pulsed optomechanical measurement we will consider the case of a resonant input pulse $\Delta=0$ as we also assume an initial mechanical state with $\expval{\Xm}=0$, this setting maximizes the amount of light that can enter the cavity.

\hypertarget{S4A}{}\subsection{A. Krauss map and heralding probability}\label{sec:S4A}
The action of a nonlinear pulsed measurement on an initial mechanical state $\rhoin$ is described by the Krauss map
\begin{eqnarray}
\rhof&=&\dfrac{\Upsilon\rhoin\Upsilon^{\dagger}}{\Ph}\label{kraussmap},
\end{eqnarray}
where $\Upsilon$ is given by Eq.~\eqref{pulsedupsilon}, $\rhof$ is the mechanical state after the measurement, and $\Ph=\tr\left(\Upsilon^{\dagger}\Upsilon\rhoin\right)$ is the heralding probability for a particular $(\Xl,\Pl)$ general-dyne outcome. Eq.~\eqref{kraussmap} implies that the position probability changes according to $\Pf(\Xm)={{F}(\Xm)\Pin(\Xm)}/{\Ph}$, where $\Pin(\Xm)=\bra{\Xm}\rhoin\ket{\Xm}$, $\Pf(\Xm)=\bra{\Xm}\rhof\ket{\Xm}$, and as $[\Upsilon,\Xm]=0$ implies $\Upsilon\ket{\Xm}=\ket{\Xm}\Upsilon(\Xm)$, the filtering function may be written as ${F}(\Xm)=\Upsilon^{\dagger}(\Xm)\Upsilon(\Xm)$, which filters the initial probability distribution to produce the final mechanical position probability. Explicitly, the filtering function is
\begin{eqnarray}
{F}&=&\frac{1}{\pi}\exp\left[{-(\Xl-a_{\alpha}{\A}(\Xm))^2-(\Pl-b_{\alpha}{\B}(\Xm))^2}\right]\label{filterNG},
\end{eqnarray}
where $\A$ and $\B$ are now functions of the variable $\Xm$ and are not operators.

\begin{figure}[h]
    \centering
    \includegraphics[width=0.7\textwidth]{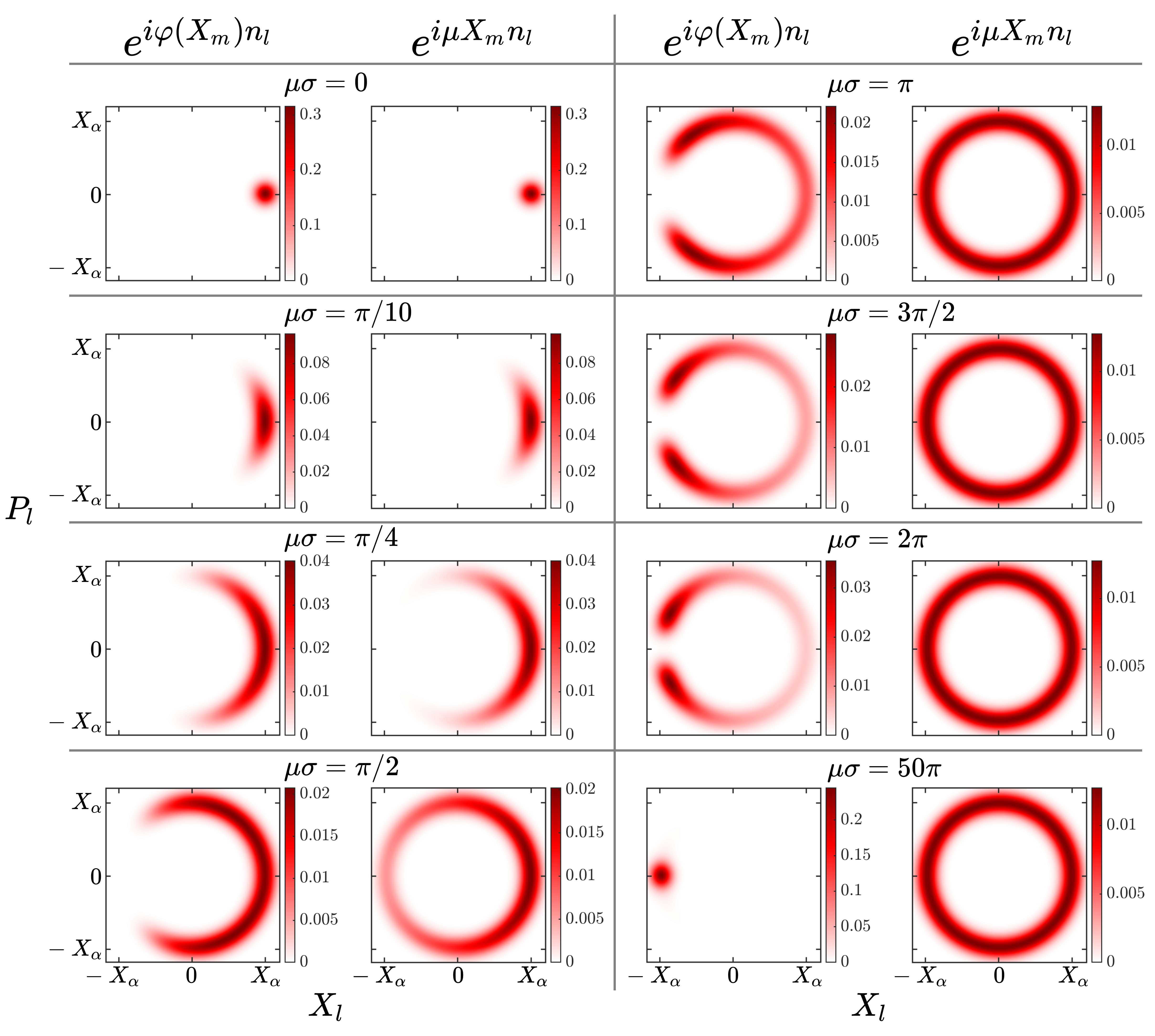}
    \caption{$Q$ functions of the optical state after the optomechanical interaction. Here, $\Xa=10$ and various values of $\mu\sigma$ are shown, where $\sigma^2$ is the initial variance of the mechanical state. $\rme^{\rmi\varphi(\Xm)\nl}$ columns: the $Q$ functions for a nonlinear optomechanical interaction that incorporates the nonlinear cavity input-output function. $\rme^{\rmi\mu\Xm\nl}$ columns: the same case but with the nonlinear optomechanical unitary approximated to first order in position $\Xm$. Note that in the expansion of $\varphi(\Xm)$, the second order term in $\Xm$ with $\Delta=0$ is zero. The $\pi$-phase limit is seen at large optical rotations, for example at $\mu\sigma=50\pi$, the optical coherent state has almost fully rotated $\pm\pi$ in phase space. An animation of the Q function with increasing $\mu\sigma$ is provided as a supplementary file.}
    \label{fig_4_supp}
\end{figure}

\hypertarget{S4B}{}\subsection{B. The Husimi-\emph{Q} function and $\pi$-phase limit}\label{sec:S4B} 

\subsubsection{Mathematical expression for the Husimi-\emph{Q} function}
The Husimi-$Q$ function of the reduced state of the optical mode $\rhol$ after the optomechanical interaction is $Q(\beta)=\frac{1}{2\pi}\bra{\beta}\rhol\ket{\beta}$,
where $\rhol=\tr_{{m}}\left(U\ket{\alpha}\bra{\alpha}\rhoin U^{\dagger}\right)$.
Therefore
\begin{eqnarray}
Q(\beta)&=&\frac{1}{2\pi}\int\rmd{\Xm}~\Pin(\Xm)|\bra{\beta}\ket{\alpha(\A+\rmi \B)}|^2\nonumber\\
&=&\frac{1}{2\pi}\int\rmd{\Xm}~\Pin(\Xm)\rme^{-|\beta-\alpha(\A+\rmi \B)|^2}\label{qeqn}.
\end{eqnarray}
Then writing the complex amplitudes as $\alpha(\A+\rmi \B)=\Xa(\A+\rmi \B)/\sqrt{2}$ and $\beta=(X_{\beta}+\rmi P_{\beta})/\sqrt{2}$ gives
\begin{eqnarray}
Q(X_{\beta},P_{\beta})&=&\frac{1}{2\pi}\int\rmd{\Xm}~\Pin(\Xm)\rme^{-\frac{1}{2}(X_{\beta}-\Xa \A)^2-\frac{1}{2}(P_{\beta}-\Xa \B)^2}.
\end{eqnarray}

Fig.~\ref{fig_4_supp} shows the $Q$ functions of the optical state after the optomechanical interactions and compares the results of this work with that induced by the unitary $\rme^{\rmi\mu\Xm\nl}$, which does not include the nonlinearity of the cavity response, demonstrating the nonlinear optomechanical phase shift. Here, we give $\mu\sigma$ in multiples of $\pi$ as when $\rme^{\rmi\mu\Xm\nl}$ is valid, $\mu\Xm$ is the optical phase and so $\mu\sigma$ represents a standard deviation in the optical phase. 

\subsubsection{Large $\mu\sigma$ and the $\pi$-phase limit}

\begin{figure}[h]
    \centering
    \includegraphics[width=\textwidth]{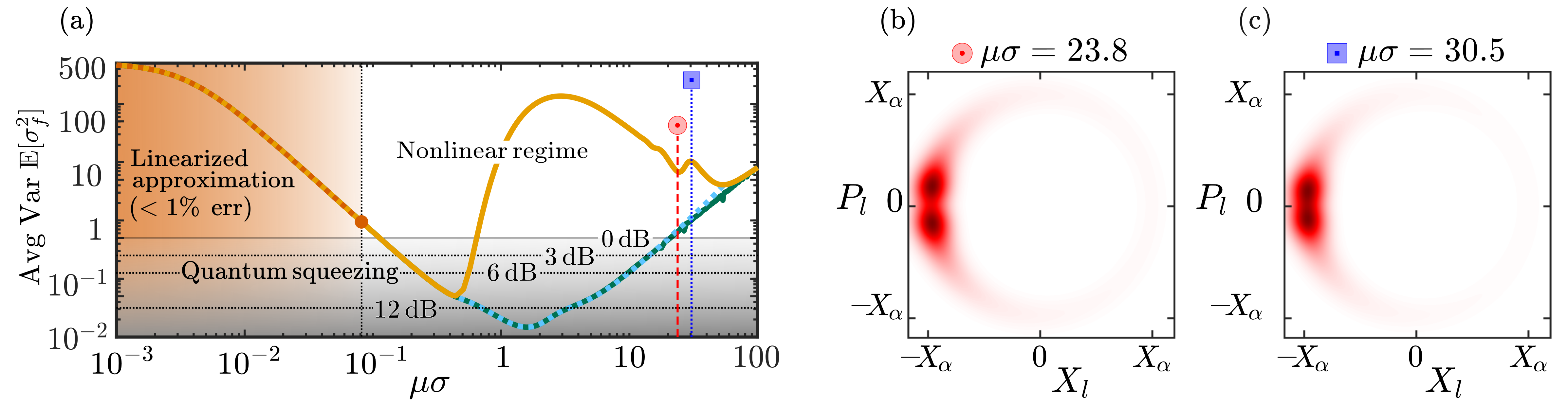}
    \caption{(a) Extended plot of the mechanical position variance averaged over all measurement outcomes $\mathds{E}[{\sigmaf^2}]$ up to $\mu\sigma=100$. Here, the $\mathds{E}[{\sigmaf^2}]$ curves for $\zeta_{\mathrm{opt}}$ and $\zeta=1$ approach each other as $\mu\sigma$ increases. Non-monotonic behaviour is also observed for the $\zeta=1$ curve between $\mu\sigma\approx20$ and $\mu\sigma\approx30$, which may be understood from the Husimi-$Q$ functions of the optical field. (b) A local minimum is located in the $\mathds{E}[{\sigmaf^2}]$ curve for $\zeta=1$  at $\mu\sigma=23.8$ (indicated by the red circle). For this value of $\mu\sigma$, a phase quadrature measurement can effectively discriminate between regions of anticlockwise and clockwise optical rotations, which are induced by positive and negative mechanical positions, respectively. (c) A local maximum is located in the $\mathds{E}[{\sigmaf^2}]$ curve for $\zeta=1$ at $\mu\sigma=30.5$ (indicated by the blue square). In this case, the regions of anticlockwise and clockwise optical rotations begin to significantly overlap and thus the phase-quadrature measurement performs more poorly.}
    \label{fig_VI_response}
\end{figure}

For any finite value of $\Xm$, in the limit $\mu\rightarrow\infty$, we have $\A\rightarrow-1$ and $\B\rightarrow0$. Hence, in this limit $\Ph={F}=\frac{1}{\pi}\exp\left[-(\Xl+a_{\alpha})^2-\Pl^2\right]$ and $\Pf(\Xm)=\Pin(\Xm)$. This means the optical coherent state has fully rotated $\pm\pi$ in phase space (see Fig.~\ref{fig_4_supp}) and no information is gained about the mechanics. More specifically, as $\mu\sigma\rightarrow\infty$ and $\Pf(\Xm)\rightarrow\Pin(\Xm)$, we have $\mathds{E}[{\sigmaf^2}]\rightarrow\sigma^2$ regardless of the value of $\zeta$. Thus, in the high $\mu\sigma$ limit all $\mathds{E}[{\sigmaf^2}]$ curves must approach each other, including the $\zeta_{\mathrm{opt}}$ and $\zeta=1$ curves as shown in the extended plot of $\mathds{E}[{\sigmaf^2}]$ in Fig.~\ref{fig_VI_response}(a). However, we remind the reader that our single cavity mode treatment is only accurate for $\mu\sigma\ll2\mathcal{F}_{c}$.

Interestingly, by extending the plot $\mathds{E}[{\sigmaf^2}]$ as in Fig.~\ref{fig_VI_response}(a), we observe non-monotonic behaviour in the $\mathds{E}[{\sigmaf^2}]$ curve for $\zeta=1$ between $\mu\sigma\approx20$ and $\mu\sigma\approx30$. This non-monotonic behaviour can be understood from the optical phase-space plots in Figs~\ref{fig_VI_response}(b) and (c). As shown in  Fig.~\ref{fig_VI_response}(b), at $\mu\sigma\approx20$ a phase-quadrature measurement is effective at discriminating between the regions of anticlockwise and clockwise optical rotation, which originate from the positive and negative values of $\Xm$ from the initial mechanical distribution, respectively. Importantly, as $\mu\sigma$ increases from $\mu\sigma\approx20$, the overlap between these two regions also increases (cf. the Supplemental animation). Therefore, a phase-quadrature measurement is less effective at discriminating between different mechanical positions and thus less effective at localizing the initial mechanical state.

\hypertarget{S4C}{}\subsection{C. Gaussian approximation to pulsed measurement}\label{sec:S4C}

\subsubsection{General-dyne coordinates}
By computing $\dfrac{\partial{F}}{\partial \Xm}|_{\Xl,\Pl}=0$ we find 
\begin{eqnarray}
b_{\alpha}\Pl \A-a_{\alpha}\Xl \B+(a_{\alpha}^2-b_{\alpha}^2)\A\B=0.\label{shortellipse}
\end{eqnarray}
We would like the expand the filtering function about its maximum to proceed with a Gaussian approximation to the general-dyne measurement. Here, we motivate the use of `general-dyne coordinates'. Let us first parametrize the $\A$ and $\B$ functions by a new variable $z$, i.e.
\begin{eqnarray}
\A(z)&=&\dfrac{1-(z/2)^2}{1+(z/2)^2},\\
\B(z)&=&\dfrac{z}{1+(z/2)^2}.
\end{eqnarray}
 Then, if we put $\Xl=a_{\alpha}\A(z)+\delta_{\Xl}$ and $\Pl=b_{\alpha}\B(z)+\delta_{\Pl}$ with $\delta=\sqrt{\delta_{\Xl}^2+\delta_{\Pl}^2}$ into Eq~\eqref{shortellipse} we find $\delta_{\Pl}=\dfrac{a_{\alpha}\B}{b_{\alpha}\A}\delta_{\Xl}$. Thus, we have two similar right-angled triangles with acute angle $\psi$ satisfying
\begin{eqnarray}
\sin(\psi)&=&\dfrac{a_{\alpha}\B}{\sqrt{b_{\alpha}^2\A^2+a_{\alpha}^2\B^2}}=\dfrac{\delta_{\Pl}}{\delta},\\
\cos(\psi)&=&\dfrac{b_{\alpha}\A}{\sqrt{b_{\alpha}^2\A^2+a_{\alpha}^2\B^2}}=\dfrac{\delta_{\Xl}}{\delta},\\
\tan(\psi)&=&\dfrac{a_{\alpha}\B}{b_{\alpha}\A}=\dfrac{\delta_{\Pl}}{\delta_{\Xl}}.
\end{eqnarray}
This motivates choosing the general-dyne coordinates $(z,\delta)$ to parametrize the measurement space
\begin{eqnarray}
\Xl=a_{\alpha}\A(z)+\delta\dfrac{b_{\alpha}\A(z)}{\sqrt{b_{\alpha}^2\A^2+a_{\alpha}^2\B^2}},\\
\Pl=b_{\alpha}\B(z)+\delta\dfrac{a_{\alpha}\B(z)}{\sqrt{b_{\alpha}^2\A^2+a_{\alpha}^2\B^2}},
\end{eqnarray}
with $z\in(-\infty,+\infty)$ and  $\delta\in[-\min(a,b),\infty)$. The Jacobian determinant for the transformation from $(\Xl,\Pl)$ space to general-dyne coordinates $(z,\delta)$ is
\begin{eqnarray}
\det\,J=\dfrac{4}{4+z^2}\left[\sqrt{a_{\alpha}^2\B^2+b_{\alpha}^2\A^2}+\delta\dfrac{a_{\alpha}b_{\alpha}}{a_{\alpha}^2\B^2+b_{\alpha}^2\A^2}\right].
\end{eqnarray} 

\subsubsection{Gaussian approximation to ${F}$, $\Pf(\Xm)$, and $\Ph$ for a $(z,\delta)$ outcome}
Inserting the general-dyne coordinates into Eq.~\eqref{filterNG} and expanding to second order about the maximum of the filtering function at $\mu\Xm=z$, $\delta=0$ gives
\begin{eqnarray}
{F}&=&\dfrac{1}{\pi}\rme^{-\delta^2-\beta(\mu\Xm-z)^2},\\
\beta&=&\dfrac{16}{(4+z^2)^2}\left[a_{\alpha}^2\B^2(z)+b_{\alpha}^2\A^2(z)+\delta\dfrac{a_{\alpha}b_{\alpha}}{\sqrt{a_{\alpha}^2\B^2(z)+b_{\alpha}^2\A^2(z)}}\right].
\end{eqnarray}
Within this Gaussian approximation, for an initial mechanical Gaussian state $\Pin(\Xm)=\rme^{-\Xm^2/2\sigma^2}/{\sqrt{2\pi\sigma^2}}$, the output position probability distribution is given by
\begin{eqnarray}
\Pf(\Xm)&=&\dfrac{1}{\sqrt{2\pi\sigmaf^2}}\exp\left[{-\dfrac{(\Xm-\bar{X})^2}{2\sigmaf^2}}\right],\\
\sigmaf^2&=&\dfrac{\sigma^2}{1+2\mu^2\sigma^2\beta},\\
\bar{X}&=&\dfrac{z}{\mu}\dfrac{2\mu^2\sigma^2\beta}{1+2\mu^2\sigma^2\beta}.
\end{eqnarray}
While the heralding probability is given by \begin{eqnarray}
\Ph&=&\int\rmd\Xm~{F}(\Xm)\Pin(\Xm)\\
&=&\dfrac{1}{\pi\sqrt{1+2\mu^2\sigma^2\beta}}\exp\left[-\delta^2-\dfrac{\beta}{1+2\mu^2\sigma^2\beta}z^2\right].
\end{eqnarray} 
Averaging over all measurement outcomes $(z,\delta)$ gives
\begin{eqnarray}
\mathds{E}[{\sigmaf^2}]=\int\rmd{z}\int\rmd{\delta}{\Ph(z,\delta)}|\det\,J|\sigmaf^2.
\end{eqnarray}

\hypertarget{S4D}{}\subsection{D. Optical loss}\label{sec:S4D}
To study the effect of optical losses on the position measurement we employ a beamsplitter model for loss after the optomechanical interaction~[{\href{https://www.google.co.uk/books/edition/Measuring_the_Quantum_State_of_Light/wmsJy1A_cyIC?hl=en&gbpv=0}{U. Leonhardt, \emph{Measuring the quantum state of light}, Vol. 22 (Cambridge university press, 1997)}}]. To this end, we introduce an extra environmental vacuum mode that interacts with the output optical mode on a beamsplitter of transmission coefficient $\eta$. We then consider a position-quadrature measurement $\bra{Z}$ made by the environment that will be traced over at a subsequent step. The measurement operator now becomes $\Upsilon'=\bra{Z}\bra{\Xl}\bra{\Pl}\Ubs B_{\eta}U\ket{\alpha}\ket{0}\ket{0}=\Theta_{Z}\Upsilon_{\eta}$ where
\begin{eqnarray}
\Upsilon_{\eta}&=&\bra{\Xl}\bra{\Pl}\Ubs\ket{\eta\alpha(\A+\rmi \B)}\ket{0}\label{krausseta}\\
\Theta_{Z}&=&\bra{Z}\ket{\sqrt{1-\eta^2}\alpha(\A+\rmi \B)},
\end{eqnarray}
and $B_{\eta}$ is the beamsplitter modelling loss. To account for loss of information we must trace over all possible measurements the environment can make $Z\in(-\infty,+\infty)$. Note that as $\Upsilon_{\eta}$ and $\Theta_{Z}$ are both functions of the position operator $\Xm$, $[\Upsilon_{\eta},\Theta_{Z}]=0$, and the measurement map may be written as
\begin{eqnarray}
\rhof'&=&\frac{1}{\Ph'}\Upsilon_{\eta}\left(\int\rmd{Z}~\Theta_{Z}\rhoin\Theta_{Z}^{\dagger}\right)\Upsilon_{\eta}^{\dagger}\label{kraussetamap},
\end{eqnarray}
with $\Ph'=\tr\left(\int\rmd{Z}~\Upsilon'^{\dagger}\Upsilon'\rhoin\right)$.

The measurement operator in Eq.~\eqref{krausseta} is simply the lossless operator $\Upsilon$ of Eq.~\eqref{pulsedupsilon} with a reduced optical amplitude $\alpha\rightarrow\eta\alpha$. Hence, we know the action of $\Upsilon_{\eta}$ on the position distribution of our initial state and so to study the effects of optical loss on the position measurement it suffices to analyze the transformation
\begin{eqnarray}
\rhoin'&=&\int\rmd{Z}~\Theta_{Z}\rhoin\Theta_{Z}^{\dagger}\label{lossmap},
\end{eqnarray}
which can be thought of as a change in the mechanical initial state $\rhoin$ due to optical loss by noting $\int\rmd{Z}~\Theta_{Z}^{\dagger}\Theta_{Z}=\mathds{1}$. As the operator $\Theta_{Z}$ is only a function of $\Xm$, $\Pin'(\Xm)=\bra{\Xm}\rhoin'\ket{\Xm}=\bra{\Xm}\rhoin\ket{\Xm}=\Pin(\Xm)$.

\hypertarget{S4E}{}\subsection{E. Phase quadrature measurements}\label{sec:S4E}

\begin{figure}[h]
    \centering
    \includegraphics[width=0.7\textwidth]{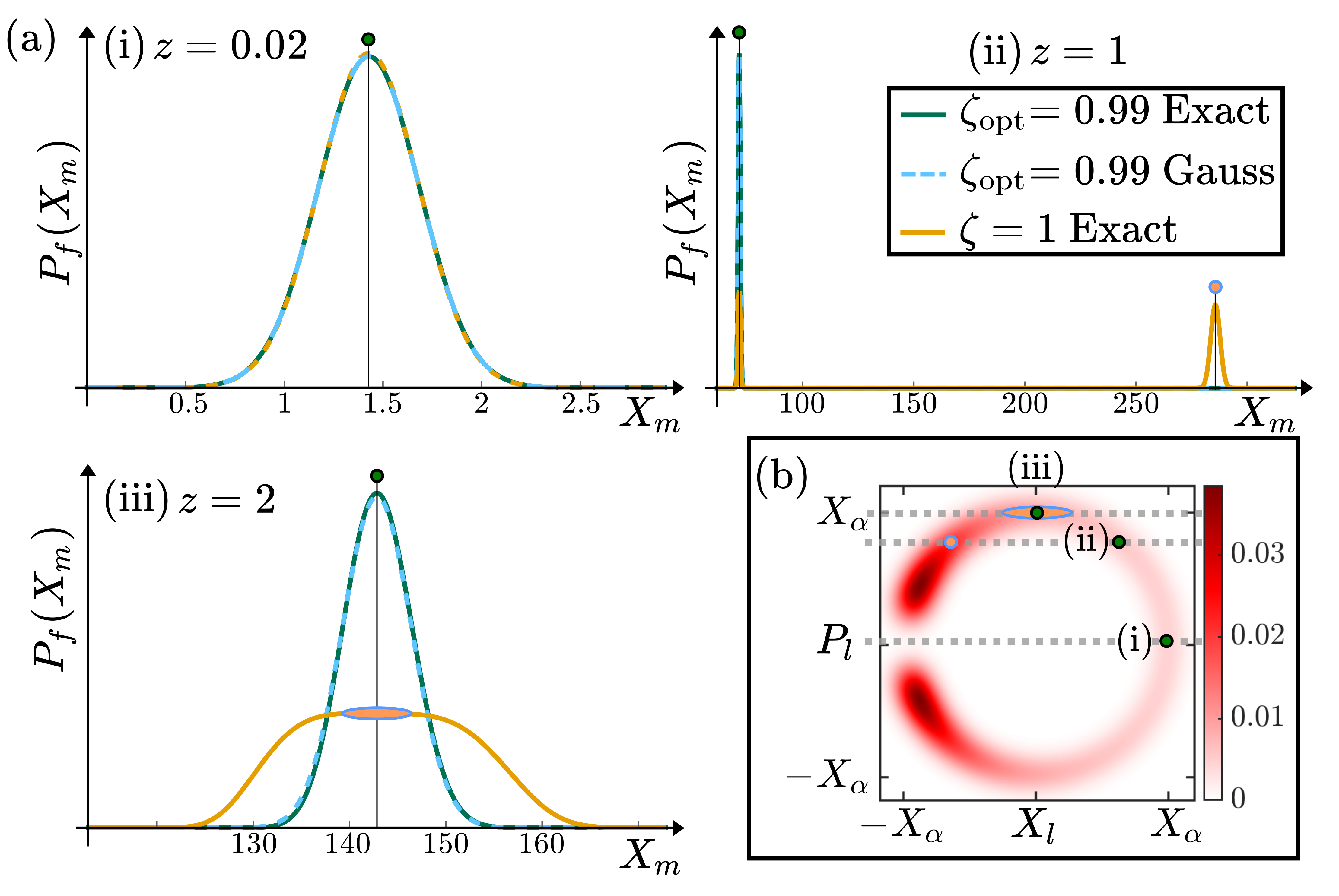}
    \caption{Comparison of phase quadrature measurements and the optimal measurement setting. (a) The output mechanical position probability distribution $P_{f}(X_m)$ for different measurement outcomes and comparing $\zeta=1$ with the optimal measurement setting $\zeta_{\mathrm{opt}}$.  Here, $\mu\sigma\approx7$ which gives $\zeta_{\mathrm{opt}}=0.99$ and we parameterize the measurement outcomes $(\Xl,\Pl)$ using the general-dyne coordinates $(z,\delta)$. Other parameters are the same as in Fig.~\ref{fig_2_main}. (i) $(z,\delta)=(0.02,0)$ or $(\Xl,\Pl)\approx(\sqrt{1-\zeta^2}\Xa,0.02\zeta\Xa)$: an outcome close to the location of the input optical coherent state $(\Xl,\Pl)=(\Xa,0)$ where $\zeta_{\mathrm{opt}}$ and $\zeta=1$ perform similarly. (ii) $(z,\delta)=(1,0)$ or $(\Xl,\Pl)=(0.6\sqrt{1-\zeta^2}\Xa,0.8\zeta\Xa)$: an outcome that produces a bimodal distribution when only the phase quadrature is measured. (iii) $(z,\delta)=(2,0)$ or $(\Xl,\Pl)=(0,\zeta\Xa)$: an outcome that produces a broad distribution when only the phase quadrature is measured. (b) The measurement outcomes (i), (ii), and (iii) are indicated by green circles on the phase-space of the output light. Meanwhile, the orange circle and ellipse indicate regions with the same $\Pl$ values as the measurement outcomes (ii) and (iii), respectively.}
    \label{fig_II_response}
\end{figure}

Despite $\zeta_{\mathrm{opt}}$ being close to 1 in the zoomed-in inset of Fig.~\ref{fig_2_main}(c), which spans $\mu\sigma=6.5$ to $\mu\sigma=10$, the $\mathds{E}[{\sigmaf^2}]$ curves corresponding to $\zeta_{\mathrm{opt}}$ and $\zeta=1$ are still very different in Fig.~\ref{fig_2_main}(b). We choose to investigate the case where $\mu\sigma\approx7$ and $\zeta_{\mathrm{opt}}=0.99$ in Fig.~\ref{fig_II_response} to explain this difference. In Fig.~\ref{fig_II_response}(a), we study the effect of the measurement on $P_{f}(\Xm)$ for three different measurement outcomes (i), (ii), (iii), which are represented by the green circles in Fig.~\ref{fig_II_response}(b). For simplicity, we choose outcomes with $\delta=0$ as $z$ is the important parameter here that quantifies the angle over which the initial coherent state ($z=0,\delta=0$) has rotated in phase space.

For measurement outcome (i), the performance of the optimal measurement and a phase quadrature measurement are very similar. The initial coherent state has only rotated a small angle, which may be well approximated by a phase displacement. For measurement outcome (ii), the phase quadrature measurement performs very poorly compared to the optimal measurement. The phase quadrature measurement is unable to localize the mechanical marginal to a single Gaussian peak and instead $P_{f}(\Xm)$ is bimodal. One peak of this bimodal distribution corresponds to the measurement outcome (ii) ($z=1$) and so is located in the same position as the narrow Gaussian produced by the optimal measurement setting $\zeta_{\mathrm{opt}}$. The other peak of this distribution corresponds to the point in phase space that has the same $\Pl$ value as $z=1$---as indicated by the orange circle in Fig.~\ref{fig_II_response}(b). The optimal measurement setting can discriminate between these two peaks by also measuring $\Xl$ quadrature. Finally, for measurement outcome (iii), the phase quadrature measurement produces a very broad $P_{f}(\Xm)$ distribution compared to the optimal setting. There is a broad region in the optical phase-space distribution that give the same measurement outcome of $\Pl=\zeta\Xa$, which can be seen by the large region that overlaps with the orange ellipse located at $\Pl=\Xa$ in Fig.~\ref{fig_II_response}(b). Within this broad region, each phase-space point in the continuum corresponds to a different value of mechanical position $\Xm$ and as the phase quadrature measurement is poor at discriminating between these points, the mechanical state is poorly localized. The quantity $\mathds{E}[{\sigmaf^2}]$ averages over all possible measurement outcomes and by measuring both the $\Xl$ and $\Pl$ quadratures, the optimal measurement setting $\zeta_{\mathrm{opt}}$ avoids the contributions from the bimodal and broad distributions of Figs~\ref{fig_II_response}(ii) and (iii).

\hypertarget{S4F}{}\subsection{F. Dependence on the detuning}\label{sec:S4F}


\begin{figure}[h]
    \centering
    \includegraphics[width=\textwidth]{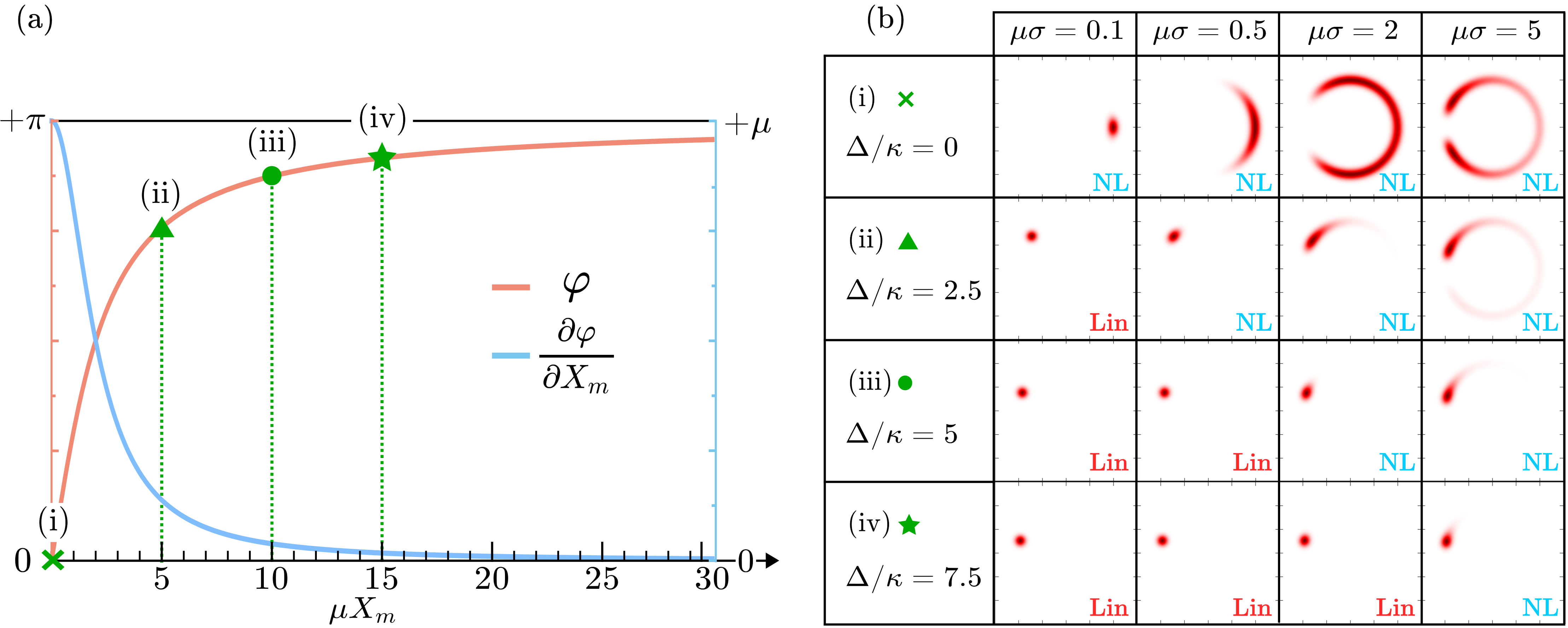}
    \caption{Dependence of the nonlinearity condition on the detuning. (a) Plot of the optical phase $\varphi(\Xm)$ and the dimensionless mechanical momentum kick per photon $\partial{\varphi}(\Xm)/\partial\Xm$. Note there is a factor of 2 difference between $\mu\Xm$ and $\Delta/\kappa$, i.e. $\Delta/\kappa\rightarrow\mu\Xm/2$---see the form of the nonlinear response function $f(\Xm)$. (b) Phase-space plots of the optical pulse after the optomechanical interaction for various values of $\mu\sigma$ and detunings $\Delta$. {Lin}: linear phase response. {NL}: nonlinear phase response. (i) $\Delta/\kappa=0$ ($\mu\Xm=0$) nonlinearities are observed for $\mu\sigma\gtrsim0.1$.  (ii) $\Delta/\kappa=2.5$ ($\mu\Xm=5$) nonlinearities are observed for $\mu\sigma\gtrsim0.5$. (iii) $\Delta/\kappa=5$ ($\mu\Xm=10$) nonlinearities are observed for $\mu\sigma\gtrsim2$. (iv) $\Delta/\kappa=7.5$ ($\mu\Xm=15$) nonlinearities are observed for $\mu\sigma\gtrsim5$.}
    \label{fig:V_response}
\end{figure}

In Fig.~\ref{fig_2_main} of the main text, we focused on the case where $\Delta=0$. However, the nonlinearity condition depends on the detuning $\Delta$ and becomes stricter as $|\Delta|$ increases away from zero, which we explore qualitatively in Fig.~\ref{fig:V_response}. In Fig.~\ref{fig:V_response}(a), we plot the optical phase response $\varphi(\Xm)$ and the mechanical momentum kick per photon $\partial{\varphi}(\Xm)/\partial\Xm$. While in Fig.~\ref{fig:V_response}(b), we plot the optical phase space of the light after the optomechanical interaction for various values of $\mu\sigma$ and 4 different detuning values (i)---(iv). A non-zero detuning $\Delta$ shifts the entire $\varphi(\Xm)$ curve along the $\mu\Xm$ axis by $-2\Delta/\kappa$, which may be seen from the form of the nonlinear response function $f(\Xm)$. Hence, a non-zero detuning $\Delta$ is equivalent to shifting the origin of the phase response $\mu\Xm=0\rightarrow\mu\Xm=2\Delta/\kappa$ and these shifts are indicated by the green markers in Fig.~\ref{fig:V_response}.

In the main text we find the nonlinearity condition at (i) $\Delta=0$ is $\mu\sigma\gtrsim0.1$. More precisely, in Fig.~\ref{fig_2_main} we define the nonlinearity condition as $\mu\sigma>0.08$ as the linearized approximation now deviates from the exact theory by $>1/\%$. This nonlinearity condition is further motivated by the experiment of Ref.~[\href{https://doi.org/10.1038/ncomms10988}{G. Brawley, \emph{et al.}, Nat. Commun. \textbf{7}, 10988 (2016)}], which, coincidentally, achieved $\mu\sigma=0.08$ and observed the quadratic and higher powers of the mechanical position $\Xm$ in the optical phase shift. In Fig.~\ref{fig:V_response}, the case where $\Delta=0$ is indicated by the green cross. At this detuning value, the optical phase is most sensitive to mechanical displacements and the mechanical momentum kick per photon is maximum. Notably, the nonlinear regime can be viewed as the regime where the optical mode experiences significant optical rotations that cannot be described by the linearized approximation. In Fig.~\ref{fig:V_response}(b), we see the optical phase space plot already shows some significant rotations at $\mu\sigma=0.1$, where mechanical information is not only encoded on solely the optical phase quadrature but also the orthogonal amplitude quadrature. Thus, we find the nonlinearity condition $\mu\sigma\gtrsim0.1$ just from observation of the phase space of the output optical pulse.

The next detuning value (ii) $\Delta/\kappa=2.5$ is indicated by the green triangle in Fig.~\ref{fig:V_response}. At $\mu\sigma=0.1$, the corresponding  phase-space plot does not show any significant optical rotations, whereas at $\mu\sigma=0.5$, the optical-phase space plot begins to show significant optical rotations due to the nonlinearities. By increasing $\Delta$ further we note that we are exploring regions of the optical phase response which are less sensitive to displacements of the mechanical position, i.e. $\partial{\varphi}(\Xm)/\partial\Xm$ is a decreasing function of $\Xm$. By again analyzing the phase-space plots of Fig.~\ref{fig:V_response}(b), we observe nonlinearities for (iii) $\Delta/\kappa=5$ (green circle) at $\mu\sigma\gtrsim2$ and for (iv) $\Delta/\kappa=7.5$ (green star) at $\mu\sigma\gtrsim5$. 
Thus, the factor by which the nonlinearity condition increases is not proportional to the difference in $\Delta/\kappa$ and this is due to the nonlinearity of the function $\varphi(\Xm)$.

\hypertarget{S5}{}\section{V. Continuous measurement}\label{sec:S5}

\hypertarget{S5A}{}\subsection{A. Agreement with the linearized regime of optomechanics and the linearized measurement rate}\label{sec:S5A}
For small optical rotations $\mu\sigma\ll1$ and zero detuning $\Delta=0$, $c=\sqrt{2{k}}f(\Xm)\approx\sqrt{2{k}}+\rmi\cl$ with $\cl=\sqrt{2\mu^2{k}}\Xm$. Using $\mathcal{H}[c+\gamma]=\mathcal{H}[c]$ and $\mathcal{D}[c+\gamma]\rho=\mathcal{D}[c]\rho+\frac{1}{2}[\gamma^{*}c-\gamma c^{\dagger},\rho]$, the nonlinear SME in Eq.~\eqref{SMEnoloss} returns the continuous SME from the linearized regime of optomechanics
\begin{eqnarray}
\rmd\rho=-\frac{\rmi}{\hbar}[H_{0},\rho]\rmd{t}+\rmi2\mu{k}[\Xm,\rho]\rmd{t}+\mathcal{D}[\cl]\rho\rmd{t}\nonumber
&&\\+\sqrt{1-\zeta^2}\mathcal{H}[\rmi\cl]\rho\rmd{W_{\Xl}}+\zeta\mathcal{H}[\cl]\rho\rmd{W_{\Pl}}.\label{SMElinearized}
\end{eqnarray}
Here, the term $\rmi2\mu{k}[\Xm,\rho]\rmd{t}$ represents a \emph{deterministic} momentum kick on the mechanical oscillator, which leads to a shift in the equilibrium mean position and is typically transformed away. Note this transformation is only valid if $\mu\sigma\ll1$.

In the linearized regime, the input optical drive is large and so the Heisenberg-Langevin equation for the intracavity field Eq.~\eqref{HLcav} can be split up into a part describing the large intracavity field $\alpha_{{c}}$ and a part describing quantum fluctuations. In this regime, the large intracavity mean field amplitude obeys 
\begin{eqnarray}
\dot{\alpha_{{c}}}=+\rmi \sqrt{2}g_{0}\alpha_{{c}}\Xm-\kappa\alpha_{{c}}+\sqrt{2\kappa}\alpha_{{in}},
\end{eqnarray}
where the input photon flux is $|\alpha_{in}|^2=2k$ as $|\alpha|^2=|\alpha_{in}|^2\rmd{t}$. Then for small optical rotations $\mu^2\sigma^2\sim{2g_{0}^2\expval{\Xm^2}}/{\kappa^2}\ll1$ the equation for this large amplitude is well approximated by
\begin{eqnarray}
\dot{\alpha_{{c}}}=-\kappa\alpha_{{c}}+\sqrt{2\kappa}\alpha_{{in}}
\end{eqnarray}
with steady state solution $\kappa|\alpha_{{c}}|^2=2|\alpha_{{in}}|^2$. Therefore, we have that ${k}=\dfrac{\kappa|\alpha_{{c}}|^2}{4}$. Using $\mu={\sqrt{8}g_{0}}/{\kappa}$ and $g=|\alpha_{{c}}|g_{0}$, we arrive at an expression for the measurement rate which is \emph{only} true in the linearized regime of optomechanics
\begin{eqnarray}
\mu^2{k}=\kl=\dfrac{2g^2}{\kappa}\label{eq:linearized_measurement_rate}.
\end{eqnarray}
However, beyond the linearized regime Eq.~\eqref{eq:linearized_measurement_rate} is \emph{not} valid. Note that many authors include an additional factor of $2$ in their definition of $\kl$ by defining $\kappa$ as an intensity quantity and if one includes optical loss into the definition, this rate becomes $\kl=2\eta g^2/\kappa$. Furthermore, in the linearized regime we can rewrite $2\mu{k}=\sqrt{2}g_{0}|\alpha_{{c}}|$ in Eq.~\eqref{SMElinearized} to write the deterministic momentum kick term in a more familiar form.

\hypertarget{S5B}{}\subsection{B. Stochastic Gaussian measurement regime}\label{sec:S5B}
We write $\mu\Xm=\mu\expval{\Xm}+\mu \Ym$, which is valid at every time step $t\rightarrow t+\rmd{t}$ in the SME. Then we may expand $c$ to first order in $\mu \Ym$ if $\expval{\mu \Ym}^2\sim\mathrm{Var}(\mu \Ym)\ll1$. Importantly, the infinitesimal phase shifts $\mu \Ym$ over each $t\rightarrow t+\rmd{t}$ may integrate to arbitrarily large optical rotations $\mu\sigma$ over a finite duration. To first order in $\mu\Ym$, 
\begin{eqnarray}
c&=&\sqrt{2{k}}f(\expval{\Xm})+\dfrac{\rmi\cl}{\left[1-\rmi\left(\frac{\mu}{2}{\expval{\Xm}}+\frac{\Delta}{\kappa}\right)\right]^2}
\end{eqnarray}
with $\cl=\cl^{\dagger}=\sqrt{2\mu^2{k}}{\Ym}$. The measurement records are therefore
\begin{eqnarray}
\rmd{y}_{\Xl}&=&\A(\expval{\Xm})\rmd{t}+\dfrac{\rmd{W}_{\Xl}}{\sqrt{8\eta_{\Xl}(1-\zeta^2){k}}},\\
\rmd{y}_{\Pl}&=&\B(\expval{\Xm})\rmd{t}+\dfrac{\rmd{W}_{\Pl}}{\sqrt{8\eta_{\Pl}\zeta{k}}},
\end{eqnarray}
as by definition $\expval{\Ym}=0$. The difference with expressions for $\rmd{y}_{\Xl}$ and $\rmd{y}_{\Pl}$ in the main text is in taking the expectation value of $\Xm$ before entering it into the functions $A$ and $B$. The measurement currents are defined in the same way as in the general nonlinear theory.

Expanding the SME of Eq.~\eqref{SMEnoloss} to first order in $\mu \Ym$ gives
\begin{eqnarray}
\rmd\rho&=&-\frac{\rmi}{\hbar}[H_{0}+H_{1},\rho]\rmd{t}+\mathcal{D}\left[\dfrac{\cl}{{1+(\frac{\mu}{2}\expval{\Xm}+\frac{\Delta}{\kappa})^2}}\right]\rho\rmd{t}\nonumber\\
&&+\sqrt{1-\zeta^2}\mathcal{H}\left[\dfrac{\rmi\cl}{\left[1-\rmi\left(\frac{\mu}{2}{\expval{\Xm}}+\frac{\Delta}{\kappa}\right)\right]^2}\right]\rho\rmd{W_{\Xl}}+\zeta\mathcal{H}\left[\dfrac{\cl}{\left[1-\rmi\left(\frac{\mu}{2}{\expval{\Xm}}+\frac{\Delta}{\kappa}\right)\right]^2}\right]\rho\rmd{W_{\Pl}},\label{SMEsmallYnoloss}
\end{eqnarray}
where the term $H_{1}$ now describes a \emph{stochastic} momentum kick, explicitly given by
\begin{eqnarray}
\frac{H_{1}}{\hbar}&=&-\dfrac{2\mu{k}}{{1+(\frac{\mu}{2}\expval{\Xm}+\frac{\Delta}{\kappa})^2}}\Ym.\label{stochastic momentum kick}
\end{eqnarray}
This momentum kick is stochastic due to the denominator in Eq.~\eqref{stochastic momentum kick}, which originates from the nonlinear cavity input-output function. Therefore to study the effect of this term, we do not transform it away by shifting the $\Xm$ and $\Pm$ coordinates. Furthermore as $\mu\expval{\Xm}/2+\Delta/\kappa\rightarrow\infty$, Eq.~\eqref{SMEsmallYnoloss} tends towards $\rmd{\rho}=-\rmi[H_{0}/\hbar,\rho]\rmd{t}$ as the light is no longer resonant with the cavity and is simply reflected at the input.

Adding mechanical decoherence and optical loss to Eq.~\eqref{SMEsmallYnoloss} and using $\mathcal{D}[c+\gamma]\rho=\mathcal{D}[c]\rho+\frac{1}{2}[\gamma^{*}c-\gamma c^{\dagger},\rho]$ gives
\begin{eqnarray}
&&\rmd\rho=-\frac{\rmi}{\hbar}[H_{0}+H_{1}+H_{2},\rho]\rmd{t}+\mathcal{D}\left[\dfrac{\cl}{{1+(\frac{\mu}{2}\expval{\Xm}+\frac{\Delta}{\kappa})^2}}\right]\rho\rmd{t}+\mathcal{D}[L_{Y}]\rho\rmd{t}\nonumber\\
&&+\sqrt{(1-\zeta^2)\eta_{\Xl}}\mathcal{H}\left[\dfrac{\rmi\cl}{\left[1-\rmi\left(\frac{\mu}{2}{\expval{\Xm}}+\frac{\Delta}{\kappa}\right)\right]^2}\right]\rho\rmd{W_{\Xl}}+\zeta\sqrt{\eta_{\Pl}}\mathcal{H}\left[\dfrac{\cl}{\left[1-\rmi\left(\frac{\mu}{2}{\expval{\Xm}}+\frac{\Delta}{\kappa}\right)\right]^2}\right]\rho\rmd{W_{\Pl}},\label{SmallYSMEloss}
\end{eqnarray}
where $H_{2}/\hbar=-\gamma\expval{\Xm}\Pm$ and $L_{Y}=\sqrt{\dfrac{4\gamma \kb T}{\hbar\wm}}\Ym+\rmi\sqrt{\dfrac{\hbar\wm \gamma}{4 \kb T}}\Pm$, which as before models quantum Brownian motion .

\subsubsection{Stochastic Riccati equations}
Here, we follow~[\href{https://doi.org/10.1103/PhysRevA.96.062131}{J.  Zhang  and  K.  M{\o}lmer,  Phys.  Rev.  A \textbf{96},  062131 (2017)}] to cast Eq.~\eqref{SmallYSMEloss} into a system of stochastic differential equations (SDEs), but we also include a linear term in the Hamiltonian, accounted for by the displacement vector $\mathbf{d}$. Defining the vector $\mathbf{r}=(\Ym,\Pm)^{\mathrm{T}}$ we can rewrite the Hamiltonian $H=H_{0}+H_{1}+H_{2}$ in Eq.~\eqref{SmallYSMEloss} as
\begin{eqnarray}
H/\hbar&=&\frac{1}{2}\mathbf{r}^{T}R\mathbf{r}+\mathbf{d}^{T}\Omega\mathbf{r},\\
R&=&\wm\mathds{1},\\
\mathbf{d}&=&\begin{pmatrix}-\gamma\expval{\Xm}\\-\wm\expval{\Xm}+\dfrac{2\mu{k}}{{1+\left(\frac{\mu}{2}\expval{\Xm}+\frac{\Delta}{\kappa}\right)^2}}\end{pmatrix},\label{stochasticd}\\
\Omega&=&\begin{pmatrix}0&&1\\-1&&0\end{pmatrix}.
\end{eqnarray}
Then writing the three independent measurement output operators in Eq.~\eqref{SmallYSMEloss} as a vector $\mathbf{c}=C\mathbf{r}$ with
\begin{eqnarray}
C&=&\begin{pmatrix}C_{11}&0\\C_{21}&0\\\ C_{31}&C_{32}\end{pmatrix},\\
C_{11}&=&\dfrac{\rmi\sqrt{2\mu^2{k}(1-\zeta^2)}}{\left[1-\rmi\left(\frac{\mu}{2}{\expval{\Xm}}+\frac{\Delta}{\kappa}\right)\right]^2},\\
C_{21}&=&\dfrac{\sqrt{2\mu^2{k}\zeta^2}}{\left[1-\rmi\left(\frac{\mu}{2}{\expval{\Xm}}+\frac{\Delta}{\kappa}\right)\right]^2},\\
{C}_{31}&=&\sqrt{\dfrac{4\gamma \kb T}{\hbar\wm}},\\
{C}_{32}&=&\rmi\sqrt{\dfrac{\hbar\wm \gamma}{4 \kb T}}.
\end{eqnarray}
Furthermore, we define the efficiency matrix $\bm{\eta}=\mathrm{diag}\left(\eta_{\Xl},\eta_{\Pl},0\right)$ and the vector of Wiener increments $\rmd\mathbf{W}=\left(\rmd{W}_{\Xl},\rmd{W}_{\Pl},0\right)^{\mathrm{T}}$. We then introduce the drift matrix $M$, the diffusion matrix $D$, and the noise matrix $N$
\begin{eqnarray}
M&=&\Omega\left({R}+\mathrm{Im}[{C}^{\dagger}(t){C}(t)]\right),\\
D&=&-\Omega\mathrm{Re}[{C}^{\dagger}(t){C}(t)]\Omega,\\
N&=&2V\mathrm{Re}[{C}^{\mathrm{T}}(t)]-\Omega\mathrm{Im}[{C}^{\mathrm{T}}(t)],\\
\end{eqnarray}
and we find that---given an initial Gaussian mechanical state---the dynamics described by Eq.~\eqref{SmallYSMEloss} can be completely described by 5 independent SDEs. Namely, two SDES describing the evolution of the first moments
\begin{eqnarray}
\rmd\rv=\left(M\rv+\mathbf{d}\right)\rmd{t}+N{\sqrt{\bm{\eta}}}\rmd{\mathbf{W}}\label{SDEmeanappendix}
\end{eqnarray}
and a stochastic Riccati matrix equation describing the evolution of the 3 independent covariance matrix elements $V$
\begin{eqnarray}
\dfrac{\rmd V}{\rmd{t}}=M V+V {M}^{\mathrm{T}}+D-N\bm{\eta}N^{\mathrm{T}}\label{Riccatieqnappendix},
\end{eqnarray}
where $V_{i,j}=\frac{1}{2}\expval{\left\{r_{i},r_{j}\right\}}-\expval{r_{i}}\expval{r_{j}}$. As opposed to linearized optomechanics, the Riccati equation is no longer deterministic. The origin of this stochasticity comes from the dependence of the measurement output operators $\mathbf{c}$ on the stochastic mean position $\expval{\Xm}$. Furthermore, note that the covariance matrix is invariant under the transformation $\Xm=\expval{\Xm}+\Ym$ at each time step, i.e. $V_{X}=\expval{\Xm^2}-\expval{\Xm}^2=\expval{\Ym^2}-\expval{\Ym}^2$ and $V_{XP}=\expval{\left\{\Xm,\Pm\right\}}/2-\expval{\Xm}\expval{\Pm}=\expval{\left\{\Ym,\Pm\right\}}/2-\expval{\Ym}\expval{\Pm}$.

\hypertarget{S5C}{}\subsection{C. Drive-locking}\label{sec:S5C}

\subsubsection{No feedback case}
\begin{figure}
    \centering{
    \includegraphics[width=0.7\textwidth]{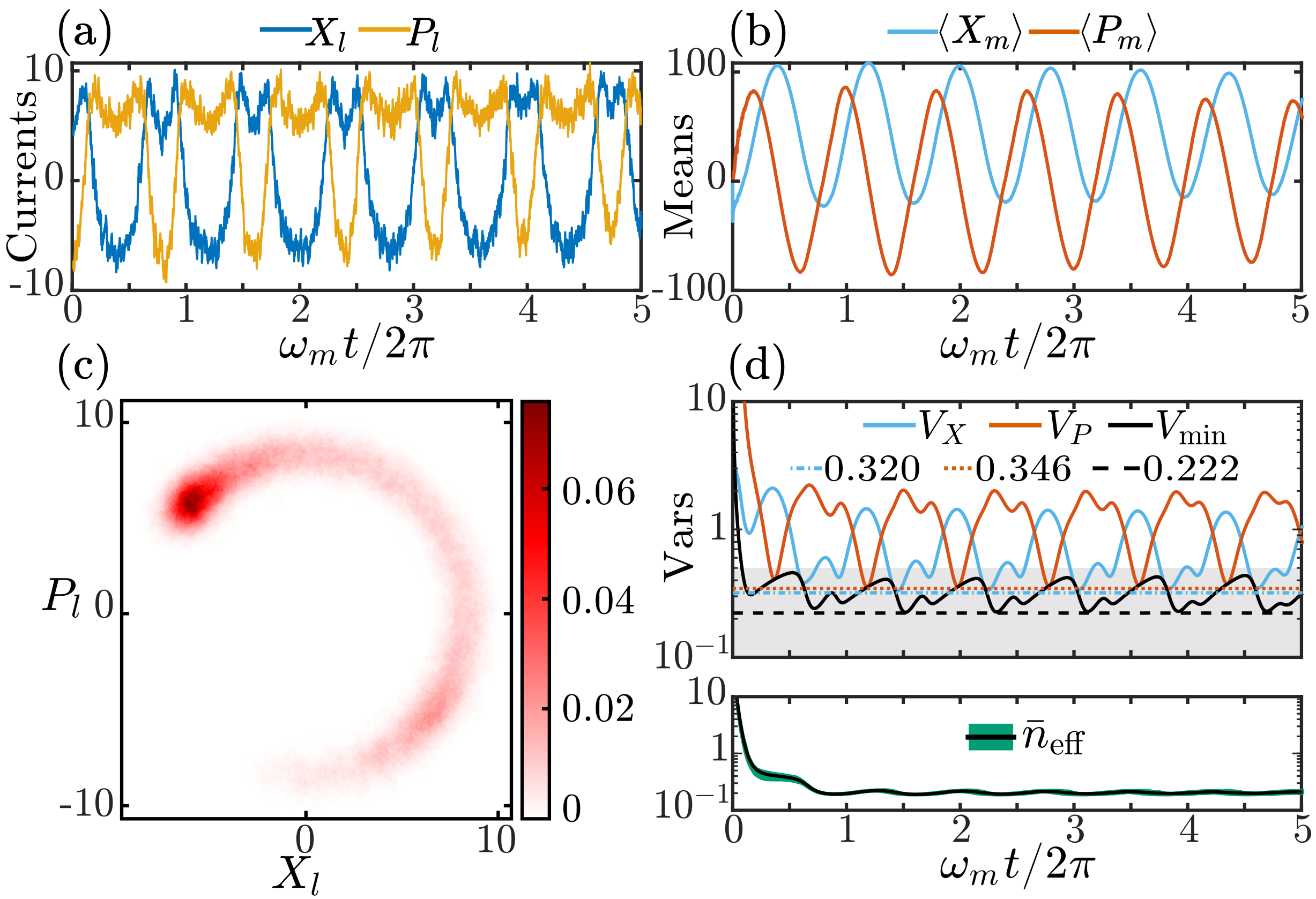}}
    \caption{Continuous position measurement in the stochastic Gaussian regime without feedback. Here, $\Xa/\sqrt{2}=10$, $\wm/2\pi=10\,\mathrm{MHz}$, $\mu=0.05$, $\mu^2k/\wm=2$, $\eta=0.7$, $\gamma/2\pi=10\,\mathrm{Hz}$, $\zeta=1/\sqrt{2}$, $T=4\,\mathrm{K}$ and the initial mechanical state is at $100\,\mathrm{mK}$. 
     The general-dyne measurement currents $\Xl$ and $\Pl$ plotted as a function of time for a given stochastic trajectory. (b) The corresponding trajectories of the mechanical means $\expval{\Xm}$ and $\expval{\Pm}$. (c) The same general-dyne currents plotted as as a $2$D histogram, normalized to one. We see that the optical phase shift averaged over this trajectory is positive. (d) Top: The position variance $V_{X}$, the momentum variance $V_{P}$, and the generalized squeezing variance $V_{\mathrm{min}}$ for the same trajectory. Here, the dashed lines correspond to the mean of the minimum of each variance quantity over 100 runs, while the grey shaded area corresponds to quantum squeezing. Bottom: The effective thermal occupation $\bar{n}_{\mathrm{eff}}$ plotted as a function of time averaged over 100 stochastic trajectories. The shaded green area indicates the upper and lower quartile range of the 100 trajectories. }\label{fig_11_supp}
\end{figure}

Choosing the same parameters as in the main text, we run the simulations for the stochastic Gaussian regime again but without feedback to the drive field, which produces the results shown in Fig.~\ref{fig_11_supp}. Comparison with Fig.~\ref{fig_3_main} confirms our choice of feedback in the main text removes the asymmetry in the optical phase space distribution and improves the mechanical squeezing. Furthermore, in Fig.~\ref{fig_11_supp} we see the evolution of the means becomes harmonic again. This is because without feedback, the intracavity power is lower and so (for these specific parameters) the higher order terms in $\Xm$ in the optical force are not present. However, in principle arbitrarily high orders in $\Xm$ are present in the optical force in the nonlinear theory for \emph{both} with and without drive-locking. Inserting Eq.~\eqref{acav} into Eq.~\eqref{HLmomentum}, we see that the dimensionless optical force is given by the right hand side of
\begin{eqnarray}
\dot{P}_{m}&=&\dfrac{\mu \ain^{\dagger}\ain}{1+\left(\frac{\mu}{2}\Xm+\frac{\Delta}{\kappa}\right)^2}.
\end{eqnarray}
By expanding this optical force in $\Xm$, we see that the contributions for a given order of $\Xm$ depend on $\Delta/\kappa$, $\mu$, and the intracavity power via the input field operators $\ain$.

\begin{figure}
    \centering{
    \includegraphics[width=0.7\textwidth]{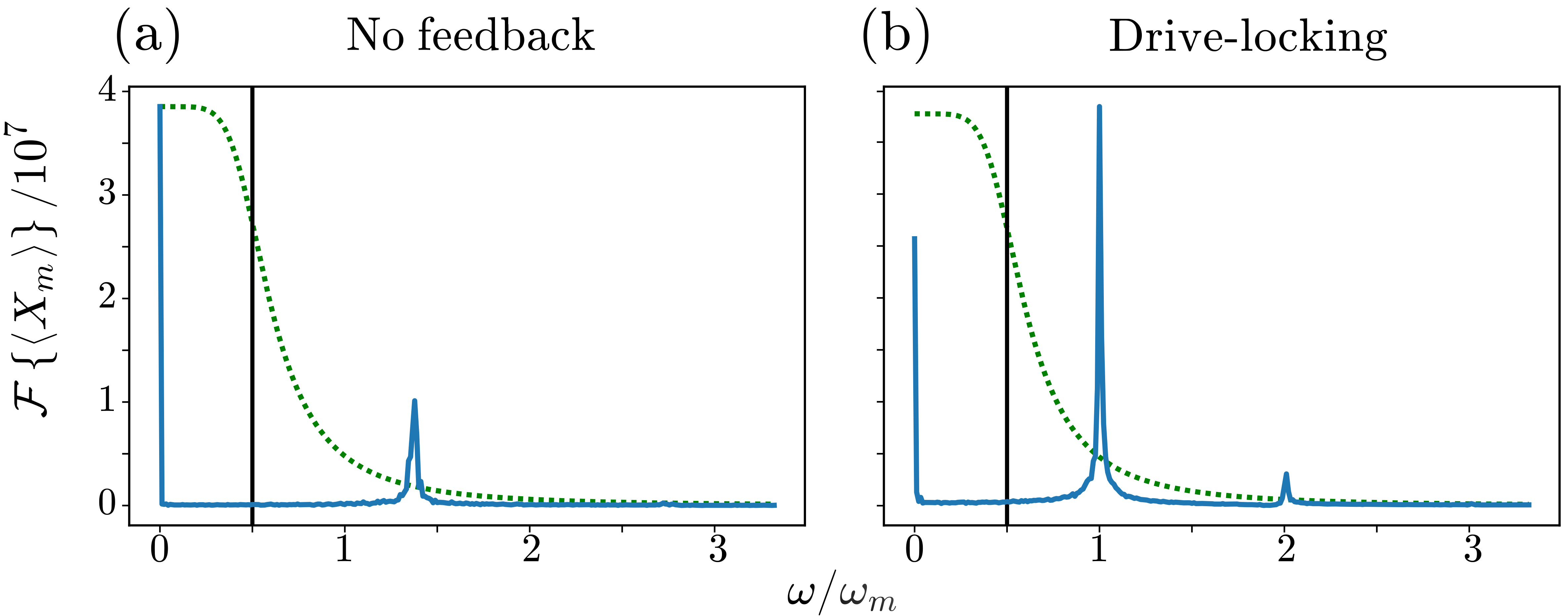}}
    \caption{Real Fourier transform of the mean mechanical position $\mathcal{F}\left\{\expval{\Xm}\right\}$ with (a) no drive feedback and with (b) drive-locking: $\Delta=-\frac{1}{2}\mu\kappa\overline{\expval{\Xm}}$. Here, $\Xa/\sqrt{2}=10$, $\wm/2\pi=10\,\mathrm{MHz}$, $\mu=0.05$, $\mu^2k/\wm=2$, $\eta=0.7$, $\gamma/2\pi=10\,\mathrm{Hz}$, $\zeta=1/\sqrt{2}$, $T=4\,\mathrm{K}$ and the initial mechanical state is a thermal state at $100\,\mathrm{mK}$. A third-order Butterworth filter is shown by the green-dashed line. Meanwhile, the frequency cutoff at $0.5\wm$ of the Butterworth is shown by the vertical black line and the mechanical frequency $\wm$ is shown by the vertical dashed line. In (b), the Butterworth is used to calculate the detuning at each timestep.  The numerical simulations last $100$ mechanical periods and the real Fourier transforms are calculated with data obtained in the last $90$ periods to avoid the transient effects associated with the response time of the filter.
    }
    \label{fig_12_supp}
\end{figure}

To gain further insight we may also take the real Fourier transform of the mean mechanical position $\mathcal{F}\left\{\expval{\Xm}\right\}$. For no feedback, this is shown in Fig.~\ref{fig_12_supp}(a) for a given trajectory. Here, we see a peak in $\mathcal{F}\left\{\expval{\Xm}\right\}$ at a higher frequency than $\wm$, which is a demonstration of the optical spring effect due to an $\Xm$-dependent term in the optical force, i.e. an $\Xm^2$-dependent term in an effective Hamiltonian. 

\subsubsection{Feedback to the drive field}
Fig.~\ref{fig_12_supp}(b) shows $\mathcal{F}\left\{\expval{\Xm}\right\}$ for a particular trajectory but now when the detuning of the drive is chosen to cancel the slowly-varying component of $\expval{\Xm}$, which symmeterizes the optical phase space distribution as in the main text. Let this low frequency component of the mean mechanical position be $\overline{\expval{\Xm}}$. Then, it follows that the portion of $\mathcal{F}\left\{\expval{\Xm}\right\}$ which is captured by the third-order Butterworth filter corresponds to $\mathcal{F}\left\{\overline{\expval{\Xm}}\right\}$. Therefore, by calculating $\mathcal{F}\left\{\expval{\Xm}\right\}$ at each time step, the low-pass filtering allows one to calculate $\overline{\expval{\Xm}}$ at each time step too. From Eq.~\eqref{SmallYSMEloss}, for example, we can identify the correct choice of feedback to cancel the low frequency components is 
\begin{eqnarray}
\Delta=-\frac{1}{2}\mu\kappa\overline{\expval{\Xm}}.
\end{eqnarray}
In this way, the drive is locked to the cavity frequency, which changes as the time averaged mean mechanical position drifts away from zero. As a reminder, $\Delta$ is defined as the detuning from the cavity resonance for $\Xm=0$. However, in our numerical simulations, or indeed an experiment, this cancellation of $\overline{\expval{\Xm}}$ must be done in a causal fashion, such that only information about $\mathcal{F}\left\{\expval{\Xm}\right\}$ at times $t'<t$ is used to compute the detuning $\Delta$ at time $t$.

Fig.~\ref{fig_12_supp}(b) contains a peak in $\mathcal{F}\left\{\expval{\Xm}\right\}$ at $\wm$, which shows that the optical spring effect is cancelled by the drive-locking. Furthermore, this figure shows a small peak in $2\wm$ as higher frequency components are now amplified, which comes from an $\Xm^2$ term in the optical force. This $2\wm$ component explains the anharmonicity in the evolution of the means in Fig.~\ref{fig_3_main}. We note that the order and cutoff frequency of the Butterworth are chosen to optimize the run time of the numerical simulations, while still demonstrating the working principle of the stochastic Gaussian regime. Choosing the optimal filtering procedure and choice of $\Delta$ depends on an experimental implementation.

\hypertarget{S5D}{}\subsection{D. Experimental parameters for sliced-photonic crystal devices}\label{sec:S5D}
\begin{figure}
    \centering{
    \includegraphics[width=\textwidth]{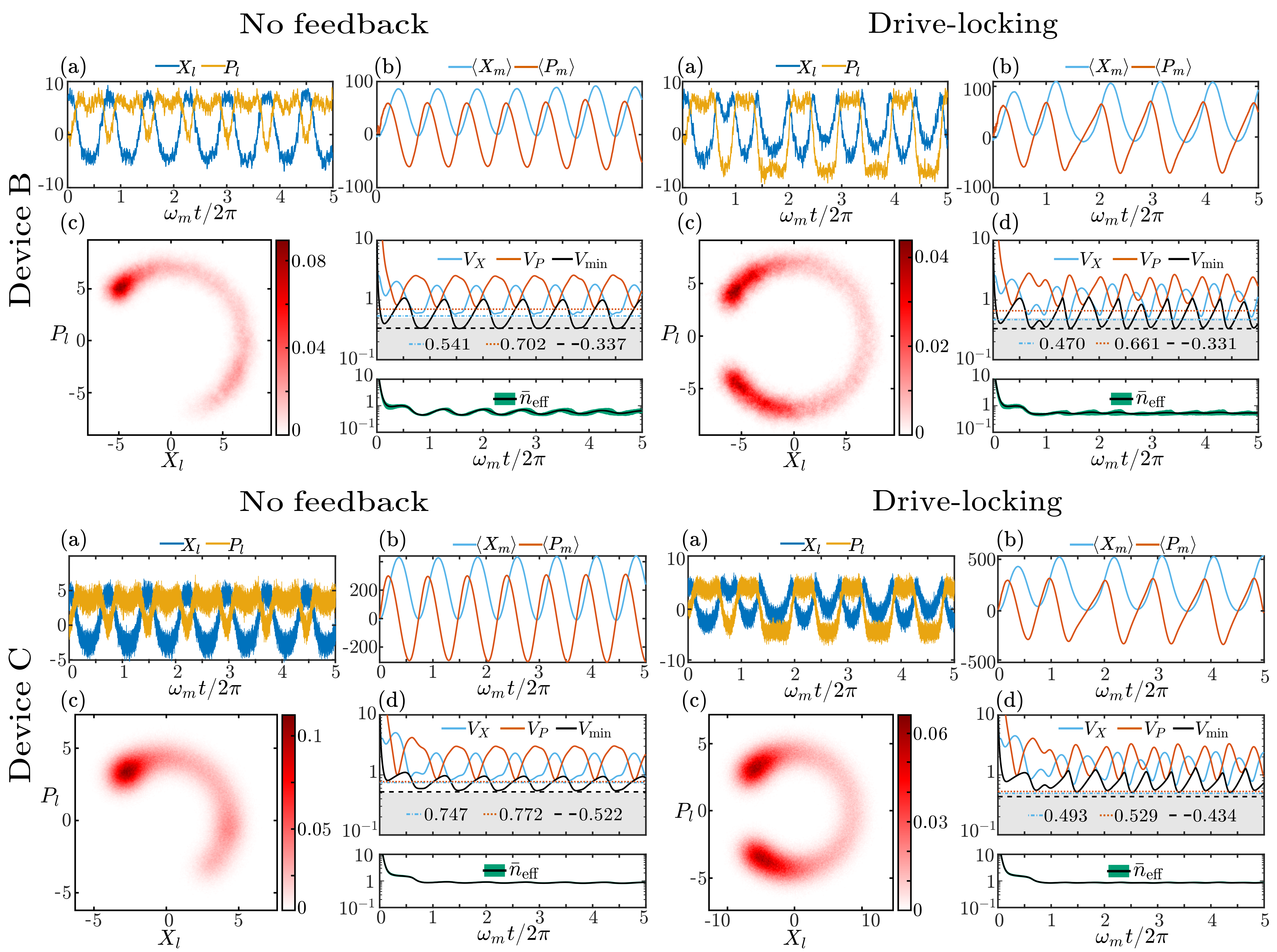}}
    \caption{ Devices B and C in the stochastic Gaussian regime. Left hand side: no drive feedback. Right hand side: drive-locking $\Delta=-\frac{1}{2}\mu\kappa\overline{\expval{\Xm}}$. Device B: $\eta=0.5$, $\gamma=100\,\mathrm{Hz}$, and $\mu=0.05$. Device C: $\eta=0.25$, $\gamma=20\,\mathrm{Hz}$, and $\mu=0.01$. In both cases $\wm/2\pi=10\,\mathrm{MHz}$, $\mu^2k/\wm=2$, $\zeta=1/\sqrt{2}$, $T=4\,\mathrm{K}$, and the initial mechanical state is a thermal state at $100\,\mathrm{mK}$. Here, the dashed lines correspond to the mean of the minimum of each variance quantity over 100 runs, each lasting 100 mechanical periods, while the grey shaded area corresponds to quantum squeezing.
    }
    \label{fig_13_supp}
\end{figure}

Here we run simulations for the stochastic Gaussian regime using different parameter sets, which are given in table~\ref{table:table1}. Device A corresponds to the parameters chosen in the main text---see Figs~\ref{fig_3_main}, \ref{fig_11_supp}, and \ref{fig_12_supp}. Devices B and C represent smaller improvements to current state-of-art parameters for sliced-photonic crystal devices, which are given in the last column of table~\ref{table:table1}. Fig.~\ref{fig_13_supp} shows the results of these simulations for devices B and C. In both cases, quantum squeezing is still possible, i.e. $V_{\mathrm{min}}<1/2$, and it is interesting to note that drive-locking is needed for the mean of the minimum value of $V_{X}$ to go below $1/2$.

\begin{table}[h!]
 \caption{\textbf{Sliced-photonic crystal parameters for devices A, B, and C and the current state-of-the-art.} Device A corresponds to improvements of the current state-of-the-art---see Figs~\ref{fig_3_main}, \ref{fig_11_supp}, and \ref{fig_12_supp}. Devices B and C---see Fig.~\ref{fig_13_supp}---represent smaller improvements. In the final column we give a parameter set for a current sliced-photonic crystal experiment [{\color{blue}{P. Neveu \emph{et al., In preparation} (2022)}}], which builds on Refs~[\href{https://doi.org/10.1038/ncomms16024}{R. Leijssen \emph{et al.}, Nat. Commun. \textbf{8}, 16024 (2017)}; \href{https://doi.org/10.1103/PhysRevLett.123.113601}{J.~T. Muhonen \emph{et al.}, Phys. Rev. Lett. \textbf{123}, 113601 (2019)}].}
\centering
\begingroup
\renewcommand{\arraystretch}{1.5}
\setlength{\tabcolsep}{1pt}
\hspace*{-\leftmargin}\begin{tabular}{c c c c c}
\hline
\hline
\textbf{Parameter} & ~~~ \textbf{Device A} ~~~ & ~~~ \textbf{Device B}~~~ & ~~~\textbf{Device C}~~~ &  ~~\textbf{Current experiment}~~ \\
\hline
Mechanical frequency $\wm/2\pi$  & $10\,\mathrm{MHz}$ & $10\,\mathrm{MHz}$ & $10\,\mathrm{MHz}$ & $3.6\,\mathrm{MHz}$ \\
       Cavity amplitude decay rate $\kappa/2\pi$ & $5\,\mathrm{GHz}$ & $5\,\mathrm{GHz}$ & $10\,\mathrm{GHz}$ &  $10\,\mathrm{GHz}$ \\
      Optomechanical coupling rate $g_{0}/2\pi$ & $90\,\mathrm{MHz}$ & $90\,\mathrm{MHz}$ & $35\,\mathrm{MHz}$ & $25\,\mathrm{MHz}$  \\
      Intrinsic mechanical decay rate $\gamma/2\pi$  & $10\,\mathrm{Hz}$ & $100\,\mathrm{Hz}$ & $20\,\mathrm{Hz}$ & $20\,\mathrm{Hz}$  \\
      Nonlinear optomechanical coupling strength $\mu$ ~~  & $0.050$ & $0.050$ & $0.010$  & $0.007$  \\
      Measurement efficiency $\eta$ & 0.70 & 0.50 & 0.25 & 0.25 \\
      \hline
      \hline
     \end{tabular}
     \endgroup
     \label{table:table1}
     \end{table}
     
     
\begin{thebibliography}{62}%
\makeatletter
\providecommand \@ifxundefined [1]{%
 \@ifx{#1\undefined}
}%
\providecommand \@ifnum [1]{%
 \ifnum #1\expandafter \@firstoftwo
 \else \expandafter \@secondoftwo
 \fi
}%
\providecommand \@ifx [1]{%
 \ifx #1\expandafter \@firstoftwo
 \else \expandafter \@secondoftwo
 \fi
}%
\providecommand \natexlab [1]{#1}%
\providecommand \enquote  [1]{``#1''}%
\providecommand \bibnamefont  [1]{#1}%
\providecommand \bibfnamefont [1]{#1}%
\providecommand \citenamefont [1]{#1}%
\providecommand \href@noop [0]{\@secondoftwo}%
\providecommand \href [0]{\begingroup \@sanitize@url \@href}%
\providecommand \@href[1]{\@@startlink{#1}\@@href}%
\providecommand \@@href[1]{\endgroup#1\@@endlink}%
\providecommand \@sanitize@url [0]{\catcode `\\12\catcode `\$12\catcode
  `\&12\catcode `\#12\catcode `\^12\catcode `\_12\catcode `\%12\relax}%
\providecommand \@@startlink[1]{}%
\providecommand \@@endlink[0]{}%
\providecommand \url  [0]{\begingroup\@sanitize@url \@url }%
\providecommand \@url [1]{\endgroup\@href {#1}{\urlprefix }}%
\providecommand \urlprefix  [0]{URL }%
\providecommand \Eprint [0]{\href }%
\providecommand \doibase [0]{http://dx.doi.org/}%
\providecommand \selectlanguage [0]{\@gobble}%
\providecommand \bibinfo  [0]{\@secondoftwo}%
\providecommand \bibfield  [0]{\@secondoftwo}%
\providecommand \translation [1]{[#1]}%
\providecommand \BibitemOpen [0]{}%
\providecommand \bibitemStop [0]{}%
\providecommand \bibitemNoStop [0]{.\EOS\space}%
\providecommand \EOS [0]{\spacefactor3000\relax}%
\providecommand \BibitemShut  [1]{\csname bibitem#1\endcsname}%
\let\auto@bib@innerbib\@empty
\bibitem [{\citenamefont {Kronwald}\ and\ \citenamefont
  {Marquardt}(2013)}]{kronwald2013optomechanically}%
  \BibitemOpen
  \bibfield  {author} {\bibinfo {author} {\bibfnamefont {A.}~\bibnamefont
  {Kronwald}}\ and\ \bibinfo {author} {\bibfnamefont {F.}~\bibnamefont
  {Marquardt}},\ }\href {https://doi.org/10.1103/PhysRevLett.111.133601}
  {\bibfield  {journal} {\bibinfo  {journal} {Phys. Rev. Lett.}\ }\textbf
  {\bibinfo {volume} {111}},\ \bibinfo {pages} {133601} (\bibinfo {year}
  {2013})}\BibitemShut {NoStop}%
\bibitem [{\citenamefont {B{\o}rkje}\ \emph {et~al.}(2013)\citenamefont
  {B{\o}rkje}, \citenamefont {Nunnenkamp}, \citenamefont {Teufel},\ and\
  \citenamefont {Girvin}}]{borkje2013signatures}%
  \BibitemOpen
  \bibfield  {author} {\bibinfo {author} {\bibfnamefont {K.}~\bibnamefont
  {B{\o}rkje}}, \bibinfo {author} {\bibfnamefont {A.}~\bibnamefont
  {Nunnenkamp}}, \bibinfo {author} {\bibfnamefont {J.~D.}\ \bibnamefont
  {Teufel}}, \ and\ \bibinfo {author} {\bibfnamefont {S.~M.}\ \bibnamefont
  {Girvin}},\ }\href {https://doi.org/10.1103/PhysRevLett.111.053603}
  {\bibfield  {journal} {\bibinfo  {journal} {Phys. Rev. Lett.}\ }\textbf
  {\bibinfo {volume} {111}},\ \bibinfo {pages} {053603} (\bibinfo {year}
  {2013})}\BibitemShut {NoStop}%
\bibitem [{\citenamefont {Lemonde}\ \emph {et~al.}(2013)\citenamefont
  {Lemonde}, \citenamefont {Didier},\ and\ \citenamefont
  {Clerk}}]{lemonde2013nonlinear}%
  \BibitemOpen
  \bibfield  {author} {\bibinfo {author} {\bibfnamefont {M.~A.}\ \bibnamefont
  {Lemonde}}, \bibinfo {author} {\bibfnamefont {N.}~\bibnamefont {Didier}}, \
  and\ \bibinfo {author} {\bibfnamefont {A.~A.}\ \bibnamefont {Clerk}},\ }\href
  {https://doi.org/10.1103/PhysRevLett.111.053602} {\bibfield  {journal}
  {\bibinfo  {journal} {Phys. Rev. Lett.}\ }\textbf {\bibinfo {volume} {111}},\
  \bibinfo {pages} {053602} (\bibinfo {year} {2013})}\BibitemShut {NoStop}%
\bibitem [{\citenamefont {Vanner}(2011)}]{vanner2011selective}%
  \BibitemOpen
  \bibfield  {author} {\bibinfo {author} {\bibfnamefont {M.~R.}\ \bibnamefont
  {Vanner}},\ }\href {https://doi.org/10.1103/PhysRevX.1.021011} {\bibfield
  {journal} {\bibinfo  {journal} {Phys. Rev. X}\ }\textbf {\bibinfo {volume}
  {1}},\ \bibinfo {pages} {021011} (\bibinfo {year} {2011})}\BibitemShut
  {NoStop}%
\bibitem [{\citenamefont {Bose}\ \emph {et~al.}(1999)\citenamefont {Bose},
  \citenamefont {Jacobs},\ and\ \citenamefont {Knight}}]{bose1999scheme}%
  \BibitemOpen
  \bibfield  {author} {\bibinfo {author} {\bibfnamefont {S.}~\bibnamefont
  {Bose}}, \bibinfo {author} {\bibfnamefont {K.}~\bibnamefont {Jacobs}}, \ and\
  \bibinfo {author} {\bibfnamefont {P.~L.}\ \bibnamefont {Knight}},\ }\href
  {https://doi.org/10.1103/PhysRevA.59.3204} {\bibfield  {journal} {\bibinfo
  {journal} {Phys. Rev. A}\ }\textbf {\bibinfo {volume} {59}},\ \bibinfo
  {pages} {3204} (\bibinfo {year} {1999})}\BibitemShut {NoStop}%
\bibitem [{\citenamefont {Ringbauer}\ \emph {et~al.}(2018)\citenamefont
  {Ringbauer}, \citenamefont {Weinhold}, \citenamefont {Howard}, \citenamefont
  {White},\ and\ \citenamefont {Vanner}}]{ringbauer2018generation}%
  \BibitemOpen
  \bibfield  {author} {\bibinfo {author} {\bibfnamefont {M.}~\bibnamefont
  {Ringbauer}}, \bibinfo {author} {\bibfnamefont {T.~J.}\ \bibnamefont
  {Weinhold}}, \bibinfo {author} {\bibfnamefont {L.~A.}\ \bibnamefont
  {Howard}}, \bibinfo {author} {\bibfnamefont {A.~G.}\ \bibnamefont {White}}, \
  and\ \bibinfo {author} {\bibfnamefont {M.~R.}\ \bibnamefont {Vanner}},\
  }\href {https://doi.org/10.1088/1367-2630/aabb8d} {\bibfield  {journal}
  {\bibinfo  {journal} {New J. Phys.}\ }\textbf {\bibinfo {volume} {20}},\
  \bibinfo {pages} {053042} (\bibinfo {year} {2018})}\BibitemShut {NoStop}%
\bibitem [{\citenamefont {Clarke}\ and\ \citenamefont
  {Vanner}(2019)}]{clarke2018growing}%
  \BibitemOpen
  \bibfield  {author} {\bibinfo {author} {\bibfnamefont {J.}~\bibnamefont
  {Clarke}}\ and\ \bibinfo {author} {\bibfnamefont {M.~R.}\ \bibnamefont
  {Vanner}},\ }\href {https://doi.org/10.1088/2058-9565/aada1d} {\bibfield
  {journal} {\bibinfo  {journal} {Quantum Sci. Technol.}\ }\textbf {\bibinfo
  {volume} {4}},\ \bibinfo {pages} {014003} (\bibinfo {year}
  {2019})}\BibitemShut {NoStop}%
\bibitem [{\citenamefont {Marshall}\ \emph {et~al.}(2003)\citenamefont
  {Marshall}, \citenamefont {Simon}, \citenamefont {Penrose},\ and\
  \citenamefont {Bouwmeester}}]{marshall2003towards}%
  \BibitemOpen
  \bibfield  {author} {\bibinfo {author} {\bibfnamefont {W.}~\bibnamefont
  {Marshall}}, \bibinfo {author} {\bibfnamefont {C.}~\bibnamefont {Simon}},
  \bibinfo {author} {\bibfnamefont {R.}~\bibnamefont {Penrose}}, \ and\
  \bibinfo {author} {\bibfnamefont {D.}~\bibnamefont {Bouwmeester}},\ }\href
  {https://doi.org/10.1103/PhysRevLett.91.130401} {\bibfield  {journal}
  {\bibinfo  {journal} {Phys. Rev. Lett.}\ }\textbf {\bibinfo {volume} {91}},\
  \bibinfo {pages} {130401} (\bibinfo {year} {2003})}\BibitemShut {NoStop}%
\bibitem [{\citenamefont {Vacanti}\ \emph {et~al.}(2008)\citenamefont
  {Vacanti}, \citenamefont {Paternostro}, \citenamefont {Palma},\ and\
  \citenamefont {Vedral}}]{vacanti2008optomechanical}%
  \BibitemOpen
  \bibfield  {author} {\bibinfo {author} {\bibfnamefont {G.}~\bibnamefont
  {Vacanti}}, \bibinfo {author} {\bibfnamefont {M.}~\bibnamefont
  {Paternostro}}, \bibinfo {author} {\bibfnamefont {G.~M.}\ \bibnamefont
  {Palma}}, \ and\ \bibinfo {author} {\bibfnamefont {V.}~\bibnamefont
  {Vedral}},\ }\href {https://doi.org/10.1088/1367-2630/10/9/095014} {\bibfield
   {journal} {\bibinfo  {journal} {New J. Phys.}\ }\textbf {\bibinfo {volume}
  {10}},\ \bibinfo {pages} {095014} (\bibinfo {year} {2008})}\BibitemShut
  {NoStop}%
\bibitem [{\citenamefont {Akram}\ \emph {et~al.}(2013)\citenamefont {Akram},
  \citenamefont {Bowen},\ and\ \citenamefont {Milburn}}]{akram2013entangled}%
  \BibitemOpen
  \bibfield  {author} {\bibinfo {author} {\bibfnamefont {U.}~\bibnamefont
  {Akram}}, \bibinfo {author} {\bibfnamefont {W.~P.}\ \bibnamefont {Bowen}}, \
  and\ \bibinfo {author} {\bibfnamefont {G.~J.}\ \bibnamefont {Milburn}},\
  }\href {https://doi.org/10.1088/1367-2630/15/9/093007} {\bibfield  {journal}
  {\bibinfo  {journal} {New J. Phys.}\ }\textbf {\bibinfo {volume} {15}},\
  \bibinfo {pages} {093007} (\bibinfo {year} {2013})}\BibitemShut {NoStop}%
\bibitem [{\citenamefont {Kanari-Naish}\ \emph {et~al.}(2022)\citenamefont
  {Kanari-Naish}, \citenamefont {Clarke}, \citenamefont {Qvarfort},\ and\
  \citenamefont {Vanner}}]{kanari2021two}%
  \BibitemOpen
  \bibfield  {author} {\bibinfo {author} {\bibfnamefont {L.~A.}\ \bibnamefont
  {Kanari-Naish}}, \bibinfo {author} {\bibfnamefont {J.}~\bibnamefont
  {Clarke}}, \bibinfo {author} {\bibfnamefont {S.}~\bibnamefont {Qvarfort}}, \
  and\ \bibinfo {author} {\bibfnamefont {M.~R.}\ \bibnamefont {Vanner}},\
  }\href {https://doi.org/10.1088/2058-9565/ac6dfd} {\bibfield  {journal}
  {\bibinfo  {journal} {Quantum Sci. Technol.}\ }\textbf {\bibinfo {volume}
  {7}},\ \bibinfo {pages} {035012} (\bibinfo {year} {2022})}\BibitemShut
  {NoStop}%
\bibitem [{\citenamefont {Rabl}(2011)}]{rabl2011photon}%
  \BibitemOpen
  \bibfield  {author} {\bibinfo {author} {\bibfnamefont {P.}~\bibnamefont
  {Rabl}},\ }\href {https://doi.org/10.1103/PhysRevLett.107.063601} {\bibfield
  {journal} {\bibinfo  {journal} {Phys. Rev. Lett.}\ }\textbf {\bibinfo
  {volume} {107}},\ \bibinfo {pages} {063601} (\bibinfo {year}
  {2011})}\BibitemShut {NoStop}%
\bibitem [{\citenamefont {Nunnenkamp}\ \emph {et~al.}(2011)\citenamefont
  {Nunnenkamp}, \citenamefont {B{\o}rkje},\ and\ \citenamefont
  {Girvin}}]{nunnenkamp2011single}%
  \BibitemOpen
  \bibfield  {author} {\bibinfo {author} {\bibfnamefont {A.}~\bibnamefont
  {Nunnenkamp}}, \bibinfo {author} {\bibfnamefont {K.}~\bibnamefont
  {B{\o}rkje}}, \ and\ \bibinfo {author} {\bibfnamefont {S.~M.}\ \bibnamefont
  {Girvin}},\ }\href {https://doi.org/10.1103/PhysRevLett.107.063602}
  {\bibfield  {journal} {\bibinfo  {journal} {Phys. Rev. Lett.}\ }\textbf
  {\bibinfo {volume} {107}},\ \bibinfo {pages} {063602} (\bibinfo {year}
  {2011})}\BibitemShut {NoStop}%
\bibitem [{\citenamefont {Brennecke}\ \emph {et~al.}(2008)\citenamefont
  {Brennecke}, \citenamefont {Ritter}, \citenamefont {Donner},\ and\
  \citenamefont {Esslinger}}]{brennecke2008cavity}%
  \BibitemOpen
  \bibfield  {author} {\bibinfo {author} {\bibfnamefont {F.}~\bibnamefont
  {Brennecke}}, \bibinfo {author} {\bibfnamefont {S.}~\bibnamefont {Ritter}},
  \bibinfo {author} {\bibfnamefont {T.}~\bibnamefont {Donner}}, \ and\ \bibinfo
  {author} {\bibfnamefont {T.}~\bibnamefont {Esslinger}},\ }\href
  {https://doi.org/10.1126/science.1163218} {\bibfield  {journal} {\bibinfo
  {journal} {Science}\ }\textbf {\bibinfo {volume} {322}},\ \bibinfo {pages}
  {235} (\bibinfo {year} {2008})}\BibitemShut {NoStop}%
\bibitem [{\citenamefont {Purdy}\ \emph {et~al.}(2010)\citenamefont {Purdy},
  \citenamefont {Brooks}, \citenamefont {Botter}, \citenamefont {Brahms},
  \citenamefont {Ma},\ and\ \citenamefont {Stamper-Kurn}}]{purdy2010tunable}%
  \BibitemOpen
  \bibfield  {author} {\bibinfo {author} {\bibfnamefont {T.~P.}\ \bibnamefont
  {Purdy}}, \bibinfo {author} {\bibfnamefont {D.~W.~C.}\ \bibnamefont
  {Brooks}}, \bibinfo {author} {\bibfnamefont {T.}~\bibnamefont {Botter}},
  \bibinfo {author} {\bibfnamefont {N.}~\bibnamefont {Brahms}}, \bibinfo
  {author} {\bibfnamefont {Z.-Y.}\ \bibnamefont {Ma}}, \ and\ \bibinfo {author}
  {\bibfnamefont {D.~M.}\ \bibnamefont {Stamper-Kurn}},\ }\href
  {https://doi.org/10.1103/PhysRevLett.105.133602} {\bibfield  {journal}
  {\bibinfo  {journal} {Phys. Rev. Lett.}\ }\textbf {\bibinfo {volume} {105}},\
  \bibinfo {pages} {133602} (\bibinfo {year} {2010})}\BibitemShut {NoStop}%
\bibitem [{\citenamefont {Brawley}\ \emph {et~al.}(2016)\citenamefont
  {Brawley}, \citenamefont {Vanner}, \citenamefont {Larsen}, \citenamefont
  {Schmid}, \citenamefont {Boisen},\ and\ \citenamefont
  {Bowen}}]{brawley2016nonlinear}%
  \BibitemOpen
  \bibfield  {author} {\bibinfo {author} {\bibfnamefont {G.~A.}\ \bibnamefont
  {Brawley}}, \bibinfo {author} {\bibfnamefont {M.~R.}\ \bibnamefont {Vanner}},
  \bibinfo {author} {\bibfnamefont {P.~E.}\ \bibnamefont {Larsen}}, \bibinfo
  {author} {\bibfnamefont {S.}~\bibnamefont {Schmid}}, \bibinfo {author}
  {\bibfnamefont {A.}~\bibnamefont {Boisen}}, \ and\ \bibinfo {author}
  {\bibfnamefont {W.~P.}\ \bibnamefont {Bowen}},\ }\href
  {https://doi.org/10.1038/ncomms10988} {\bibfield  {journal} {\bibinfo
  {journal} {Nat. Commun.}\ }\textbf {\bibinfo {volume} {7}},\ \bibinfo {pages}
  {10988} (\bibinfo {year} {2016})}\BibitemShut {NoStop}%
\bibitem [{\citenamefont {Leijssen}\ \emph {et~al.}(2017)\citenamefont
  {Leijssen}, \citenamefont {La~Gala}, \citenamefont {Freisem}, \citenamefont
  {Muhonen},\ and\ \citenamefont {Verhagen}}]{leijssen2017nonlinear}%
  \BibitemOpen
  \bibfield  {author} {\bibinfo {author} {\bibfnamefont {R.}~\bibnamefont
  {Leijssen}}, \bibinfo {author} {\bibfnamefont {G.~R.}\ \bibnamefont
  {La~Gala}}, \bibinfo {author} {\bibfnamefont {L.}~\bibnamefont {Freisem}},
  \bibinfo {author} {\bibfnamefont {J.~T.}\ \bibnamefont {Muhonen}}, \ and\
  \bibinfo {author} {\bibfnamefont {E.}~\bibnamefont {Verhagen}},\ }\href
  {https://doi.org/10.1038/ncomms16024} {\bibfield  {journal} {\bibinfo
  {journal} {Nat. Commun.}\ }\textbf {\bibinfo {volume} {8}},\ \bibinfo {pages}
  {16024} (\bibinfo {year} {2017})}\BibitemShut {NoStop}%
\bibitem [{\citenamefont {Fedorov}\ \emph {et~al.}(2020)\citenamefont
  {Fedorov}, \citenamefont {Beccari}, \citenamefont {Arabmoheghi},
  \citenamefont {Wilson}, \citenamefont {Engelsen},\ and\ \citenamefont
  {Kippenberg}}]{fedorov2020thermal}%
  \BibitemOpen
  \bibfield  {author} {\bibinfo {author} {\bibfnamefont {S.~A.}\ \bibnamefont
  {Fedorov}}, \bibinfo {author} {\bibfnamefont {A.}~\bibnamefont {Beccari}},
  \bibinfo {author} {\bibfnamefont {A.}~\bibnamefont {Arabmoheghi}}, \bibinfo
  {author} {\bibfnamefont {D.~J.}\ \bibnamefont {Wilson}}, \bibinfo {author}
  {\bibfnamefont {N.~J.}\ \bibnamefont {Engelsen}}, \ and\ \bibinfo {author}
  {\bibfnamefont {T.~J.}\ \bibnamefont {Kippenberg}},\ }\href
  {https://doi.org/10.1364/OPTICA.402449} {\bibfield  {journal} {\bibinfo
  {journal} {Optica}\ }\textbf {\bibinfo {volume} {7}},\ \bibinfo {pages}
  {1609} (\bibinfo {year} {2020})}\BibitemShut {NoStop}%
\bibitem [{\citenamefont {Krause}\ \emph {et~al.}(2012)\citenamefont {Krause},
  \citenamefont {Winger}, \citenamefont {Blasius}, \citenamefont {Lin},\ and\
  \citenamefont {Painter}}]{krause2012high}%
  \BibitemOpen
  \bibfield  {author} {\bibinfo {author} {\bibfnamefont {A.~G.}\ \bibnamefont
  {Krause}}, \bibinfo {author} {\bibfnamefont {M.}~\bibnamefont {Winger}},
  \bibinfo {author} {\bibfnamefont {T.~D.}\ \bibnamefont {Blasius}}, \bibinfo
  {author} {\bibfnamefont {Q.}~\bibnamefont {Lin}}, \ and\ \bibinfo {author}
  {\bibfnamefont {O.}~\bibnamefont {Painter}},\ }\href
  {https://doi.org/10.1038/nphoton.2012.245} {\bibfield  {journal} {\bibinfo
  {journal} {Nat. Photonics}\ }\textbf {\bibinfo {volume} {6}},\ \bibinfo
  {pages} {768} (\bibinfo {year} {2012})}\BibitemShut {NoStop}%
\bibitem [{\citenamefont {Guzm{\'a}n~Cervantes}\ \emph
  {et~al.}(2014)\citenamefont {Guzm{\'a}n~Cervantes}, \citenamefont
  {Kumanchik}, \citenamefont {Pratt},\ and\ \citenamefont
  {Taylor}}]{guzman2014high}%
  \BibitemOpen
  \bibfield  {author} {\bibinfo {author} {\bibfnamefont {F.}~\bibnamefont
  {Guzm{\'a}n~Cervantes}}, \bibinfo {author} {\bibfnamefont {L.}~\bibnamefont
  {Kumanchik}}, \bibinfo {author} {\bibfnamefont {J.}~\bibnamefont {Pratt}}, \
  and\ \bibinfo {author} {\bibfnamefont {J.~M.}\ \bibnamefont {Taylor}},\
  }\href {https://doi.org/10.1063/1.4881936} {\bibfield  {journal} {\bibinfo
  {journal} {Appl. Phys. Lett.}\ }\textbf {\bibinfo {volume} {104}},\ \bibinfo
  {pages} {221111} (\bibinfo {year} {2014})}\BibitemShut {NoStop}%
\bibitem [{\citenamefont {Chaste}\ \emph {et~al.}(2012)\citenamefont {Chaste},
  \citenamefont {Eichler}, \citenamefont {Moser}, \citenamefont {Ceballos},
  \citenamefont {Rurali},\ and\ \citenamefont
  {Bachtold}}]{chaste2012nanomechanical}%
  \BibitemOpen
  \bibfield  {author} {\bibinfo {author} {\bibfnamefont {J.}~\bibnamefont
  {Chaste}}, \bibinfo {author} {\bibfnamefont {A.}~\bibnamefont {Eichler}},
  \bibinfo {author} {\bibfnamefont {J.}~\bibnamefont {Moser}}, \bibinfo
  {author} {\bibfnamefont {G.}~\bibnamefont {Ceballos}}, \bibinfo {author}
  {\bibfnamefont {R.}~\bibnamefont {Rurali}}, \ and\ \bibinfo {author}
  {\bibfnamefont {A.}~\bibnamefont {Bachtold}},\ }\href
  {https://doi.org/10.1038/nnano.2012.42} {\bibfield  {journal} {\bibinfo
  {journal} {Nat. Nanotechnol.}\ }\textbf {\bibinfo {volume} {7}},\ \bibinfo
  {pages} {301} (\bibinfo {year} {2012})}\BibitemShut {NoStop}%
\bibitem [{\citenamefont {Ranjit}\ \emph {et~al.}(2016)\citenamefont {Ranjit},
  \citenamefont {Cunningham}, \citenamefont {Casey},\ and\ \citenamefont
  {Geraci}}]{ranjit2016zeptonewton}%
  \BibitemOpen
  \bibfield  {author} {\bibinfo {author} {\bibfnamefont {G.}~\bibnamefont
  {Ranjit}}, \bibinfo {author} {\bibfnamefont {M.}~\bibnamefont {Cunningham}},
  \bibinfo {author} {\bibfnamefont {K.}~\bibnamefont {Casey}}, \ and\ \bibinfo
  {author} {\bibfnamefont {A.~A.}\ \bibnamefont {Geraci}},\ }\href
  {https://doi.org/10.1103/PhysRevA.93.053801} {\bibfield  {journal} {\bibinfo
  {journal} {Phys. Rev. A}\ }\textbf {\bibinfo {volume} {93}},\ \bibinfo
  {pages} {053801} (\bibinfo {year} {2016})}\BibitemShut {NoStop}%
\bibitem [{\citenamefont {Rugar}\ \emph {et~al.}(2004)\citenamefont {Rugar},
  \citenamefont {Budakian}, \citenamefont {Mamin},\ and\ \citenamefont
  {Chui}}]{rugar2004single}%
  \BibitemOpen
  \bibfield  {author} {\bibinfo {author} {\bibfnamefont {D.}~\bibnamefont
  {Rugar}}, \bibinfo {author} {\bibfnamefont {R.}~\bibnamefont {Budakian}},
  \bibinfo {author} {\bibfnamefont {H.~J.}\ \bibnamefont {Mamin}}, \ and\
  \bibinfo {author} {\bibfnamefont {B.~W.}\ \bibnamefont {Chui}},\ }\href
  {https://doi.org/10.1038/nature02658} {\bibfield  {journal} {\bibinfo
  {journal} {Nature}\ }\textbf {\bibinfo {volume} {430}},\ \bibinfo {pages}
  {329} (\bibinfo {year} {2004})}\BibitemShut {NoStop}%
\bibitem [{\citenamefont {Longo}\ \emph {et~al.}(2013)\citenamefont {Longo},
  \citenamefont {Alonso-Sarduy}, \citenamefont {Rio}, \citenamefont {Bizzini},
  \citenamefont {Trampuz}, \citenamefont {Notz}, \citenamefont {Dietler},\ and\
  \citenamefont {Kasas}}]{longo2013rapid}%
  \BibitemOpen
  \bibfield  {author} {\bibinfo {author} {\bibfnamefont {G.}~\bibnamefont
  {Longo}}, \bibinfo {author} {\bibfnamefont {L.}~\bibnamefont
  {Alonso-Sarduy}}, \bibinfo {author} {\bibfnamefont {L.~M.}\ \bibnamefont
  {Rio}}, \bibinfo {author} {\bibfnamefont {A.}~\bibnamefont {Bizzini}},
  \bibinfo {author} {\bibfnamefont {A.}~\bibnamefont {Trampuz}}, \bibinfo
  {author} {\bibfnamefont {J.}~\bibnamefont {Notz}}, \bibinfo {author}
  {\bibfnamefont {G.}~\bibnamefont {Dietler}}, \ and\ \bibinfo {author}
  {\bibfnamefont {S.}~\bibnamefont {Kasas}},\ }\href
  {https://doi.org/10.1038/nnano.2013.120} {\bibfield  {journal} {\bibinfo
  {journal} {Nat. Nanotechnol.}\ }\textbf {\bibinfo {volume} {8}},\ \bibinfo
  {pages} {522} (\bibinfo {year} {2013})}\BibitemShut {NoStop}%
\bibitem [{\citenamefont {Abbott}\ \emph {et~al.}(2016)\citenamefont {Abbott},
  \citenamefont {Abbott}, \citenamefont {Abbott}, \citenamefont {Abernathy},
  \citenamefont {Acernese}, \citenamefont {Ackley}, \citenamefont {Adams},
  \citenamefont {Adams}, \citenamefont {Addesso}, \citenamefont {Adhikari}
  \emph {et~al.}}]{abbott2016observation}%
  \BibitemOpen
  \bibfield  {author} {\bibinfo {author} {\bibfnamefont {B.~P.}\ \bibnamefont
  {Abbott}}, \bibinfo {author} {\bibfnamefont {R.}~\bibnamefont {Abbott}},
  \bibinfo {author} {\bibfnamefont {T.~D.}\ \bibnamefont {Abbott}}, \bibinfo
  {author} {\bibfnamefont {M.~R.}\ \bibnamefont {Abernathy}}, \bibinfo {author}
  {\bibfnamefont {F.}~\bibnamefont {Acernese}}, \bibinfo {author}
  {\bibfnamefont {K.}~\bibnamefont {Ackley}}, \bibinfo {author} {\bibfnamefont
  {C.}~\bibnamefont {Adams}}, \bibinfo {author} {\bibfnamefont
  {T.}~\bibnamefont {Adams}}, \bibinfo {author} {\bibfnamefont
  {P.}~\bibnamefont {Addesso}}, \bibinfo {author} {\bibfnamefont
  {R.}~\bibnamefont {Adhikari}},  \emph {et~al.},\ }\href
  {https://doi.org/10.1103/PhysRevLett.116.061102} {\bibfield  {journal}
  {\bibinfo  {journal} {Phys. Rev. Lett.}\ }\textbf {\bibinfo {volume} {116}},\
  \bibinfo {pages} {061102} (\bibinfo {year} {2016})}\BibitemShut {NoStop}%
\bibitem [{\citenamefont {Moore}\ and\ \citenamefont
  {Geraci}(2021)}]{moore2021searching}%
  \BibitemOpen
  \bibfield  {author} {\bibinfo {author} {\bibfnamefont {D.~C.}\ \bibnamefont
  {Moore}}\ and\ \bibinfo {author} {\bibfnamefont {A.~A.}\ \bibnamefont
  {Geraci}},\ }\href {https://doi.org/10.1088/2058-9565/abcf8a} {\bibfield
  {journal} {\bibinfo  {journal} {Quantum Sci. Technol.}\ }\textbf {\bibinfo
  {volume} {6}},\ \bibinfo {pages} {014008} (\bibinfo {year}
  {2021})}\BibitemShut {NoStop}%
\bibitem [{\citenamefont {Carney}\ \emph {et~al.}(2021)\citenamefont {Carney},
  \citenamefont {Krnjaic}, \citenamefont {Moore}, \citenamefont {Regal},
  \citenamefont {Afek}, \citenamefont {Bhave}, \citenamefont {Brubaker},
  \citenamefont {Corbitt}, \citenamefont {Cripe}, \citenamefont {Crisosto}
  \emph {et~al.}}]{carney2021mechanical}%
  \BibitemOpen
  \bibfield  {author} {\bibinfo {author} {\bibfnamefont {D.}~\bibnamefont
  {Carney}}, \bibinfo {author} {\bibfnamefont {G.}~\bibnamefont {Krnjaic}},
  \bibinfo {author} {\bibfnamefont {D.~C.}\ \bibnamefont {Moore}}, \bibinfo
  {author} {\bibfnamefont {C.~A.}\ \bibnamefont {Regal}}, \bibinfo {author}
  {\bibfnamefont {G.}~\bibnamefont {Afek}}, \bibinfo {author} {\bibfnamefont
  {S.}~\bibnamefont {Bhave}}, \bibinfo {author} {\bibfnamefont
  {B.}~\bibnamefont {Brubaker}}, \bibinfo {author} {\bibfnamefont
  {T.}~\bibnamefont {Corbitt}}, \bibinfo {author} {\bibfnamefont
  {J.}~\bibnamefont {Cripe}}, \bibinfo {author} {\bibfnamefont
  {N.}~\bibnamefont {Crisosto}},  \emph {et~al.},\ }\href
  {https://doi.org/10.1088/2058-9565/abcfcd} {\bibfield  {journal} {\bibinfo
  {journal} {Quantum Sci. Technol.}\ }\textbf {\bibinfo {volume} {6}},\
  \bibinfo {pages} {024002} (\bibinfo {year} {2021})}\BibitemShut {NoStop}%
\bibitem [{\citenamefont {Geraci}\ \emph {et~al.}(2008)\citenamefont {Geraci},
  \citenamefont {Smullin}, \citenamefont {Weld}, \citenamefont {Chiaverini},\
  and\ \citenamefont {Kapitulnik}}]{geraci2008improved}%
  \BibitemOpen
  \bibfield  {author} {\bibinfo {author} {\bibfnamefont {A.~A.}\ \bibnamefont
  {Geraci}}, \bibinfo {author} {\bibfnamefont {S.~J.}\ \bibnamefont {Smullin}},
  \bibinfo {author} {\bibfnamefont {D.~M.}\ \bibnamefont {Weld}}, \bibinfo
  {author} {\bibfnamefont {J.}~\bibnamefont {Chiaverini}}, \ and\ \bibinfo
  {author} {\bibfnamefont {A.}~\bibnamefont {Kapitulnik}},\ }\href
  {https://doi.org/10.1103/PhysRevD.78.022002} {\bibfield  {journal} {\bibinfo
  {journal} {Phys. Rev. D}\ }\textbf {\bibinfo {volume} {78}},\ \bibinfo
  {pages} {022002} (\bibinfo {year} {2008})}\BibitemShut {NoStop}%
\bibitem [{\citenamefont {Braginsky}\ and\ \citenamefont
  {Khalili}(1995)}]{braginsky1995quantum}%
  \BibitemOpen
  \bibfield  {author} {\bibinfo {author} {\bibfnamefont {V.~B.}\ \bibnamefont
  {Braginsky}}\ and\ \bibinfo {author} {\bibfnamefont {F.~Y.}\ \bibnamefont
  {Khalili}},\ }\href@noop {} {\emph {\bibinfo {title} {Quantum
  {M}easurement}}}\ (\bibinfo  {publisher} {Cambridge University Press},\
  \bibinfo {year} {1995})\BibitemShut {NoStop}%
\bibitem [{\citenamefont {Murch}\ \emph {et~al.}(2008)\citenamefont {Murch},
  \citenamefont {Moore}, \citenamefont {Gupta},\ and\ \citenamefont
  {Stamper-Kurn}}]{murch2008observation}%
  \BibitemOpen
  \bibfield  {author} {\bibinfo {author} {\bibfnamefont {K.~W.}\ \bibnamefont
  {Murch}}, \bibinfo {author} {\bibfnamefont {K.~L.}\ \bibnamefont {Moore}},
  \bibinfo {author} {\bibfnamefont {S.}~\bibnamefont {Gupta}}, \ and\ \bibinfo
  {author} {\bibfnamefont {D.~M.}\ \bibnamefont {Stamper-Kurn}},\ }\href
  {https://doi.org/10.1038/nphys965} {\bibfield  {journal} {\bibinfo  {journal}
  {Nat. Phys.}\ }\textbf {\bibinfo {volume} {4}},\ \bibinfo {pages} {561}
  (\bibinfo {year} {2008})}\BibitemShut {NoStop}%
\bibitem [{\citenamefont {Purdy}\ \emph {et~al.}(2013)\citenamefont {Purdy},
  \citenamefont {Peterson},\ and\ \citenamefont
  {Regal}}]{purdy2013observation}%
  \BibitemOpen
  \bibfield  {author} {\bibinfo {author} {\bibfnamefont {T.~P.}\ \bibnamefont
  {Purdy}}, \bibinfo {author} {\bibfnamefont {R.~W.}\ \bibnamefont {Peterson}},
  \ and\ \bibinfo {author} {\bibfnamefont {C.~A.}\ \bibnamefont {Regal}},\
  }\href {https://doi.org/10.1126/science.1231282} {\bibfield  {journal}
  {\bibinfo  {journal} {Science}\ }\textbf {\bibinfo {volume} {339}},\ \bibinfo
  {pages} {801} (\bibinfo {year} {2013})}\BibitemShut {NoStop}%
\bibitem [{\citenamefont {Cripe}\ \emph {et~al.}(2019)\citenamefont {Cripe},
  \citenamefont {Aggarwal}, \citenamefont {Lanza}, \citenamefont {Libson},
  \citenamefont {Singh}, \citenamefont {Heu}, \citenamefont {Follman},
  \citenamefont {Cole}, \citenamefont {Mavalvala},\ and\ \citenamefont
  {Corbitt}}]{cripe2019measurement}%
  \BibitemOpen
  \bibfield  {author} {\bibinfo {author} {\bibfnamefont {J.}~\bibnamefont
  {Cripe}}, \bibinfo {author} {\bibfnamefont {N.}~\bibnamefont {Aggarwal}},
  \bibinfo {author} {\bibfnamefont {R.}~\bibnamefont {Lanza}}, \bibinfo
  {author} {\bibfnamefont {A.}~\bibnamefont {Libson}}, \bibinfo {author}
  {\bibfnamefont {R.}~\bibnamefont {Singh}}, \bibinfo {author} {\bibfnamefont
  {P.}~\bibnamefont {Heu}}, \bibinfo {author} {\bibfnamefont {D.}~\bibnamefont
  {Follman}}, \bibinfo {author} {\bibfnamefont {G.~D.}\ \bibnamefont {Cole}},
  \bibinfo {author} {\bibfnamefont {N.}~\bibnamefont {Mavalvala}}, \ and\
  \bibinfo {author} {\bibfnamefont {T.}~\bibnamefont {Corbitt}},\ }\href
  {https://doi.org/10.1038/s41586-019-1051-4} {\bibfield  {journal} {\bibinfo
  {journal} {Nature}\ }\textbf {\bibinfo {volume} {568}},\ \bibinfo {pages}
  {364} (\bibinfo {year} {2019})}\BibitemShut {NoStop}%
\bibitem [{\citenamefont {Doherty}\ and\ \citenamefont
  {Jacobs}(1999)}]{doherty1999feedback}%
  \BibitemOpen
  \bibfield  {author} {\bibinfo {author} {\bibfnamefont {A.~C.}\ \bibnamefont
  {Doherty}}\ and\ \bibinfo {author} {\bibfnamefont {K.}~\bibnamefont
  {Jacobs}},\ }\href {https://doi.org/10.1103/PhysRevA.60.2700} {\bibfield
  {journal} {\bibinfo  {journal} {Phys. Rev. A}\ }\textbf {\bibinfo {volume}
  {60}},\ \bibinfo {pages} {2700} (\bibinfo {year} {1999})}\BibitemShut
  {NoStop}%
\bibitem [{\citenamefont {Mancini}\ \emph {et~al.}(1998)\citenamefont
  {Mancini}, \citenamefont {Vitali},\ and\ \citenamefont
  {Tombesi}}]{mancini1998optomechanical}%
  \BibitemOpen
  \bibfield  {author} {\bibinfo {author} {\bibfnamefont {S.}~\bibnamefont
  {Mancini}}, \bibinfo {author} {\bibfnamefont {D.}~\bibnamefont {Vitali}}, \
  and\ \bibinfo {author} {\bibfnamefont {P.}~\bibnamefont {Tombesi}},\ }\href
  {https://doi.org/10.1103/PhysRevLett.80.688} {\bibfield  {journal} {\bibinfo
  {journal} {Phys. Rev. Lett.}\ }\textbf {\bibinfo {volume} {80}},\ \bibinfo
  {pages} {688} (\bibinfo {year} {1998})}\BibitemShut {NoStop}%
\bibitem [{\citenamefont {Cohadon}\ \emph {et~al.}(1999)\citenamefont
  {Cohadon}, \citenamefont {Heidmann},\ and\ \citenamefont
  {Pinard}}]{cohadon1999cooling}%
  \BibitemOpen
  \bibfield  {author} {\bibinfo {author} {\bibfnamefont {P.~F.}\ \bibnamefont
  {Cohadon}}, \bibinfo {author} {\bibfnamefont {A.}~\bibnamefont {Heidmann}}, \
  and\ \bibinfo {author} {\bibfnamefont {M.}~\bibnamefont {Pinard}},\ }\href
  {https://doi.org/10.1103/PhysRevLett.83.3174} {\bibfield  {journal} {\bibinfo
   {journal} {Phys. Rev. Lett.}\ }\textbf {\bibinfo {volume} {83}},\ \bibinfo
  {pages} {3174} (\bibinfo {year} {1999})}\BibitemShut {NoStop}%
\bibitem [{\citenamefont {Wilson}\ \emph {et~al.}(2015)\citenamefont {Wilson},
  \citenamefont {Sudhir}, \citenamefont {Piro}, \citenamefont {Schilling},
  \citenamefont {Ghadimi},\ and\ \citenamefont
  {Kippenberg}}]{wilson2015measurement}%
  \BibitemOpen
  \bibfield  {author} {\bibinfo {author} {\bibfnamefont {D.~J.}\ \bibnamefont
  {Wilson}}, \bibinfo {author} {\bibfnamefont {V.}~\bibnamefont {Sudhir}},
  \bibinfo {author} {\bibfnamefont {N.}~\bibnamefont {Piro}}, \bibinfo {author}
  {\bibfnamefont {R.}~\bibnamefont {Schilling}}, \bibinfo {author}
  {\bibfnamefont {A.}~\bibnamefont {Ghadimi}}, \ and\ \bibinfo {author}
  {\bibfnamefont {T.~J.}\ \bibnamefont {Kippenberg}},\ }\href
  {https://doi.org/10.1038/nature14672} {\bibfield  {journal} {\bibinfo
  {journal} {Nature}\ }\textbf {\bibinfo {volume} {524}},\ \bibinfo {pages}
  {325} (\bibinfo {year} {2015})}\BibitemShut {NoStop}%
\bibitem [{\citenamefont {Rossi}\ \emph {et~al.}(2018)\citenamefont {Rossi},
  \citenamefont {Mason}, \citenamefont {Chen}, \citenamefont {Tsaturyan},\ and\
  \citenamefont {Schliesser}}]{rossi2018measurement}%
  \BibitemOpen
  \bibfield  {author} {\bibinfo {author} {\bibfnamefont {M.}~\bibnamefont
  {Rossi}}, \bibinfo {author} {\bibfnamefont {D.}~\bibnamefont {Mason}},
  \bibinfo {author} {\bibfnamefont {J.}~\bibnamefont {Chen}}, \bibinfo {author}
  {\bibfnamefont {Y.}~\bibnamefont {Tsaturyan}}, \ and\ \bibinfo {author}
  {\bibfnamefont {A.}~\bibnamefont {Schliesser}},\ }\href
  {https://doi.org/10.1038/s41586-018-0643-8} {\bibfield  {journal} {\bibinfo
  {journal} {Nature}\ }\textbf {\bibinfo {volume} {563}},\ \bibinfo {pages}
  {53} (\bibinfo {year} {2018})}\BibitemShut {NoStop}%
\bibitem [{\citenamefont {Braginsky}\ \emph {et~al.}(1980)\citenamefont
  {Braginsky}, \citenamefont {Vorontsov},\ and\ \citenamefont
  {Thorne}}]{braginsky1980quantum}%
  \BibitemOpen
  \bibfield  {author} {\bibinfo {author} {\bibfnamefont {V.~B.}\ \bibnamefont
  {Braginsky}}, \bibinfo {author} {\bibfnamefont {Y.~I.}\ \bibnamefont
  {Vorontsov}}, \ and\ \bibinfo {author} {\bibfnamefont {K.~S.}\ \bibnamefont
  {Thorne}},\ }\href {https://doi.org/10.1126/science.209.4456.547} {\bibfield
  {journal} {\bibinfo  {journal} {Science}\ }\textbf {\bibinfo {volume}
  {209}},\ \bibinfo {pages} {547} (\bibinfo {year} {1980})}\BibitemShut
  {NoStop}%
\bibitem [{\citenamefont {Clerk}\ \emph {et~al.}(2008)\citenamefont {Clerk},
  \citenamefont {Marquardt},\ and\ \citenamefont {Jacobs}}]{clerk2008back}%
  \BibitemOpen
  \bibfield  {author} {\bibinfo {author} {\bibfnamefont {A.~A.}\ \bibnamefont
  {Clerk}}, \bibinfo {author} {\bibfnamefont {F.}~\bibnamefont {Marquardt}}, \
  and\ \bibinfo {author} {\bibfnamefont {K.}~\bibnamefont {Jacobs}},\ }\href
  {https://doi.org/10.1088/1367-2630/10/9/095010} {\bibfield  {journal}
  {\bibinfo  {journal} {New J. Phys.}\ }\textbf {\bibinfo {volume} {10}},\
  \bibinfo {pages} {095010} (\bibinfo {year} {2008})}\BibitemShut {NoStop}%
\bibitem [{\citenamefont {Suh}\ \emph {et~al.}(2014)\citenamefont {Suh},
  \citenamefont {Weinstein}, \citenamefont {Lei}, \citenamefont {Wollman},
  \citenamefont {Steinke}, \citenamefont {Meystre}, \citenamefont {Clerk},\
  and\ \citenamefont {Schwab}}]{suh2014mechanically}%
  \BibitemOpen
  \bibfield  {author} {\bibinfo {author} {\bibfnamefont {J.}~\bibnamefont
  {Suh}}, \bibinfo {author} {\bibfnamefont {A.~J.}\ \bibnamefont {Weinstein}},
  \bibinfo {author} {\bibfnamefont {C.~U.}\ \bibnamefont {Lei}}, \bibinfo
  {author} {\bibfnamefont {E.~E.}\ \bibnamefont {Wollman}}, \bibinfo {author}
  {\bibfnamefont {S.~K.}\ \bibnamefont {Steinke}}, \bibinfo {author}
  {\bibfnamefont {P.}~\bibnamefont {Meystre}}, \bibinfo {author} {\bibfnamefont
  {A.~A.}\ \bibnamefont {Clerk}}, \ and\ \bibinfo {author} {\bibfnamefont
  {K.~C.}\ \bibnamefont {Schwab}},\ }\href
  {https://doi.org/10.1126/science.1253258} {\bibfield  {journal} {\bibinfo
  {journal} {Science}\ }\textbf {\bibinfo {volume} {344}},\ \bibinfo {pages}
  {1262} (\bibinfo {year} {2014})}\BibitemShut {NoStop}%
\bibitem [{\citenamefont {Shomroni}\ \emph {et~al.}(2019)\citenamefont
  {Shomroni}, \citenamefont {Qiu}, \citenamefont {Malz}, \citenamefont
  {Nunnenkamp},\ and\ \citenamefont {Kippenberg}}]{shomroni2019optical}%
  \BibitemOpen
  \bibfield  {author} {\bibinfo {author} {\bibfnamefont {I.}~\bibnamefont
  {Shomroni}}, \bibinfo {author} {\bibfnamefont {L.}~\bibnamefont {Qiu}},
  \bibinfo {author} {\bibfnamefont {D.}~\bibnamefont {Malz}}, \bibinfo {author}
  {\bibfnamefont {A.}~\bibnamefont {Nunnenkamp}}, \ and\ \bibinfo {author}
  {\bibfnamefont {T.~J.}\ \bibnamefont {Kippenberg}},\ }\href
  {https://doi.org/10.1038/s41467-019-10024-3} {\bibfield  {journal} {\bibinfo
  {journal} {Nat. Commun.}\ }\textbf {\bibinfo {volume} {10}},\ \bibinfo
  {pages} {2086} (\bibinfo {year} {2019})}\BibitemShut {NoStop}%
\bibitem [{\citenamefont {Braginsky}\ \emph {et~al.}(1978)\citenamefont
  {Braginsky}, \citenamefont {Vorontsov},\ and\ \citenamefont
  {Khalili}}]{braginskii1978optimal}%
  \BibitemOpen
  \bibfield  {author} {\bibinfo {author} {\bibfnamefont {V.~B.}\ \bibnamefont
  {Braginsky}}, \bibinfo {author} {\bibfnamefont {Y.~I.}\ \bibnamefont
  {Vorontsov}}, \ and\ \bibinfo {author} {\bibfnamefont {F.~Y.}\ \bibnamefont
  {Khalili}},\ }\href {http://persianka.viperson.ru/data/200908/1.pdf}
  {\bibfield  {journal} {\bibinfo  {journal} {JETP Lett.}\ }\textbf {\bibinfo
  {volume} {27}} (\bibinfo {year} {1978})}\BibitemShut {NoStop}%
\bibitem [{\citenamefont {Vanner}\ \emph {et~al.}(2011)\citenamefont {Vanner},
  \citenamefont {Pikovski}, \citenamefont {Cole}, \citenamefont {Kim},
  \citenamefont {Brukner}, \citenamefont {Hammerer}, \citenamefont {Milburn},\
  and\ \citenamefont {Aspelmeyer}}]{vanner2011pulsed}%
  \BibitemOpen
  \bibfield  {author} {\bibinfo {author} {\bibfnamefont {M.~R.}\ \bibnamefont
  {Vanner}}, \bibinfo {author} {\bibfnamefont {I.}~\bibnamefont {Pikovski}},
  \bibinfo {author} {\bibfnamefont {G.~D.}\ \bibnamefont {Cole}}, \bibinfo
  {author} {\bibfnamefont {M.~S.}\ \bibnamefont {Kim}}, \bibinfo {author}
  {\bibfnamefont {{\v{C}}.}~\bibnamefont {Brukner}}, \bibinfo {author}
  {\bibfnamefont {K.}~\bibnamefont {Hammerer}}, \bibinfo {author}
  {\bibfnamefont {G.~J.}\ \bibnamefont {Milburn}}, \ and\ \bibinfo {author}
  {\bibfnamefont {M.}~\bibnamefont {Aspelmeyer}},\ }\href
  {https://doi.org/10.1073/pnas.1105098108} {\bibfield  {journal} {\bibinfo
  {journal} {Proc. Natl. Acad. Sci.}\ }\textbf {\bibinfo {volume} {108}},\
  \bibinfo {pages} {16182} (\bibinfo {year} {2011})}\BibitemShut {NoStop}%
\bibitem [{\citenamefont {Vanner}\ \emph {et~al.}(2013)\citenamefont {Vanner},
  \citenamefont {Hofer}, \citenamefont {Cole},\ and\ \citenamefont
  {Aspelmeyer}}]{vanner2013cooling}%
  \BibitemOpen
  \bibfield  {author} {\bibinfo {author} {\bibfnamefont {M.~R.}\ \bibnamefont
  {Vanner}}, \bibinfo {author} {\bibfnamefont {J.}~\bibnamefont {Hofer}},
  \bibinfo {author} {\bibfnamefont {G.~D.}\ \bibnamefont {Cole}}, \ and\
  \bibinfo {author} {\bibfnamefont {M.}~\bibnamefont {Aspelmeyer}},\ }\href
  {https://doi.org/10.1038/ncomms3295} {\bibfield  {journal} {\bibinfo
  {journal} {Nat. Commun.}\ }\textbf {\bibinfo {volume} {4}},\ \bibinfo {pages}
  {2295} (\bibinfo {year} {2013})}\BibitemShut {NoStop}%
\bibitem [{\citenamefont {Muhonen}\ \emph {et~al.}(2019)\citenamefont
  {Muhonen}, \citenamefont {La~Gala}, \citenamefont {Leijssen},\ and\
  \citenamefont {Verhagen}}]{muhonen2019state}%
  \BibitemOpen
  \bibfield  {author} {\bibinfo {author} {\bibfnamefont {J.~T.}\ \bibnamefont
  {Muhonen}}, \bibinfo {author} {\bibfnamefont {G.~R.}\ \bibnamefont
  {La~Gala}}, \bibinfo {author} {\bibfnamefont {R.}~\bibnamefont {Leijssen}}, \
  and\ \bibinfo {author} {\bibfnamefont {E.}~\bibnamefont {Verhagen}},\ }\href
  {https://doi.org/10.1103/PhysRevLett.123.113601} {\bibfield  {journal}
  {\bibinfo  {journal} {Phys. Rev. Lett.}\ }\textbf {\bibinfo {volume} {123}},\
  \bibinfo {pages} {113601} (\bibinfo {year} {2019})}\BibitemShut {NoStop}%
\bibitem [{\citenamefont {Walker}\ and\ \citenamefont
  {Carroll}(1986)}]{walker1986multiport}%
  \BibitemOpen
  \bibfield  {author} {\bibinfo {author} {\bibfnamefont {N.}~\bibnamefont
  {Walker}}\ and\ \bibinfo {author} {\bibfnamefont {J.~E.}\ \bibnamefont
  {Carroll}},\ }\href {https://doi.org/10.1007/BF02032562} {\bibfield
  {journal} {\bibinfo  {journal} {Opt. Quantum Electron.}\ }\textbf {\bibinfo
  {volume} {18}},\ \bibinfo {pages} {355} (\bibinfo {year} {1986})}\BibitemShut
  {NoStop}%
\bibitem [{\citenamefont {Furusawa}\ \emph {et~al.}(1998)\citenamefont
  {Furusawa}, \citenamefont {S{\o}rensen}, \citenamefont {Braunstein},
  \citenamefont {Fuchs}, \citenamefont {Kimble},\ and\ \citenamefont
  {Polzik}}]{furusawa1998unconditional}%
  \BibitemOpen
  \bibfield  {author} {\bibinfo {author} {\bibfnamefont {A.}~\bibnamefont
  {Furusawa}}, \bibinfo {author} {\bibfnamefont {J.~L.}\ \bibnamefont
  {S{\o}rensen}}, \bibinfo {author} {\bibfnamefont {S.~L.}\ \bibnamefont
  {Braunstein}}, \bibinfo {author} {\bibfnamefont {C.~A.}\ \bibnamefont
  {Fuchs}}, \bibinfo {author} {\bibfnamefont {H.~J.}\ \bibnamefont {Kimble}}, \
  and\ \bibinfo {author} {\bibfnamefont {E.~S.}\ \bibnamefont {Polzik}},\
  }\href {https://doi.org/10.1126/science.282.5389.706} {\bibfield  {journal}
  {\bibinfo  {journal} {Science}\ }\textbf {\bibinfo {volume} {282}},\ \bibinfo
  {pages} {706} (\bibinfo {year} {1998})}\BibitemShut {NoStop}%
\bibitem [{\citenamefont {Vidrighin}\ \emph {et~al.}(2014)\citenamefont
  {Vidrighin}, \citenamefont {Donati}, \citenamefont {Genoni}, \citenamefont
  {Jin}, \citenamefont {Kolthammer}, \citenamefont {Kim}, \citenamefont
  {Datta}, \citenamefont {Barbieri},\ and\ \citenamefont
  {Walmsley}}]{vidrighin2014joint}%
  \BibitemOpen
  \bibfield  {author} {\bibinfo {author} {\bibfnamefont {M.~D.}\ \bibnamefont
  {Vidrighin}}, \bibinfo {author} {\bibfnamefont {G.}~\bibnamefont {Donati}},
  \bibinfo {author} {\bibfnamefont {M.~G.}\ \bibnamefont {Genoni}}, \bibinfo
  {author} {\bibfnamefont {X.-M.}\ \bibnamefont {Jin}}, \bibinfo {author}
  {\bibfnamefont {W.~S.}\ \bibnamefont {Kolthammer}}, \bibinfo {author}
  {\bibfnamefont {M.~S.}\ \bibnamefont {Kim}}, \bibinfo {author} {\bibfnamefont
  {A.}~\bibnamefont {Datta}}, \bibinfo {author} {\bibfnamefont
  {M.}~\bibnamefont {Barbieri}}, \ and\ \bibinfo {author} {\bibfnamefont
  {I.~A.}\ \bibnamefont {Walmsley}},\ }\href
  {https://doi.org/10.1038/ncomms4532} {\bibfield  {journal} {\bibinfo
  {journal} {Nat. Commun.}\ }\textbf {\bibinfo {volume} {5}},\ \bibinfo {pages}
  {3532} (\bibinfo {year} {2014})}\BibitemShut {NoStop}%
\bibitem [{sup()}]{supp}%
  \BibitemOpen
  \href@noop {} {}\bibinfo {note} {See the Supplemental Material for further
  theoretical details and an animation of the optical Husimi-$Q$
  function.}\BibitemShut {Stop}%
\bibitem [{\citenamefont {Pikovski}\ \emph {et~al.}(2012)\citenamefont
  {Pikovski}, \citenamefont {Vanner}, \citenamefont {Aspelmeyer}, \citenamefont
  {Kim},\ and\ \citenamefont {Brukner}}]{pikovski2012probing}%
  \BibitemOpen
  \bibfield  {author} {\bibinfo {author} {\bibfnamefont {I.}~\bibnamefont
  {Pikovski}}, \bibinfo {author} {\bibfnamefont {M.~R.}\ \bibnamefont
  {Vanner}}, \bibinfo {author} {\bibfnamefont {M.}~\bibnamefont {Aspelmeyer}},
  \bibinfo {author} {\bibfnamefont {M.~S.}\ \bibnamefont {Kim}}, \ and\
  \bibinfo {author} {\bibfnamefont {{\v{C}}.}~\bibnamefont {Brukner}},\ }\href
  {https://doi.org/10.1038/nphys2262} {\bibfield  {journal} {\bibinfo
  {journal} {Nat. Phys.}\ }\textbf {\bibinfo {volume} {8}},\ \bibinfo {pages}
  {393} (\bibinfo {year} {2012})}\BibitemShut {NoStop}%
\bibitem [{\citenamefont {Wang}\ and\ \citenamefont
  {Safavi-Naeini}(2017)}]{wang2017enhancing}%
  \BibitemOpen
  \bibfield  {author} {\bibinfo {author} {\bibfnamefont {Z.}~\bibnamefont
  {Wang}}\ and\ \bibinfo {author} {\bibfnamefont {A.~H.}\ \bibnamefont
  {Safavi-Naeini}},\ }\href {https://doi.org/10.1038/ncomms15886} {\bibfield
  {journal} {\bibinfo  {journal} {Nat. Commun.}\ }\textbf {\bibinfo {volume}
  {8}},\ \bibinfo {pages} {15886} (\bibinfo {year} {2017})}\BibitemShut
  {NoStop}%
\bibitem [{\citenamefont {Leonhardt}(1997)}]{leonhardt1997measuring}%
  \BibitemOpen
  \bibfield  {author} {\bibinfo {author} {\bibfnamefont {U.}~\bibnamefont
  {Leonhardt}},\ }\href@noop {} {\emph {\bibinfo {title} {Measuring the
  {Q}uantum {S}tate of {L}ight}}}\ (\bibinfo  {publisher} {Cambridge
  {U}niversity {P}ress},\ \bibinfo {year} {1997})\BibitemShut {NoStop}%
\bibitem [{foo()}]{footnoteWindow}%
  \BibitemOpen
  \href@noop {} {}\bibinfo {note} {Note that an outcome window may be utilized
  to further reduce the variance at the cost of introducing a finite
  window-dependent heralding probability.}\BibitemShut {Stop}%
\bibitem [{\citenamefont {Schlosshauer}(2007)}]{schlosshauer2007decoherence}%
  \BibitemOpen
  \bibfield  {author} {\bibinfo {author} {\bibfnamefont {M.~A.}\ \bibnamefont
  {Schlosshauer}},\ }\href@noop {} {\emph {\bibinfo {title} {Decoherence: and
  the {Q}uantum-to-{C}lassical {T}ransition}}}\ (\bibinfo  {publisher}
  {Springer Science \& Business Media},\ \bibinfo {year} {2007})\BibitemShut
  {NoStop}%
\bibitem [{\citenamefont {Jacobs}\ and\ \citenamefont
  {Steck}(2006)}]{jacobs2006straightforward}%
  \BibitemOpen
  \bibfield  {author} {\bibinfo {author} {\bibfnamefont {K.}~\bibnamefont
  {Jacobs}}\ and\ \bibinfo {author} {\bibfnamefont {D.~A.}\ \bibnamefont
  {Steck}},\ }\href {https://doi.org/10.1080/00107510601101934} {\bibfield
  {journal} {\bibinfo  {journal} {Contemp. Phys.}\ }\textbf {\bibinfo {volume}
  {47}},\ \bibinfo {pages} {279} (\bibinfo {year} {2006})}\BibitemShut
  {NoStop}%
\bibitem [{\citenamefont {Jacobs}(2014)}]{jacobs2014quantum}%
  \BibitemOpen
  \bibfield  {author} {\bibinfo {author} {\bibfnamefont {K.}~\bibnamefont
  {Jacobs}},\ }\href@noop {} {\emph {\bibinfo {title} {Quantum measurement
  theory and its applications}}}\ (\bibinfo  {publisher} {Cambridge University
  Press},\ \bibinfo {year} {2014})\BibitemShut {NoStop}%
\bibitem [{\citenamefont {Wiseman}\ and\ \citenamefont
  {Milburn}(2009)}]{wiseman2009quantum}%
  \BibitemOpen
  \bibfield  {author} {\bibinfo {author} {\bibfnamefont {H.~M.}\ \bibnamefont
  {Wiseman}}\ and\ \bibinfo {author} {\bibfnamefont {G.~J.}\ \bibnamefont
  {Milburn}},\ }\href@noop {} {\emph {\bibinfo {title} {Quantum {M}easurement
  and {C}ontrol}}}\ (\bibinfo  {publisher} {Cambridge {U}niversity {P}ress},\
  \bibinfo {year} {2009})\BibitemShut {NoStop}%
\bibitem [{\citenamefont {Caves}\ and\ \citenamefont
  {Milburn}(1987)}]{caves1987quantum}%
  \BibitemOpen
  \bibfield  {author} {\bibinfo {author} {\bibfnamefont {C.~M.}\ \bibnamefont
  {Caves}}\ and\ \bibinfo {author} {\bibfnamefont {G.~J.}\ \bibnamefont
  {Milburn}},\ }\href {https://doi.org/10.1103/PhysRevA.36.5543} {\bibfield
  {journal} {\bibinfo  {journal} {Phys. Rev. A}\ }\textbf {\bibinfo {volume}
  {36}},\ \bibinfo {pages} {5543} (\bibinfo {year} {1987})}\BibitemShut
  {NoStop}%
\bibitem [{\citenamefont {Rossi}\ \emph {et~al.}(2019)\citenamefont {Rossi},
  \citenamefont {Mason}, \citenamefont {Chen},\ and\ \citenamefont
  {Schliesser}}]{rossi2019observing}%
  \BibitemOpen
  \bibfield  {author} {\bibinfo {author} {\bibfnamefont {M.}~\bibnamefont
  {Rossi}}, \bibinfo {author} {\bibfnamefont {D.}~\bibnamefont {Mason}},
  \bibinfo {author} {\bibfnamefont {J.}~\bibnamefont {Chen}}, \ and\ \bibinfo
  {author} {\bibfnamefont {A.}~\bibnamefont {Schliesser}},\ }\href
  {https://doi.org/10.1103/PhysRevLett.123.163601} {\bibfield  {journal}
  {\bibinfo  {journal} {Phys. Rev. Lett.}\ }\textbf {\bibinfo {volume} {123}},\
  \bibinfo {pages} {163601} (\bibinfo {year} {2019})}\BibitemShut {NoStop}%
\bibitem [{\citenamefont {Meng}\ \emph {et~al.}(2020)\citenamefont {Meng},
  \citenamefont {Brawley}, \citenamefont {Bennett}, \citenamefont {Vanner},\
  and\ \citenamefont {Bowen}}]{meng2020mechanical}%
  \BibitemOpen
  \bibfield  {author} {\bibinfo {author} {\bibfnamefont {C.}~\bibnamefont
  {Meng}}, \bibinfo {author} {\bibfnamefont {G.~A.}\ \bibnamefont {Brawley}},
  \bibinfo {author} {\bibfnamefont {J.~S.}\ \bibnamefont {Bennett}}, \bibinfo
  {author} {\bibfnamefont {M.~R.}\ \bibnamefont {Vanner}}, \ and\ \bibinfo
  {author} {\bibfnamefont {W.~P.}\ \bibnamefont {Bowen}},\ }\href
  {https://doi.org/10.1103/PhysRevLett.125.043604} {\bibfield  {journal}
  {\bibinfo  {journal} {Phys. Rev. Lett.}\ }\textbf {\bibinfo {volume} {125}},\
  \bibinfo {pages} {043604} (\bibinfo {year} {2020})}\BibitemShut {NoStop}%
\bibitem [{\citenamefont {Zhang}\ and\ \citenamefont
  {M{\o}lmer}(2017)}]{zhang2017prediction}%
  \BibitemOpen
  \bibfield  {author} {\bibinfo {author} {\bibfnamefont {J.}~\bibnamefont
  {Zhang}}\ and\ \bibinfo {author} {\bibfnamefont {K.}~\bibnamefont
  {M{\o}lmer}},\ }\href {https://doi.org/10.1103/PhysRevA.96.062131} {\bibfield
   {journal} {\bibinfo  {journal} {Phys. Rev. A}\ }\textbf {\bibinfo {volume}
  {96}},\ \bibinfo {pages} {062131} (\bibinfo {year} {2017})}\BibitemShut
  {NoStop}%
\bibitem [{\citenamefont {Jacobs}(2010)}]{jacobs2010stochastic}%
  \BibitemOpen
  \bibfield  {author} {\bibinfo {author} {\bibfnamefont {K.}~\bibnamefont
  {Jacobs}},\ }\href@noop {} {\emph {\bibinfo {title} {Stochastic {P}rocesses
  for {P}hysicists: {U}nderstanding {N}oisy {S}ystems}}}\ (\bibinfo
  {publisher} {Cambridge University Press},\ \bibinfo {year}
  {2010})\BibitemShut {NoStop}%
\end{thebibliography}
\end{document}